\documentclass[%
 reprint,
unsortedaddress,
nofootinbib,
 amsmath,amssymb,
 pre,
nobalancelastpage,
]{revtex4-2}
\usepackage{enumitem}
\usepackage{graphicx}
\usepackage{dcolumn}
\usepackage{bm}
\usepackage{hyperref}
\hypersetup{
    colorlinks=true,
    linkcolor=blue,
    filecolor=magenta,
    urlcolor=cyan,
}

\usepackage[UKenglish]{babel}
\usepackage[utf8]{inputenc}
\usepackage{mathtools}
\usepackage{amsfonts}
\usepackage{dcolumn}
\usepackage{booktabs}
\usepackage{cleveref}

\newcommand{\Figref}[1]{{Fig.~\ref{#1}}}
\newcommand{\Eqnref}[1]{{Eq.~\ref{#1}}}
\newcommand{\Tblref}[1]{{Table~\ref{#1}}}

\usepackage[position=bottom]{subfig}
\begin{document}


\title{Map Equation Centrality: Community-aware Centrality based on the Map Equation}

\author{Christopher Blöcker}
 \email{christopher.blocker@umu.se}
\affiliation{%
 Integrated Science Lab, Department of Physics, Ume{\aa} University, SE-901 87 Ume{\aa}, Sweden
}%
\author{Juan Carlos Nieves}%
 \email{juan.carlos.nieves@umu.se}
\affiliation{%
 Department of Computing Science, Ume{\aa} University, SE-901 87 Ume{\aa}, Sweden
}%
\author{Martin Rosvall}%
 \email{martin.rosvall@umu.se}
\affiliation{%
 Integrated Science Lab, Department of Physics, Ume{\aa} University, SE-901 87 Ume{\aa}, Sweden
}%




\date{\today}

\begin{abstract}
To measure node importance, network scientists employ centrality scores that typically take a microscopic or macroscopic perspective, relying on node features or global network structure.
However, traditional centrality measures such as degree centrality, betweenness centrality, or PageRank neglect the community structure found in real-world networks.
To study node importance based on network flows from a mesoscopic perspective, we analytically derive a community-aware information-theoretic centrality score based on network flow and the coding principles behind the map equation: map equation centrality.
Map equation centrality measures how much further we can compress the network's modular description by not coding for random walker transitions to the respective node, using an adapted coding scheme and determining node importance from a network flow-based point of view.
The information-theoretic centrality measure can be determined from a node's local network context alone because changes to the coding scheme only affect other nodes in the same module.
Map equation centrality is agnostic to the chosen network flow model and allows researchers to select the model that best reflects the dynamics of the process under study.
Applied to synthetic networks, we highlight how our approach enables a more fine-grained differentiation between nodes than node-local or network-global measures.
Predicting influential nodes for two different dynamical processes on real-world networks with traditional and other community-aware centrality measures, we find that activating nodes based on map equation centrality scores tends to create the largest cascades in a linear threshold model.
\end{abstract}

\maketitle


\section{Introduction}
\label{sec:introduction}
Networks are simple yet powerful representations of how things connect: the world wide web captures connections between websites, and social networks describe relationships between persons.
So-called centrality measures determine node importance and enable us to rank nodes, compare them with each other, and find the most important ones.
Real-world applications are manifold and include identifying the most popular websites, which components in an infrastructure network have the most impact when they fail, and who drives disease spreading in a social network.

Classical centrality measures consider node importance on a microscopic scale at the node level or on a macroscopic scale at the network level.
For example, degree centrality defines a node's importance proportional to its degree, and betweenness centrality calculates node importance as the number of shortest paths that pass through it \cite{Koschutzki2005}.
Eigenvector centrality-based measures, such as Katz centrality \cite{katz1953new} and PageRank \cite{pagerank-gleich}, implement a reputation system and derive a node's importance from how important its neighbours are, leading to a system of recursive equations.
However, real-world networks often exhibit community structure.
Loosely speaking, they contain groups of nodes, so-called communities, with more connections within groups than between.
But precise definitions of what constitutes a community differ depending on context and assumptions, resulting in a manifold of justifiable characterisations \cite{FORTUNATO201075}.
Classical centrality measures neglect the mesoscopic scale of communities and can often not distinguish between nodes with the same features or nodes embedded in similar network regions.
For example, degree centrality assigns the same score to same-degree nodes, and PageRank cannot distinguish between nodes receiving the same amount of support.

To address this issue, network scientists have developed community-aware centrality scores that typically define node importance in terms of intra-community and inter-community link patterns.
They commonly evaluate their effectiveness in a disease spreading setting where the objective is to contain an epidemic by immunising a limited fraction of the population \cite{Cherifi2019, masuda2009immunization, ghalmane2019immunization, community-aware-centrality-evaluation}.
For example, community-based betweenness centrality considers only shortest paths with endpoints in different communities \cite{kitromilidis2018community}.
Community hub-bridge calculates a node's importance as the sum of its intra-community and inter-community links, weighted by the size of the node's community and the number of communities it connects to, respectively \cite{ghalmane2019immunization}.
Community-based centrality determines a node's importance as the number of connections it has to other communities, weighted by the communities' relative sizes \cite{zhao2015community}.
Masuda proposed a measure based on eigenvector centrality that quantifies a node's centrality in terms of its contribution to the connectivity between modules, giving higher importance to those nodes that, if removed, would fragment the network more \cite{masuda2009immunization}.
Modular centrality defines a generic framework that operates on top of classical centrality measures to retrofit them with community awareness.
It decomposes the network into local, intra-community, and global, inter-community parts and represents a node's centrality as a combination of a local and global component \cite{ghalmane2019centrality}.
In networks with overlapping community structures, nodes that belong to several communities may have high influence despite having a low degree because they act as bridges between communities \cite{10.1145/3184558.3191566}.
For epidemic settings, a node's number of community memberships, sometimes called membership centrality, is typically at least as good an estimator of influence as global centrality measures \cite{Hebert-Dufresne2013}.
Assuming a network's overlapping community structure is known, random walk-based approaches can be employed to extract high-degree nodes from overlapping regions \cite{TAGHAVIAN2017148}.
Overlapping modular centrality generalises modular centrality and takes into account the possibly multiple community memberships that nodes have, resulting in increased influence in the local parts of a network \cite{Ghalmane2019}.
Recently, modularity vitality has been proposed \cite{modularity-vitality} based on the community-detection approach known as modularity \cite{PhysRevE.69.026113} and generalised to overlapping communities \cite{10.1145/3487351.3488277}.

We focus on non-overlapping community structure and derive a centrality score from the information-theoretic community-detection method known as the map equation \cite{Rosvall1118} analytically: \emph{map equation centrality}.
Deriving community-aware centrality scores from a community-detection approach provides more clarity and precision because the resulting measures adopt the same assumptions regarding what constitutes communities as the underlying community-detection approach.
The map equation framework uses random walks to model network flows and identifies communities as those network regions where a random walker tends to spend a long time before switching to a different region. Therefore, map equation centrality determines node importance from a flow-based perspective.
Using a toy example, we highlight how map equation centrality exploits community structure to distinguish between nodes where classical centrality measures fail.
To understand how map equation centrality is affected by randomness in the link patterns of a network, we generate an Lancichinetti–Fortunato–Radicchi (LFR) network with planted community structure, rewire different fractions of the links, and compare the resulting community structures and node centralities with the ground truth.
To evaluate the performance of map equation centrality, we apply it to twelve empirical networks to identify influential nodes.
Like in previous work on centrality scores, we contrast our predictions with the spreading power of nodes obtained from simulations of a Susceptible-Infected-Recovered (SIR) disease-spreading model \cite{ghalmane2019immunization,community-aware-centrality-evaluation} and the adoptions of ideas modelled by the linear threshold model \cite{10.1007/978-3-030-93409-5_29}.
For comparison, we include degree centrality as a local measure, betweenness centrality as a global measure, as well as three other community-aware centrality measures in our evaluation.
We find that map equation centrality performs amongst the best in half of the networks in the SIR setting and tends to outperform the baseline measures in the linear threshold setting.

\section{The map equation framework}
\begin{figure*}
    \centering
    \subfloat[\label{fig:communication-game-one-level}]{%
        \includegraphics[width=.9\columnwidth]{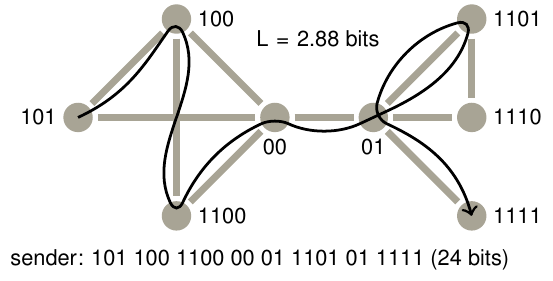}
    }\qquad
    \subfloat[\label{fig:communication-game-two-level}]{%
        \includegraphics[width=.9\columnwidth]{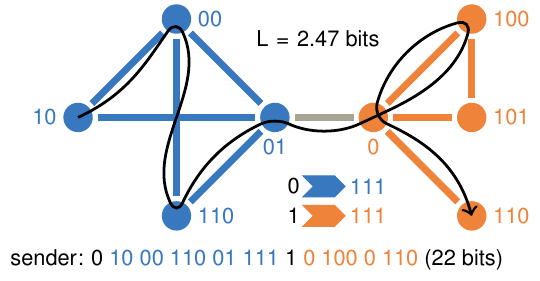}
    }
    \caption{
        A communication game on a network where colours indicate module assignments and node labels show codewords. The black trace shows a possible node sequence during a random walk; the corresponding sequence of codewords to describe the walk is shown in the bottom. The average per-step codelength is shown as $L$.
        \textbf{(a)} The one-level partition where all nodes are in the same module and there is only one codebook.
        \textbf{(b)} Nodes are split into two modules with one codebook per module and an additional index-level codebook, indicated by coloured arrows. Module entry and exit codewords are shown on the left and right of the arrows, respectively. The codelength $L$ is reduced because the partition captures those areas where the random walker tends to stay for a longer time.}
    \label{fig:communication-game}
\end{figure*}
\noindent
The map equation \cite{Rosvall1118} is a flow-based information-theoretic objective function for community detection.
It takes a network $G = \left(V, E, \delta\right)$, possibly weighted and/or directed, and a partition $\mathsf{M}$ of the network's nodes into modules as input, and measures how well the partition captures the network's community structure.
Here, $V$ is the set of nodes, $E \subseteq V \times V$ is the set of links, and $\delta \colon E \to \mathbb{R}^+$ is a function that assigns weights to the links.
A partition $\mathsf{M}$ is a split of the network's nodes into disjoint, possibly nested sets.

Conceptually, the map equation models network flow with a random walk on the network and calculates how many bits are required, on average, to encode one random-walker step.
To explain the inner workings of the map equation, we consider a communication game where the sender updates the receiver about the location of a random walker on a network.
We assume that, when at node $u$, the probability that the random walker chooses an outgoing link $e = \left(u,v\right) \in E$ is proportional to the link's weight, $\delta\left(e\right)$.

In the simplest case, when there is only one module that contains all nodes, we assign unique codewords to the nodes according to a Huffman code based on the nodes' visit rates at ergodicity.
We refer to such a partition as the one-level partition and denote it as $\mathsf{M}_1$.
When the random walker takes a step, the sender communicates one codeword to the receiver (\Figref{fig:communication-game-one-level}).
According to Shannon's source coding theorem \cite{Shannon1948}, the lower bound for the per-step codelength, $L$, is precisely the entropy of the nodes' visit rates,
\begin{equation}
    L\left(G, \mathsf{M}_1\right) = H\left(P\right) = - \sum_{u \in V} p_u \log_2 p_u,
    \label{eqn:map-equation-one-level}
\end{equation}
where $H$ is the Shannon entropy, $P$ is the set of node visit rates, and $p_u$ is the visit rate of node $u$.

In undirected networks, we calculate the node visit rates analytically as $p_u = \frac{s_u}{\sum_{v \in V} s_v}$, where $s_u = \sum_{v \in V}\delta\left(\left(u,v\right)\right)$ is the strength of node $u$.
In directed networks, we obtain the visit rates numerically as the stationary distribution of a random walk on the network.
The Perron-Frobenius theorem guarantees the existence of such an ergodic distribution in strongly connected networks; to ensure ergodicity in weakly-connected networks, there are different options.
\nobreak{PageRank} relaxes these dynamics by introducing uniform node teleportation, letting the random walker teleport to a node selected uniformly at random at some small rate \cite{pagerank-gleich}, introducing a teleportation parameter.
To reduce the effect of this parameter, an alternative is so-called unrecorded link teleportation \cite{PhysRevE.85.056107}, a similar approach where the random walker teleports, at some small rate, to links proportionally to their weight.

\begin{figure*}
    \centering
    \subfloat[\label{fig:silencing-same-code}]{%
        \includegraphics[width=.9\columnwidth]{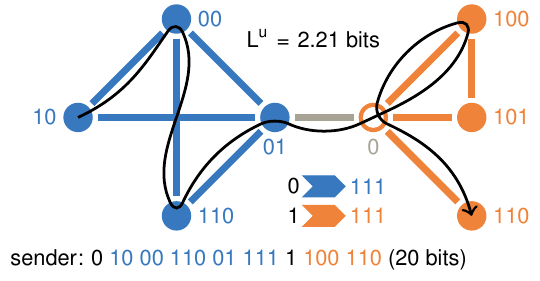}
    }\qquad
    \subfloat[\label{fig:silencing-new-code}]{%
        \includegraphics[width=.9\columnwidth]{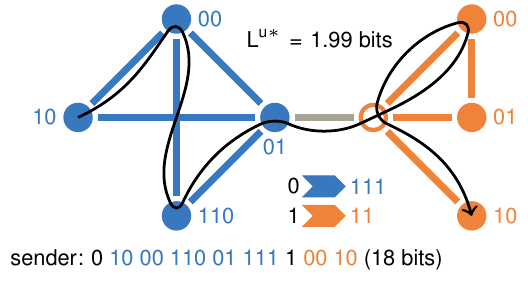}
    }
    \caption{
        Two options for describing a random walk when a node is silenced, with the silenced node shown as a ring. In both cases, the sender still communicates module entries through the silenced node.
        \textbf{(a)} Using the same code as before: when the random walker visits the silenced node, the sender does not use the corresponding node-visit codeword.
        \textbf{(b)} Designing a new code: the silenced node does not receive a codeword and visits to that node cannot be encoded.}
    \label{fig:silencing}
\end{figure*}

In networks with community structure, we can achieve shorter codelengths than with the one-level partition.
Splitting the nodes into modules allows us to assign unique codewords within modules, and re-use codewords across modules.
However, we need to pay for this by encoding transitions between modules: we introduce one designated exit codeword per module, as well as an index-level codebook for encoding transitions into modules.
Now, the sender communicates one codeword for transitions within modules, and three codewords for transitions between modules, that is one module exit codeword from the old module codebook, one module entry codeword from the index-level codebook, and one codeword from the new module codebook to visit a node in the new module (\Figref{fig:communication-game}).
The codelength for such a two-level map is given by the sum of the index-level entropy and the module-level entropies, weighted by the rate at which each codebook is used,
\begin{equation}
    L\left(G, \mathsf{M}\right) = q H\left(Q\right) + \sum_{\mathsf{m} \in \mathsf{M}} p_\mathsf{m} H\left(P_\mathsf{m}\right).
    \label{eqn:map-equation-two-level}
\end{equation}
Here, $P_\mathsf{m} = \left\{ p_u~|~u \in \mathsf{m} \right\} \cup \left\{ \mathsf{m}_\text{exit} \right\}$ is the set of node visit rates for module $\mathsf{m}$, including the module exit rate for module $\mathsf{m}$, $\mathsf{m}_\text{exit}$, and $p_\mathsf{m} = \sum_{p \in P_\mathsf{m}} p$ is the rate at which the sender uses the codebook for module $\mathsf{m}$.
$Q = \left\{ \mathsf{m}_\text{enter}~|~ \mathsf{m} \in \mathsf{M} \right\}$ is the set of module entry rates, and $q = \sum_{q_\mathsf{m} \in Q} q_\mathsf{m}$ is the rate at which the sender uses the index-level codebook.

When a partition reflects the structure of the network well and groups those nodes together where the random walker stays for a longer time, transitions between modules occur at a low frequency, overall compressing the average per-step codelength.
Thus, finding the optimal partition according to the map equation becomes a search problem.
Through recursion, we can generalise this approach to partitions nested at arbitrary depth and reduce the codelength even further in networks with hierarchical community structure.

\section{Map Equation Centrality}
To define our community-aware centrality score, map equation centrality, we take inspiration from the concept of network vitality.
Given a function $f$ that operates on networks and calculates a numerical value, the vitality $\mu\left(G, u\right)$ with respect to a node $u$ is defined as
\begin{equation}
  \mu\left(G, u\right) = f \left(G\right) - f \left(G - \left\{u\right\}\right),
  \label{eqn:vitality}
\end{equation}
where $G - \left\{u\right\}$ denotes $G$ with $u$ removed \cite{Koschutzki2005}.
But because removing a node and its incident links from the network would disrupt the network's community structure and change the nodes' visit rates, instead, we keep the network unchanged and only omit $u$ when describing the community structure -- we call this \emph{silencing a node}.
We realise silencing with the Vickrey-Clarke-Groves (VCG) principle for setting prices in multi-item auctions, such as AdWords auctions, a generalisation of second-price sealed-bid auctions for single items where the bidder who submits the highest bid for an item receives the item for the value of the second-highest bid \cite{vickrey1961counterspeculation}.
The VCG mechanism determines the price that bidder $b$ has to pay for item $i$ as the marginal harm caused to other bidders who, because of $b$'s existence, receive an item $j \neq i$ that they value lower than $i$.
The price that $b$ pays for $i$ is ``the difference between the optimal valuation achievable by allocating everyone except person $b$ to all the positions and the optimal valuation obtainable by allocating everyone except person $b$ to all positions other than i'' \cite{leonard1983elicitation}.
Specifically, $b$'s price for $i$ does not depend on $b$'s own wealth but is determined by the collective marginal harm caused to the remaining bidders.
Following the same idea, we define a node $u$'s importance as the collective marginal harm it causes to the remaining nodes in terms of codeword length, that is, by how many bits the codeword lengths for the remaining nodes could be reduced if $u$ was silenced.


In terms of the communication game, silencing a node means that, when the random walker visits a silenced node $u$, the sender does not communicate the codeword for visiting $u$ to the receiver (\Figref{fig:silencing-same-code}).
But this is inefficient because node $u$ has a codeword that is never used, that is, the sender uses more bits than necessary to describe the random walk.
Instead, we can design a new coding scheme without assigning a codeword to $u$ and, thereby, compress the description of the random walk (\Figref{fig:silencing-new-code}).
Map equation centrality is always positive because silencing a node deletes its codeword from the coding scheme, making it possible to assign shorter codewords to the remaining nodes.
Following the VCG principle, we define the centrality of node $u$ as the difference between the original, inefficient code---we call it $L^u$---and the updated, efficient code---we call it $L^{u*}$,
\begin{equation}
  \lambda\left(G, \mathsf{M}, u\right) = L^u \left(G, \mathsf{M}\right) - L^{u*} \left(G, \mathsf{M}\right).
  \label{eqn:map-equation-centrality}
\end{equation}
Paraphrasing the VCG principle, map equation centrality for node $u$ is the codelength difference between the optimal coding scheme that assigns codewords to all nodes but never uses the codeword for node $u$ and the optimal coding scheme that assigns codewords to all nodes but $u$.
We derive expressions for $L^u$ and $L^{u*}$ from the map equation, and, for clarity, begin with one-level partitions, then moving on to two-level and hierarchical partitions.

First, we consider the case where we use the old coding scheme.
We obtain the codelength resulting from silencing $u$ from \Eqnref{eqn:map-equation-one-level} by removing $u$ from the summation,
\begin{equation}
    L^u \left(G, \mathsf{M}_1\right) = - \sum_{\mathclap{v \in V, v \neq u}} p_v \log_2 p_v.
    \label{eqn:silenced-one-level-same-code}
\end{equation}
Designing a new coding scheme without a codeword for $u$ changes the codeword lengths for the rest of the nodes.
Before, the codeword length for some node $v$ was given by its visit rate as $\log_2 p_v$, but now that $u$ does not receive a codeword anymore, we need to re-normalise accordingly.
The new codeword length for node $v \neq u$ is $\log_2 \frac{p_v}{1-p_u}$, and for $u$ it is zero, resulting in a codelength of
\begin{equation}
L^{u*} \left(G,\mathsf{M}_1\right) = - \sum_{\mathclap{v \in V, v \neq u}} p_v \log_2 \frac{p_v}{1 - p_u}.
  \label{eqn:silenced-one-level-new-code}
\end{equation}
Plugging \Eqnref{eqn:silenced-one-level-same-code} and \Eqnref{eqn:silenced-one-level-new-code} into \Eqnref{eqn:map-equation-centrality}, we get $u$'s contribution to the codelength in the one-level partition $\mathsf{M}_1$,
\begin{align}
\lambda\left(G, \mathsf{M}_1, u\right) & = L^u\left(G,\mathsf{M}_1\right) - L^{u*}\left(G,\mathsf{M}_1\right) \nonumber \\
& = - \left(1 - p_u\right) \log_2 \left(1 - p_u\right).
  \label{eqn:map-equation-centrality-one-level}
\end{align}

We move on to derive the same quantities for two-level partitions $\mathsf{M}$.
Again, we begin by considering the resulting codelength when silencing node $u$ but using the old coding scheme.
Then, we design a new coding scheme that does not assigning a codeword to $u$, and calculate the difference between the two coding schemes to obtain $u$'s contribution.
For clearer derivations, we distinguish explicitly between $\mathsf{m}_u$, the module that contains $u$, and the rest of the modules by rewriting the map equation (\Eqnref{eqn:map-equation-two-level}),
\begin{align}
& L\left(G, \mathsf{M}\right) = q H\left(Q\right) + \sum_{\mathsf{m} \in \mathsf{M}} p_\mathsf{m} H\left(P_\mathsf{m}\right) \label{eqn:map-equation-rewrite} \\
& = \overbracket{q H\left(Q\right)}^{\text{index level}} + \overbracket{\sum_{\mathclap{\mathsf{m} \in \mathsf{M}, \mathsf{m} \neq \mathsf{m}_u}} p_\mathsf{m} H\left(P_\mathsf{m}\right)}^{\text{modules without $u$}} - \overbracket{\sum_{p \in P_{\mathsf{m}_u}} p \log_2 \frac{p}{p_{\mathsf{m}_u}}}^{\text{module with $u$}}. \nonumber
\end{align}

From \Eqnref{eqn:map-equation-rewrite}, it becomes clear that silencing node $u$ in a two-level partition only affects the module that contains $u$ because a codeword for $u$ only exists in the context of $\mathsf{m}_u$, but not in other modules.
The codelength for a two-level partition $\mathsf{M}$, using the old coding scheme while $u$ is silenced is
\begin{align}
  & L^u\left(G,\mathsf{M}\right) = \label{eqn:silence-two-level-same-code} \\
  & \quad \overbracket{q H\left(Q\right)}^{\text{index level}} + \overbracket{\sum_{\mathclap{\mathsf{m} \in \mathsf{M}, \mathsf{m} \neq \mathsf{m}_u}} p_\mathsf{m} H\left(P_\mathsf{m}\right)}^{\text{modules without $u$}} - \overbracket{\sum_{\mathclap{p \in P_{\mathsf{m}_u} \setminus \left\{ p_u\right\}}} p \log_2 \frac{p}{p_{\mathsf{m}_u}}}^{\text{module with $u$}}.
  \nonumber
\end{align}

Because of the modular structure of the coding scheme, when designing a new code, only codewords for nodes in module $\mathsf{m}_u$ are affected while other modules and the index level remain unaffected.
The new codebook usage rate for module $\mathsf{m}_u$ is $p_{\mathsf{m}_u} - p_u$, which is also the term we use for re-normalising the node visit rates for nodes in $\mathsf{m}_u$.
That is, the new rate at which the codeword for $v \in \mathsf{m}_u$ with $v \neq u$ is used is $\frac{p_v}{p_{\mathsf{m}_u} - p_u}$, and the module exit codeword is used at rate $\frac{{\mathsf{m}_u}_\text{exit}}{p_{\mathsf{m}_u} - p_u}$.
The new codelength for $\mathsf{M}$ is
\begin{align}
  & L^{u*}\left(G, \mathsf{M}\right) = \label{eqn:silenced-two-level-new-code} \\
  & \overbracket{q H\left(Q\right)}^{\text{index level}} + \overbracket{\sum_{\mathclap{\mathsf{m} \in \mathsf{M}, \mathsf{m} \neq \mathsf{m}_u}} p_\mathsf{m} H\left(P_\mathsf{m}\right)}^{\text{modules without $u$}} - \overbracket{\sum_{\mathclap{p \in P_{\mathsf{m}_u} \setminus \left\{p_u\right\}}} p \log_2 \frac{p}{p_{\mathsf{m}_u} - p_u}}^{\text{module with $u$}}. \nonumber
\end{align}

Plugging \Eqnref{eqn:silence-two-level-same-code} and \Eqnref{eqn:silenced-two-level-new-code} into \Eqnref{eqn:map-equation-centrality}, we get $u$'s contribution to the two-level codelength where the terms for the index level and those modules that do not contain $u$ cancel out,
\begin{align}
  \lambda\left(G, \mathsf{M}, u\right) & = L^{u}\left(G, \mathsf{M}\right) - L^{u*}\left(G, \mathsf{M}\right) \nonumber \\
  & = - \sum_{\mathclap{p \in P_{\mathsf{m}_u} \setminus p_u}} p \log_2 \frac{p_{\mathsf{m}_u} - p_u }{p_{\mathsf{m}_u}} \nonumber \\
  & = - \left(p_{\mathsf{m}_u} - p_u\right) \log_2 \frac{p_{\mathsf{m}_u} - p_u}{p_{\mathsf{m}_u}}
  \label{eqn:map-equation-centrality-two-level}
\end{align}
For the one-level partition $\mathsf{M}_1$, the expression in \Eqnref{eqn:map-equation-centrality-two-level} reduces to \Eqnref{eqn:map-equation-centrality-one-level} because all nodes are in the same module and, consequently, $p_{\mathsf{m}_u} = 1$.

Through recursion, we can extended map equation centrality to hierarchical partitions with more than two levels.
In fact, since silencing a node $u$ only affects module $\mathsf{m}_u$, \Eqnref{eqn:map-equation-centrality-two-level} can be used to calculate centralities for nodes in modules that are nested deeper in the module hierarchy of a network.
Further, we can extend map equation centrality to silencing a set of nodes by adjusting \Eqnref{eqn:silence-two-level-same-code} and \Eqnref{eqn:silenced-two-level-new-code} (see appendix A), leading to
\begin{equation}
  \lambda\left(G, \mathsf{M}, U\right) = - \sum_{\mathclap{\mathsf{m} \in \mathsf{M}, \mathsf{m} \cap U \neq \emptyset}} \left(p_\mathsf{m} - p_{\mathsf{m} \cap U}\right) \log_2 \frac{p_\mathsf{m} - p_{\mathsf{m} \cap U}}{p_\mathsf{m}}.
  \label{eqn:map-equation-centrality-sets}
\end{equation}
Here, $U$ is the set of nodes that are silenced, and $p_{\mathsf{m} \cap U} = \sum_{u \in \mathsf{m} \cap U} p_u$ is the sum of visit rates for the silenced nodes in module $\mathsf{m}$.
Moreover, map equation centrality is agnostic to the chosen flow model and can be used with standard PageRank, unrecorded link teleportation, or other suitable flow models that may be chosen based on the dynamic process that is analysed.
Map equation centrality can be generalised to overlapping communities through memory networks \cite{edler2017algorithms} using trajectory data to determine link weights.

Map equation centrality relates to the Kullback-Leibler divergence, also known as relative entropy, and defined as $D_{KL} \left( P || Q \right) = - \sum_{x \in X} p\left(x\right) \log_2 \frac{q\left(x\right)}{p\left(x\right)}$, where $X$ is a set of events, and $P$ and $Q$ are probability distributions over $X$.
The KL divergence quantifies the expected number of extra bits that are required to encode a sequence of events with true distribution $P$, assuming that we use a code optimised for $Q$.
In this light, the importance of a node $u$ is the Kullback-Leibler divergence between encoding visits in module $\mathsf{m}_u$ with true codebook usage rate $p_{\mathsf{m}_u}$ and silencing $u$, resulting in a new codebook usage rate after silencing of $p_{\mathsf{m_u}} - p_u$.
Because no other modules than $\mathsf{m}_u$ contribute to our score, $u$'s importance under map equation centrality is fully determined by its own visit rate $p_u$ and its modular context through $p_{\mathsf{m}_u}$.

\section{Application to Synthetic and Empirical Networks}
We have implemented map equation centrality in Infomap, a fast and greedy optimisation algorithm for the map equation with an open source implementation available on GitHub\footnote{\url{https://github.com/mapequation/infomap}} \cite{infomap}.
In a network with $n$ nodes, Infomap detects communities and computes codeword usage rates for all nodes and codebook usage rates for all modules in time $\mathcal{O}\left(n \log n\right)$ \cite{edler2017algorithms}.
With this information available, traversing the network partition and computing map equation centrality scores for all $n$ nodes takes time $\mathcal{O}\left(n\right)$.
Detecting communities and computing map equation centrality scores combined takes time $\mathcal{O}\left(n \log n\right)$.

To evaluate map equation centrality, we apply it to synthetic and empirical networks.
First, using a toy example, we highlight how map equation centrality overcomes traditional centrality scores' inability to distinguish between same-feature nodes when adopting a local or global point of view.
Second, we generate an LFR network with strong community structure and measure how the ranking of nodes according to map equation centrality changes as we rewire different fractions of the network's links.
Third, we evaluate map equation centrality alongside two traditional and three community-aware centrality scores on a set of empirical social, biological, web, co-authorship, and infrastructure networks using two different spreading processes, (i) the linear threshold model and (ii)  the Susceptible-Infected-Recovered (SIR) disease spreading model.
We explore the centrality scores through the lens of two different spreading processes because they highlight various aspects.
Neither of them is more valid than the other but they are simply tools for comparison of different use cases.
In both cases, we test two different flow models as a basis for community detection with Infomap: (a) unrecorded link teleportation \cite{PhysRevE.85.056107}, and (b) recorded node teleportation, corresponding to standard PageRank with teleportation rate $0.15$ \cite{pagerank-gleich}.
In principle, one could define further domain-specific flow models, determine node visit rates through simulations, and use them as an input for Infomap.
For reproducibility, we provide our code for evaluation in a GitHub repository\footnote{\url{https://github.com/mapequation/map-equation-centrality}}.

\begin{table}
    \centering
    \caption{
        Rounded centrality scores for the toy network: degree centrality (DC), betweenness centrality (BC), PageRank without teleportation (PR), and map equation centrality  ($\lambda$) for the one-level partition $\mathsf{M}_1$, a sub-optimal partition $\mathsf{M}_\text{sub}$, and the optimal partition $\mathsf{M}_\text{opt}$, shown in (\Figref{fig:k4-s3}).
    }
    \begin{tabular}{c c c c c c c}
        \toprule
        u & DC & BC & PR & $\lambda(\mathsf{M}_1)$ & $\lambda(\mathsf{M}_\text{sub})$ & $\lambda(\mathsf{M}_\text{opt})$ \\
        \midrule
        1 & 0.43 & 0.02 & 0.15 & 0.20 & 0.16 & 0.184 \\
        2 & 0.29 & 0    & 0.10 & 0.14 & 0.12 & 0.130 \\
        3 & 0.29 & 0    & 0.10 & 0.14 & 0.12 & 0.130 \\
        4 & 0.57 & 0.60 & 0.20 & 0.26 & 0.24 & \textbf{0.228} \\
        \rule{0pt}{3ex}\noindent
        5 & 0.57 & 0.67 & 0.20 & 0.26 & 0.24 & \textbf{0.212} \\
        6 & 0.29 & 0    & 0.10 & 0.14 & 0.13 & 0.127 \\
        7 & 0.29 & 0    & 0.10 & 0.14 & 0.13 & 0.127 \\
        8 & 0.14 & 0    & 0.05 & 0.07 & 0.07 & 0.068 \\
        \bottomrule
    \end{tabular}
    \label{tab:k4-s3}
\end{table}

\subsection{Toy Example: How Map Equation Centrality Discerns Same-Feature Nodes}
\begin{figure}
    \centering
    \subfloat[\label{fig:k4-s3-pagerank}]{%
        \includegraphics[width=.45\columnwidth]{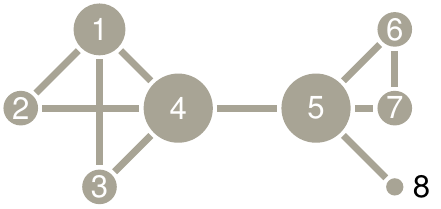}
    }\hfill
    \subfloat[\label{fig:k4-s3-map-equation-centrality}]{%
        \includegraphics[width=.45\columnwidth]{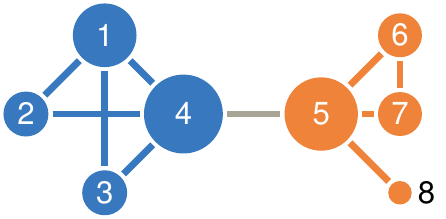}
    }
    \caption{
        Illustration of the centrality scores from \Tblref{tab:k4-s3}. Node colours indicate community assignments, node diameter is proportional to \textbf{(a)} degree centrality and PageRank without teleportation, and \textbf{(b)} map equation centrality.}
    \label{fig:k4-s3}
\end{figure}
\noindent
We use a small, undirected network with eight nodes and ten links (\Figref{fig:k4-s3}), and use \texttt{networkx} \cite{networkx} and Infomap to calculate centrality scores for its nodes (\Tblref{tab:k4-s3}).
The optimal way to partition the network, by design and recovered by Infomap, is to group the nodes into two communities as indicated by colours (\Figref{fig:k4-s3-map-equation-centrality}).

We find that neither degree centrality nor map equation centrality when based on the one-level partition $\mathsf{M}_1 = \left\{1,2,3,4,5,6,7,8\right\}$ can distinguish between nodes with the same degree (\Figref{fig:k4-s3-pagerank}, \Tblref{tab:k4-s3}).
This is because using $\mathsf{M}_1$ turns map equation centrality into a global approach, ignoring the network's mesoscopic community structure.
However, when using the sub-optimal two-level partition $\mathsf{M}_\text{sub} = \left\{\left\{1,2,3\right\}, \left\{4,5,6,7,8\right\}\right\}$ with codelength $3.24$~bits, or the optimal two-level partition $\mathsf{M}_\text{opt} = \left\{\left\{1,2,3,4\right\}, \left\{5,6,7,8\right\}\right\}$ with codelength $2.47$~bits (\Figref{fig:k4-s3-map-equation-centrality}), map equation centrality distinguishes between same-degree nodes that are embedded in different modules while same-degree nodes in the same module remain indistinguishable (\Figref{fig:k4-s3-map-equation-centrality}, \Tblref{tab:k4-s3}).
We explain this by interpreting \Eqnref{eqn:map-equation-centrality-two-level}: the importance of a node $u$ is determined by its visit rate, $p_u$, as well as the codebook usage rate of its module, $p_{\mathsf{m}_u}$, that is, modules with a higher codebook usage rate boost the importance of their member nodes to a higher degree than modules with a lower codebook usage rate.

\subsection{Synthetic Network: Behaviour of Map Equation Centrality under Link Rewiring}
\begin{figure}
    \centering
    \includegraphics[width=.75\columnwidth]{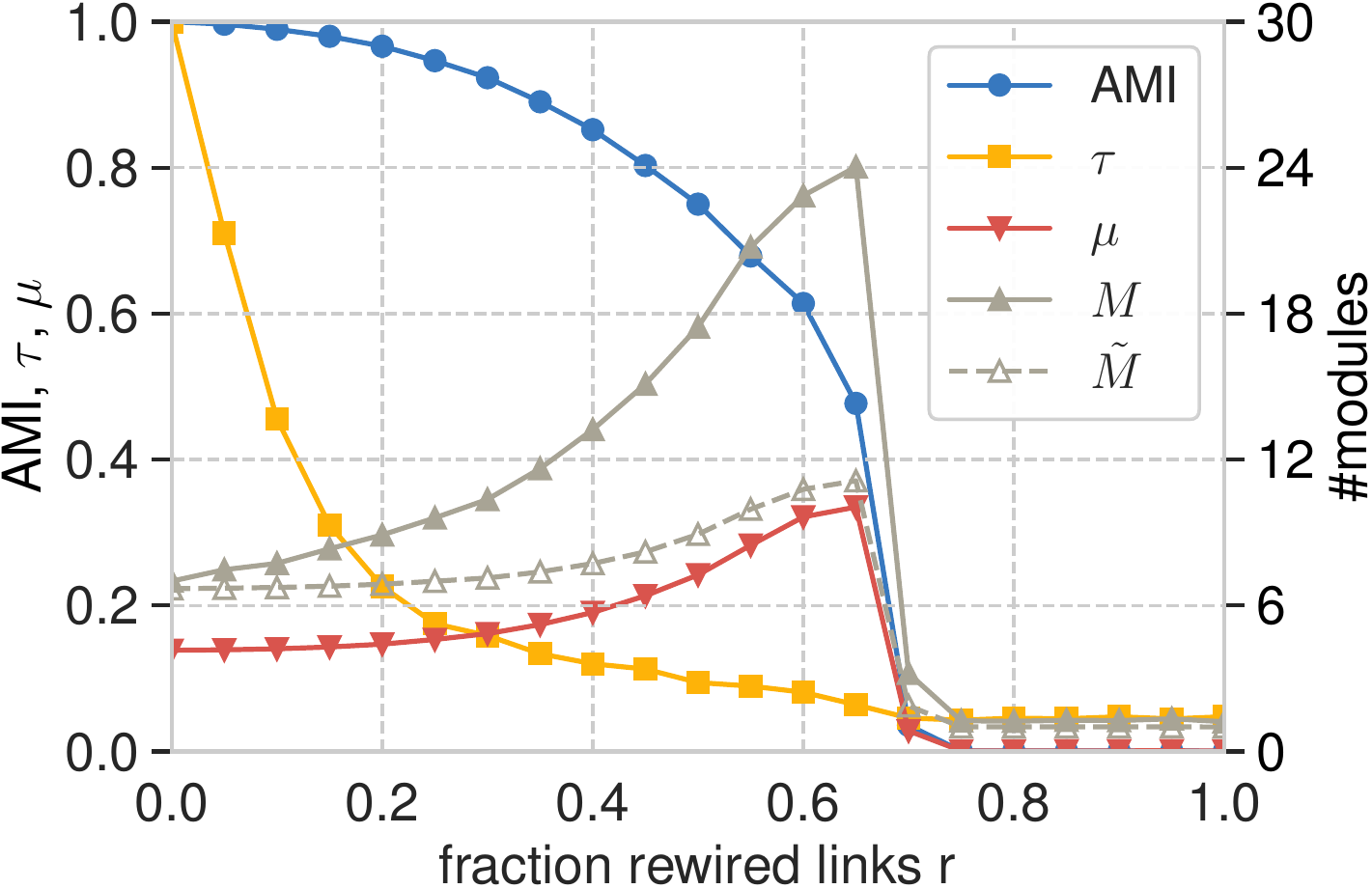}
    \caption{Results under rewiring of an LFR network. For each fraction of rewired links, $r$, we infer the community structure with Infomap and compute map equation centrality scores. We report AMI with the ground truth partition, correlation with the node ranking under the ground truth partition, $\tau$, the mixing, $\mu$, as well as the number of detected communities $M$ and effective number of communities $\tilde{M}$. The reported values are averages over 100 rewirings for each $r$.}
    \label{fig:lfr}
\end{figure}

We generate an LFR network \cite{PhysRevE.78.046110} with 1,000 nodes, average degree $k = 10$, minimum community size 100, node degree exponent $\gamma = 2.5$, community size exponent $\beta = 1.5$, and mixing parameter $\mu = 0.1$.
The resulting network has 7 communities, and, using those communities, we calculate map equation centrality scores for all nodes.
We then rewire an $r$-fraction of the network's links and use Infomap to detect communities $\mathsf{M}$ in the rewired network and Kendall's $\tau$ coefficient to measure how the nodes' ranking has changed.
With adjusted mutual information (AMI), we estimate the agreement between the new communities and the ground truth community structure.
We also compute the effective number of communities as the perplexity over the relative modules' sizes, $\tilde{M} = 2^{H\left(\mathsf{M}\right)}$, where $H\left(\mathsf{M}\right) = \sum_{\mathsf{m} \in \mathsf{M}} \frac{\left|\mathsf{m}\right|}{N} \log_2 \frac{\left|\mathsf{m}\right|}{N} $ is the Shannon entropy of the relative module sizes, $N$ is the number of nodes in the networks, and $\left|\mathsf{m}\right|$ is the number of nodes in module $\mathsf{m}$.
The effective number of modules is the number of same-size modules with the same entropy into which the nodes would be partitioned.
An effective number of modules close to the actual number of detected modules indicates that the detected communities have similar size, whereas a much smaller number of effective modules indicates partition with a smaller number of large modules and a larger number of small modules.
For robust results, we repeat the rewiring for each $r$ 100 times and report average values for AMI, $\tau$, resulting mixing $\mu$, the number of communities $M$, and the number of effective communities $\tilde{M}$; the results are shown in \Figref{fig:lfr}.

Overall, we see that small amounts of noise caused by rewiring can affect the node ranking to a larger extent despite a relatively stable number of communities with high AMI values.

\subsection{Datasets and Methods}
\begin{table*}
    \centering
    \caption{
        Details for eight empirical networks: their number of nodes, $N$, number of links, $\left|E\right|$, average degree, $k$, epidemic threshold, $p_\text{th}$; the number of communities inferred with Infomap, $M$, the effective number of communities $\tilde{M}$, and mixing, $\mu$, both for link and node teleportation. Directed networks are marked with *.
    }
    \begin{tabular}{l rrrr rrr c rrr}
        \toprule
        & & & & & \multicolumn{3}{c}{Link teleportation} && \multicolumn{3}{c}{Node teleportation} \\
        \cmidrule{6-8} \cmidrule{10-12}
        Network & $N$ & $\left|E\right|$ & k & $p_\text{th}$ & $M$ & $\tilde{M}$ & $\mu$ && $M$ & $\tilde{M}$ & $\mu$ \\
        
        \midrule
        
        Facebook friends                 &    329 &   1,954 & 11.9 & 0.048 &   21 & 13 & 0.129 &&   22 & 13 & 0.115 \\
        Copenhagen                       &    800 &   6,429 & 16.1 & 0.038 &   37 & 29 & 0.499 &&   36 & 29 & 0.499 \\
        Uni email\textsuperscript{*}     &  1,133 &   5,452 & 19.2 & 0.027 &   50 & 32 & 0.406 &&   55 & 34 & 0.409 \\
        
        \rule{0pt}{3ex}\noindent
        
        Polblogs\textsuperscript{*}      &  1,222 &  19,024 & 31.1 & 0.010 &   88 & 6 & 0.164 &&   51 & 6 & 0.177 \\
        Interactome yeast                &  1,458 &   1,993 &  2.7 & 0.161 &  166 & 142 & 0.237 &&  178 & 154 & 0.247 \\
        Ego Facebook                     &  4,039 &  88,234 & 43.7 & 0.009 &   77 & 32 & 0.082 &&   86 & 35 & 0.083 \\
        
        \rule{0pt}{3ex}\noindent
        
        Power                            &  4,941 &   6,594 &  2.7 & 0.348 &  428 & 377 & 0.163 &&  465 & 416 & 0.177 \\
        Facebook organizations           &  5,524 &  94,219 & 34.1 & 0.016 &   53 & 30 & 0.360 &&   62 & 40 & 0.391 \\
        Physics collaborations           &  8,798 &  27,416 &  6.2 & 0.066 &  610 & 491 & 0.218 &&  656 & 537 & 0.227 \\
        
        \rule{0pt}{3ex}\noindent
        
        Google\textsuperscript{*}        & 15,763 & 171,206 & 21.7 & 0.001 &  597 & 260 & 0.470 &&  600 & 225 & 0.518 \\
        PGP\textsuperscript{*}           & 39,796 & 301,498 & 15.2 & 0.010 & 2,851 & 1,529 & 0.285 && 3,285 & 1,843 & 0.305 \\
        Facebook wall\textsuperscript{*} & 43,953 & 271,375 & 12.3 & 0.028 & 2,995 & 1,375 & 0.493 && 3,228 & 1,747 & 0.519 \\
        \bottomrule
    \end{tabular}
    \label{tab:empirical-networks}
\end{table*}
\noindent
We use twelve real-world networks, retrieved from \nobreak{netzschleuder} \cite{netzschleuder}, to evaluate map equation centrality's performance.
Seven of the networks are undirected while five are directed.
\begin{description}
    \item[Facebook friends] undirected network of Facebook friendships, recorded in April 2014, where a link between users A and B means that they are friends on Facebook \cite{PhysRevE.96.042307}.
    
    \item[Copenhagen] undirected network of Facebook friendships between university students from Copenhagen where a link between users A and B means that they are friends on Facebook \cite{sapiezynski2019interaction}.
    
    \item[Uni email] directed network of email exchanges at the Rovira i Virgili University in Spain, recorded in 2003, where a link from user A to user B means that user A has sent an email to user B \cite{PhysRevE.68.065103}.

    \item[Polblogs] directed network of U.S. political blog websites, recorded in 2004, where a link from blog A to B means that A has a hyperlink to B \cite{10.1145/1134271.1134277}.

    \item[Interactome yeast] undirected network of yeast proteins where a link between proteins A and B means that they interact with each other \cite{coulomb2005gene}.
    
    \item[Ego Facebook] undirected network of Facebook friendships, recorded in 2021, where a link between users A and B means they are friends on Facebook \cite{mcauley2014discovering}.

    \item[Power] undirected network of the power grid in the western U.S. where nodes represent generators, transformers, and substations, and they are connected by a link if a high-voltage transmission line runs between them \cite{Watts1998}.
    
    \item[Facebook organizations] undirected network of Facebook friendships between users working at the same organization, a link between users A and B means that they are friends on Facebook \cite{fire2016organization}.
    
    \item[Physics collaborations] undirected co-authorship network between researchers who have a preprint on arXiv, recorded in May 2014, where a link between researcher A and B means that they have written an arXiv preprint together \cite{PhysRevX.5.011027}.
    
    \item[Google] directed network of hyperlinks between internal websites at Google, recorded in 2004. A link from page A to B means that there is a hyperlink from A to B \cite{Palla_2007}.
    
    \item[PGP] directed network of users in the Pretty-Good-Privacy (PGP) web of trust, recorded in November 2009. A link from user A to user B means that user A trusts user B \cite{richters2011trust}.
    
    \item[Facebook wall] directed network of interactions between Facebook users, recorded in 2009, where a link from user A to user B means that user A has posted on user B's wall \cite{viswanath2009evolution}.
\end{description}
\Tblref{tab:empirical-networks} provides details about the networks' size, average node degree, epidemic threshold; their number of communities as detected with Infomap, effective number of communities, and mixing, both for link and node teleportation.
Since estimating nodes' spreading power with the SIR simulation as well as the linear threshold model disregard link weights, we treat all networks as unweighted.

\begin{figure*}
    \centering
    \subfloat[\label{fig:lt-0.5-facebook-friends}]{%
        \includegraphics[width=.23\textwidth]{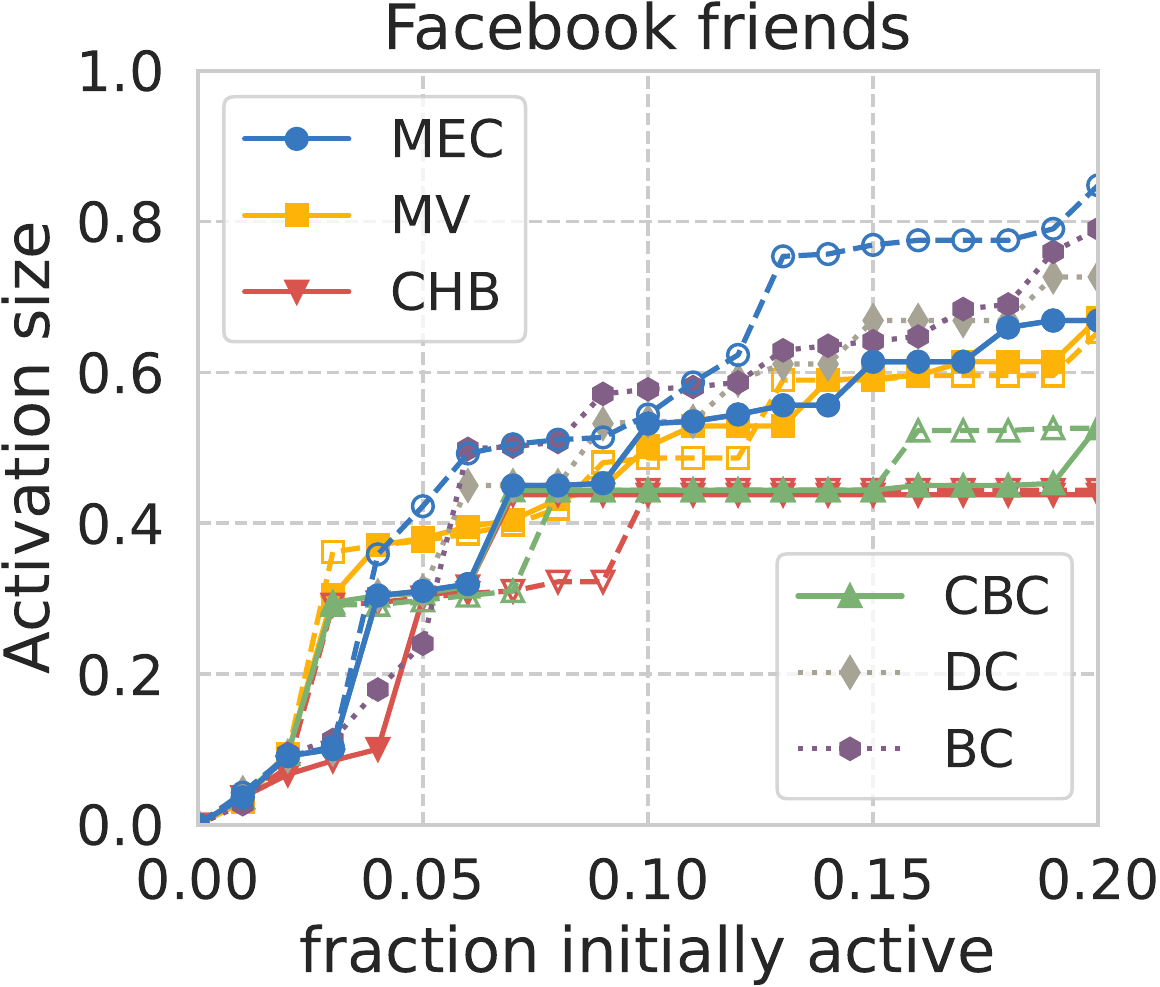}
    }\hfill
    \subfloat[\label{fig:lt-0.5-copenhagen}]{%
        \includegraphics[width=.23\textwidth]{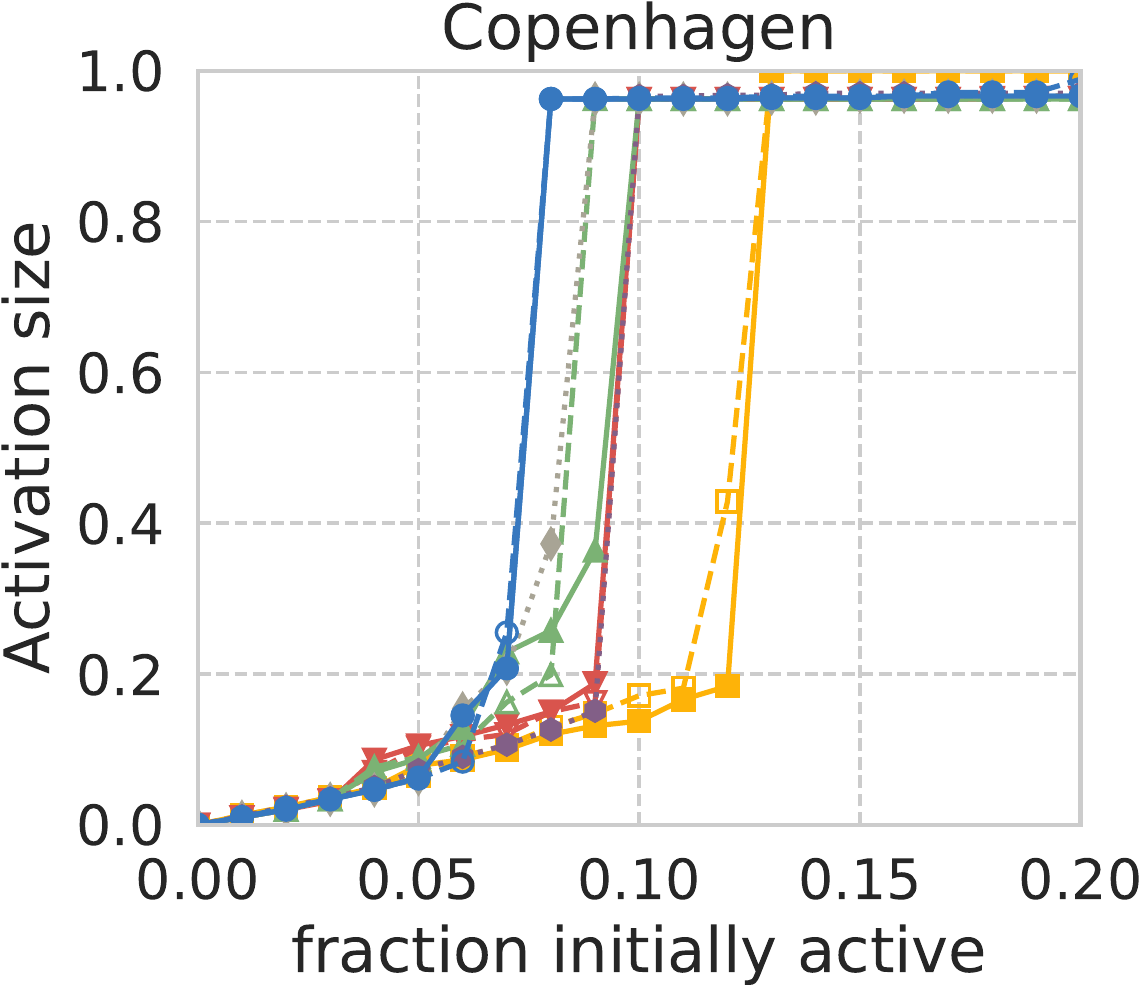}
    }\hfill
    \subfloat[\label{fig:lt-0.5-uni-email}]{%
        \includegraphics[width=.23\textwidth]{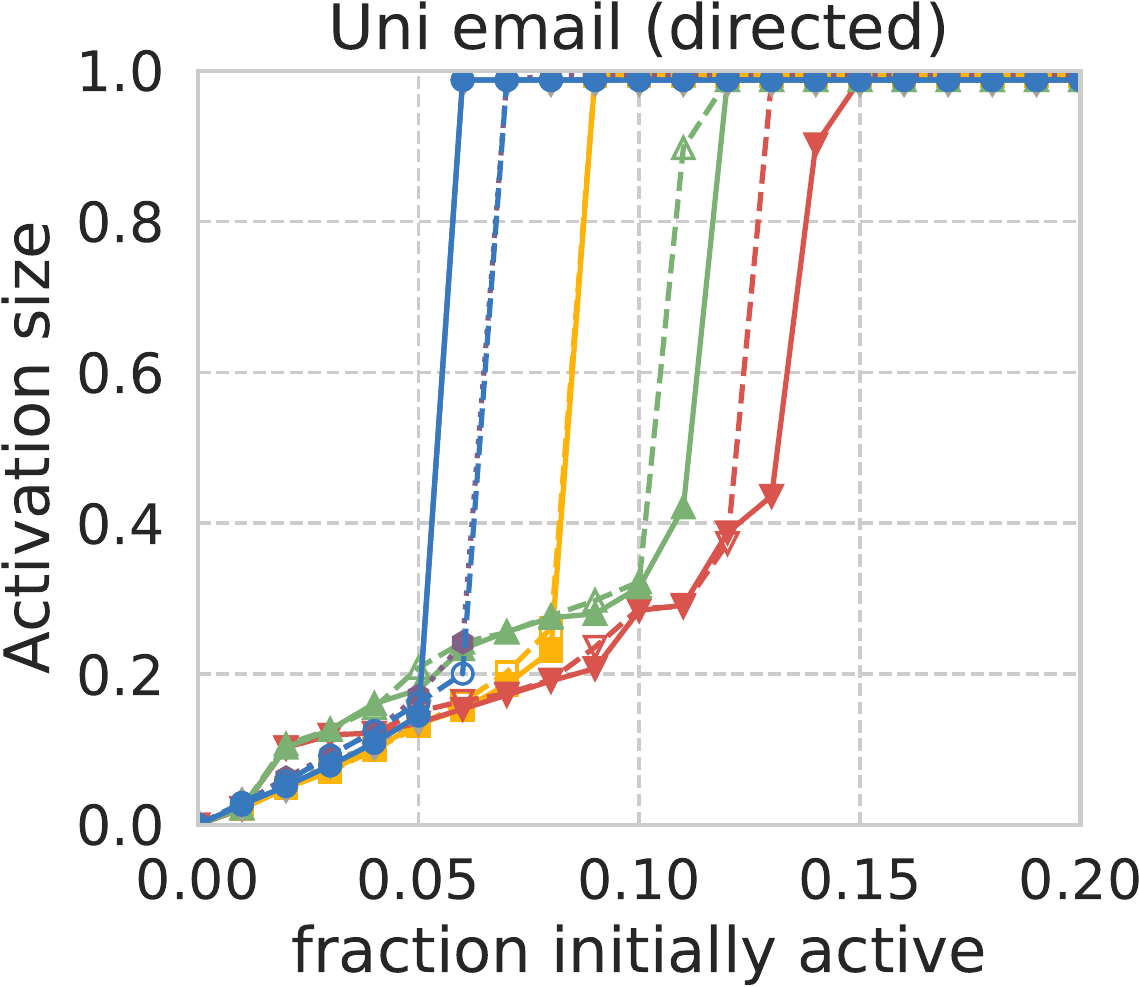}
    }\hfill
    \subfloat[\label{fig:lt-0.5-polblogs}]{%
        \includegraphics[width=.23\textwidth]{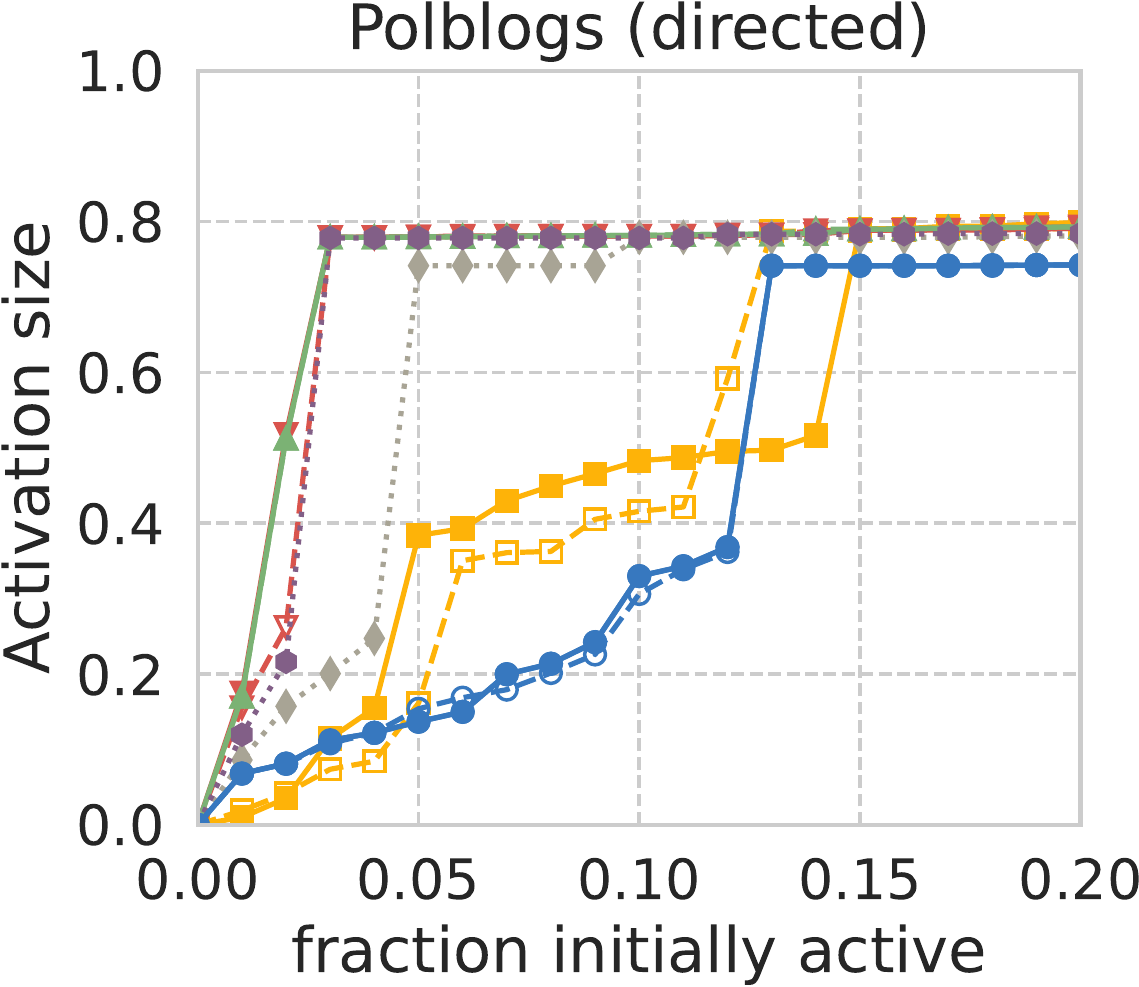}
    }\hfill
    \subfloat[\label{fig:lt-0.5-interactome-yeast}]{%
        \includegraphics[width=.23\textwidth]{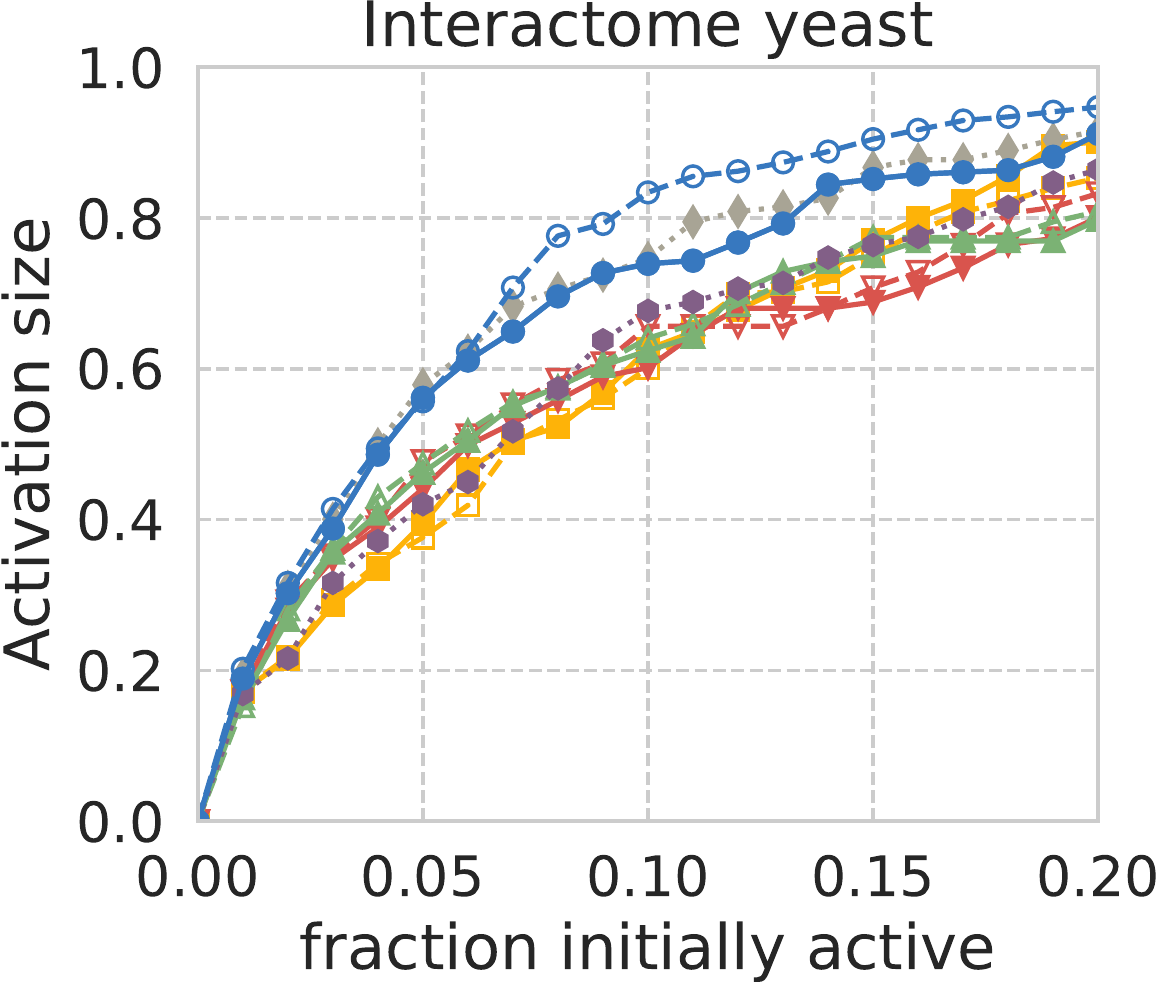}
    }\hfill
    \subfloat[\label{fig:lt-0.5-ego-facebook}]{%
        \includegraphics[width=.23\textwidth]{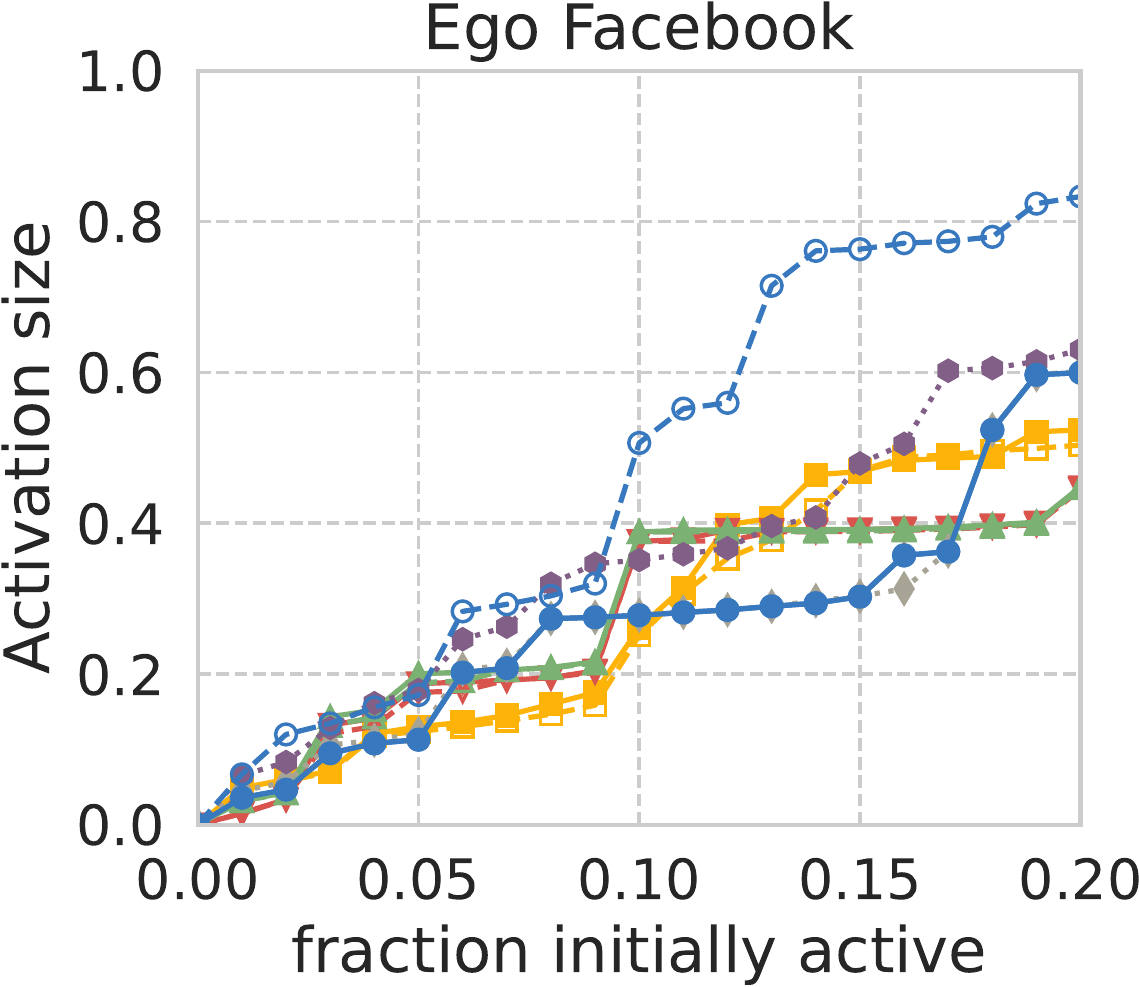}
    }\hfill
    \subfloat[\label{fig:lt-0.5-power}]{%
        \includegraphics[width=.23\textwidth]{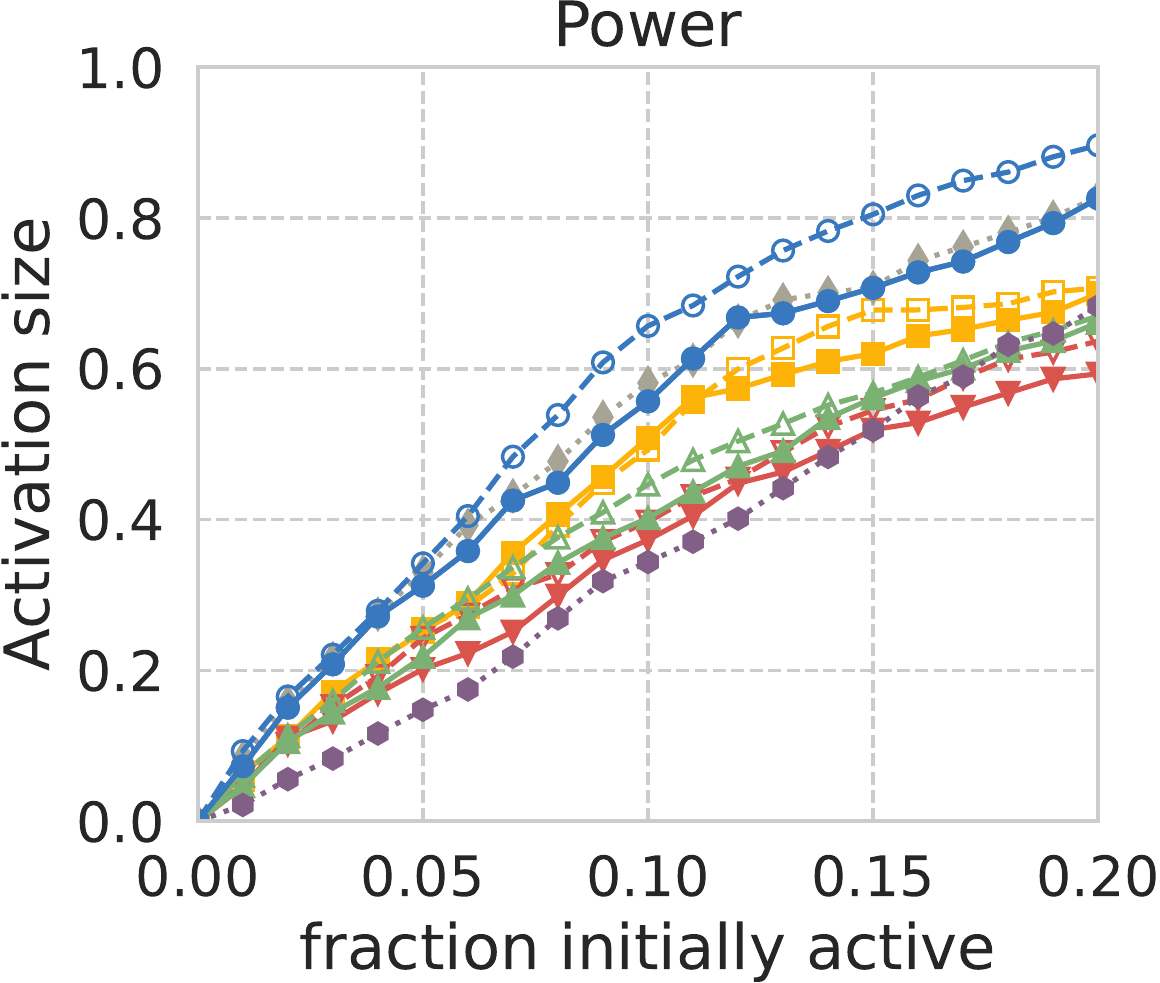}
    }\hfill
    \subfloat[\label{fig:lt-0.5-facebook-organizations}]{%
        \includegraphics[width=.23\textwidth]{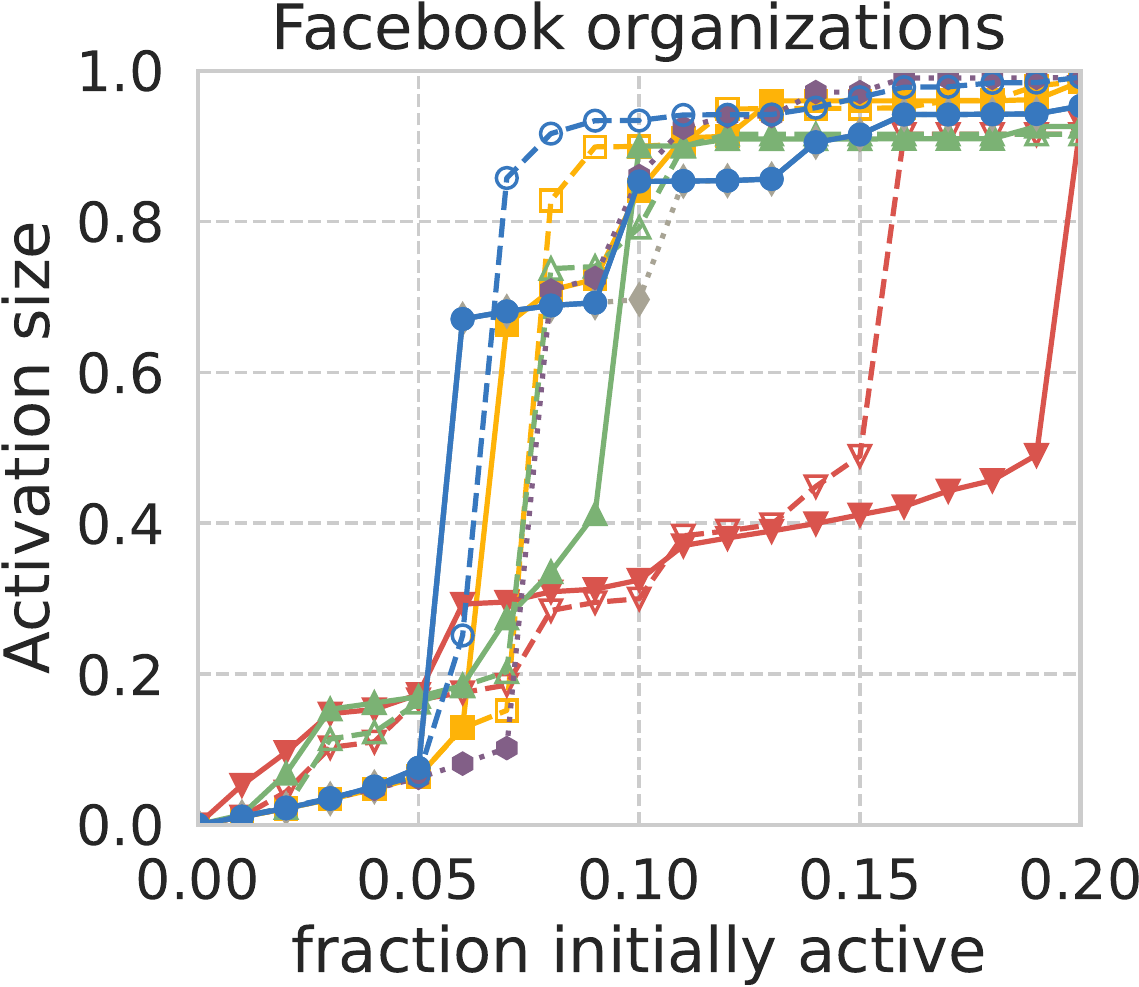}
    }\hfill
    \subfloat[\label{fig:lt-0.5-physics-collaborations}]{%
        \includegraphics[width=.23\textwidth]{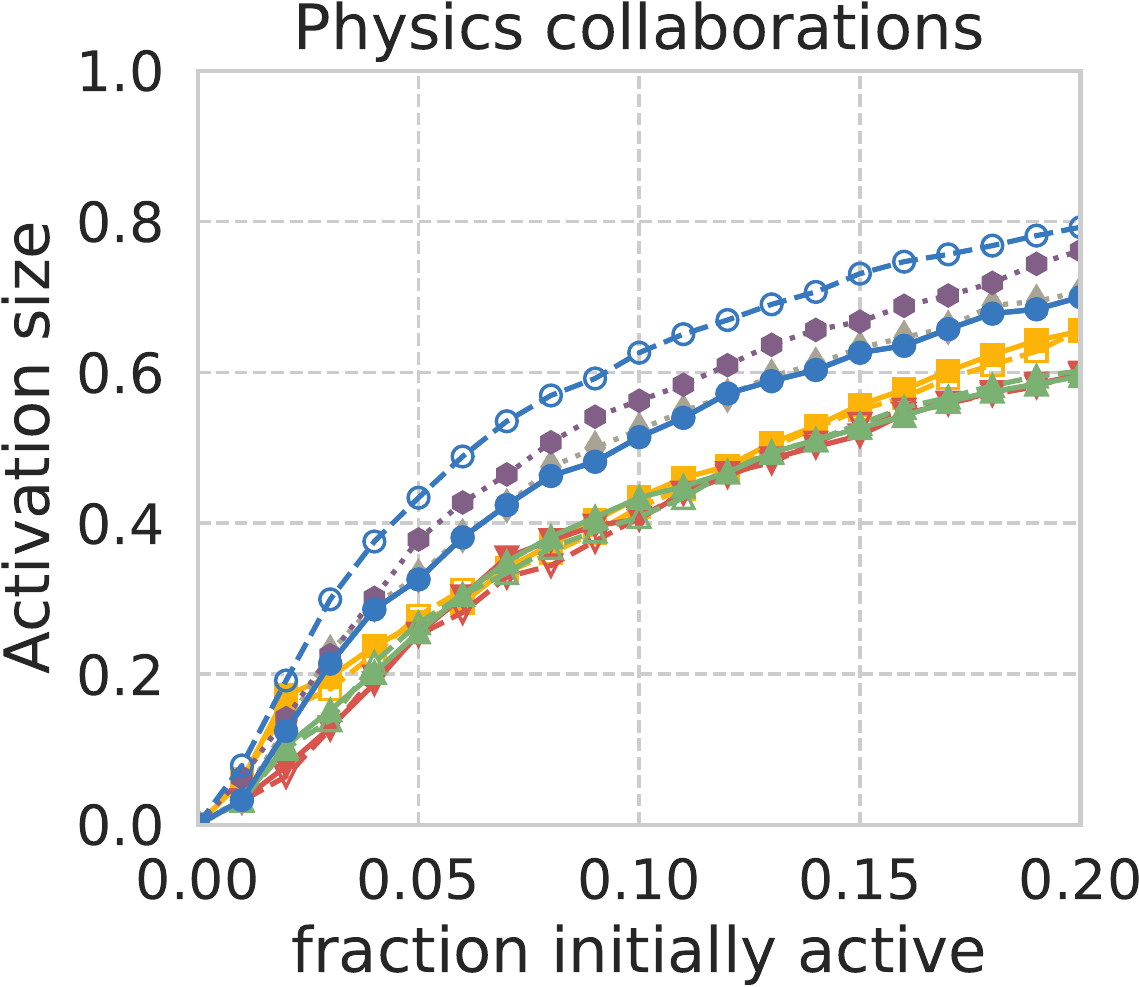}
    }\hfill
    \subfloat[\label{fig:lt-0.5-google}]{%
        \includegraphics[width=.23\textwidth]{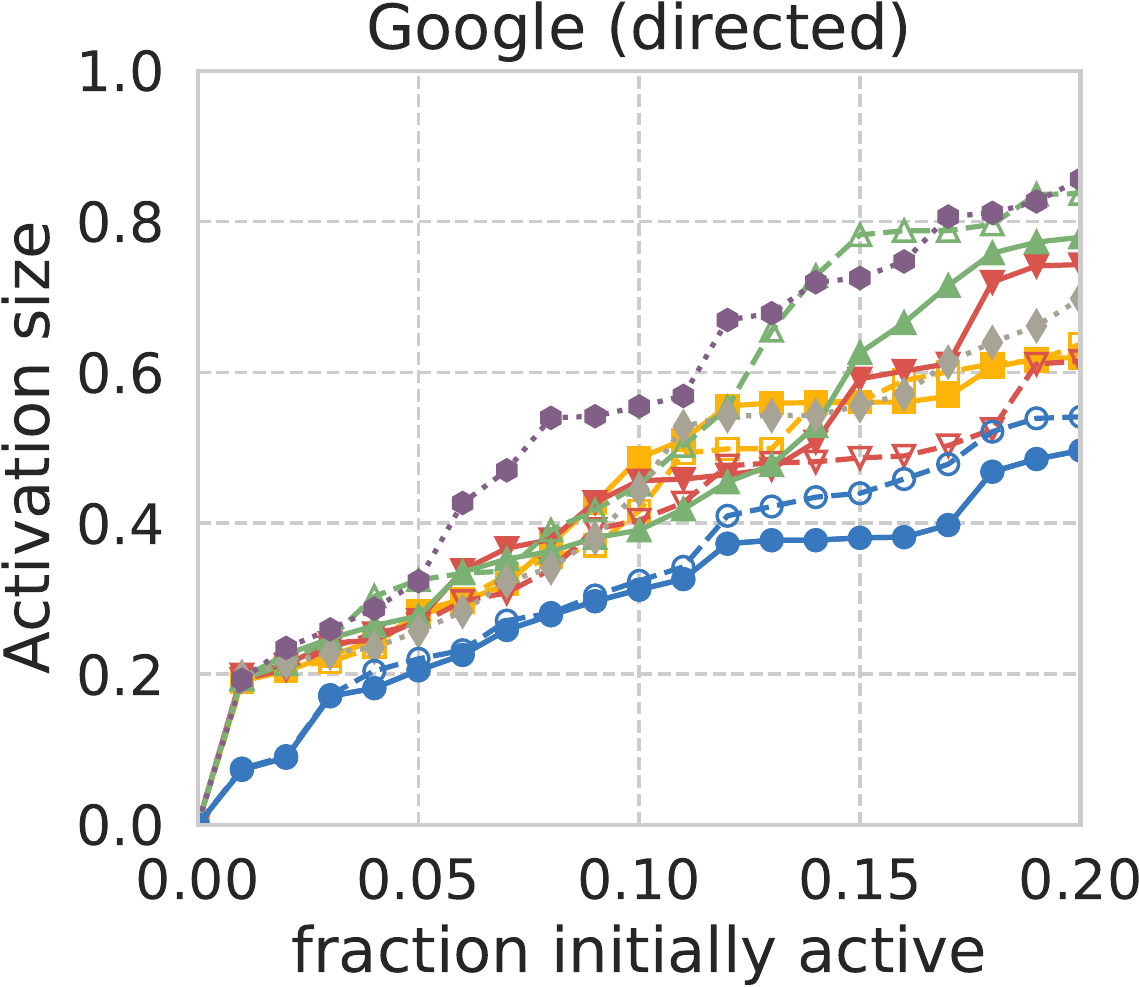}
    }\hfill
    \subfloat[\label{fig:lt-0.5-pgp}]{%
        \includegraphics[width=.23\textwidth]{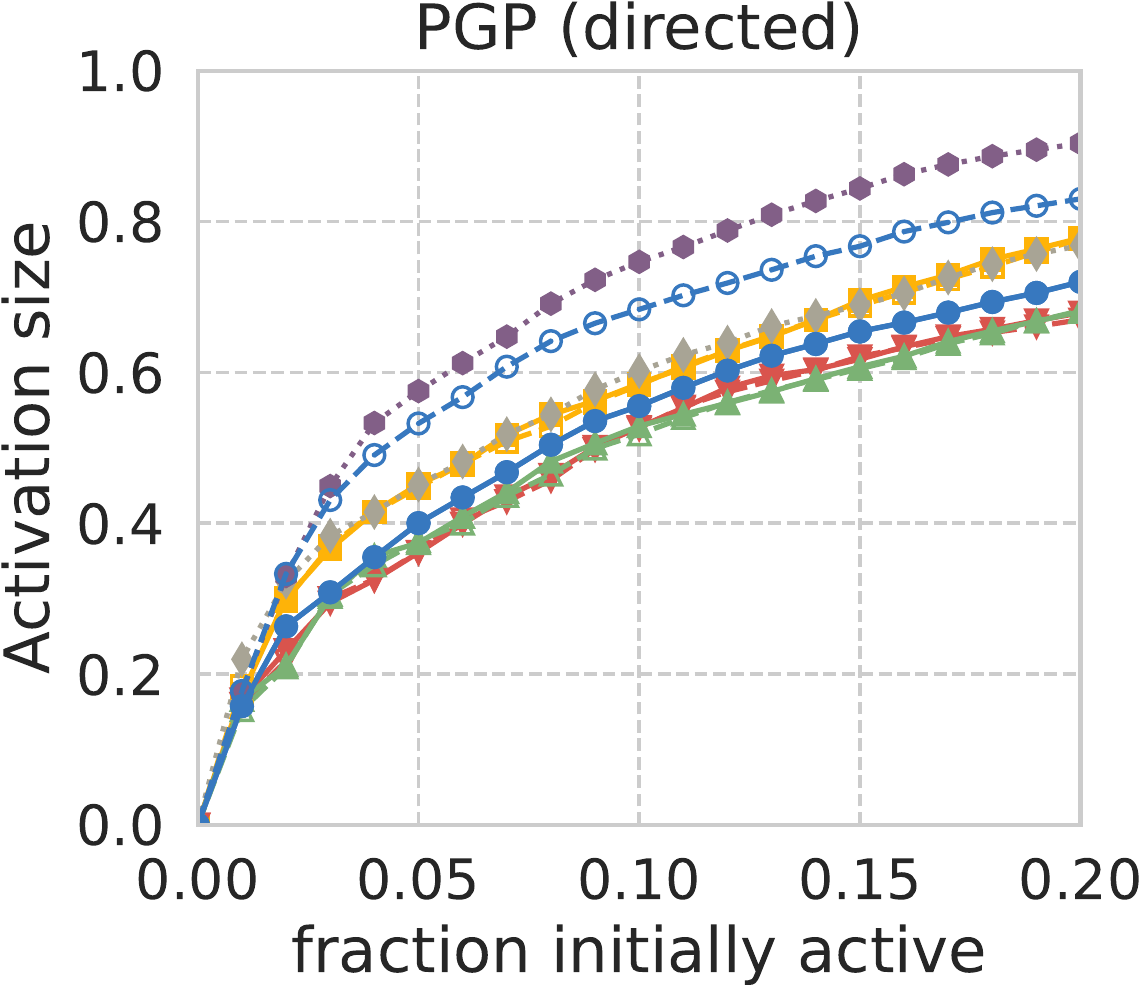}
    }\hfill
    \subfloat[\label{fig:lt-0.5-facebook-wall}]{%
        \includegraphics[width=.23\textwidth]{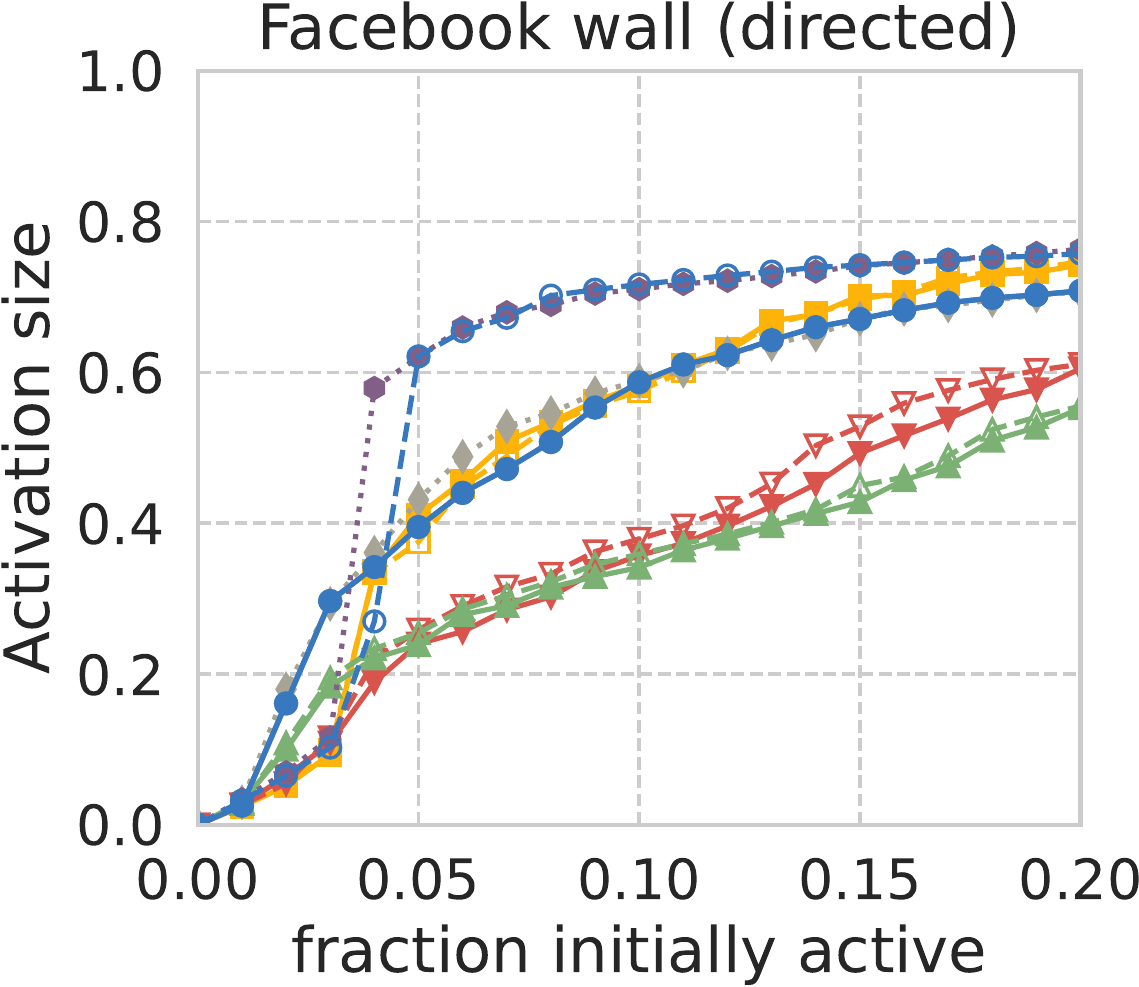}
    }
    \caption{Activation size for map equation centrality (MEC), modularity vitality (MV), community hub-bridge (CHB), community-based centrality (CBC), degree centrality (DC), and betweenness centrality (BC) in twelve empirical networks under the linear threshold model with threshold $t = 0.5$. Community structures are identified with Infomap; solid lines use the unrecorded link teleportation flow model, dashed lines use recorded node teleportation.}
    \label{fig:empirical-results-lt-0.5}
\end{figure*}

To infer the networks' community structure, we select the solution with the shortest codelength from 1000 Infomap runs, both using unrecorded link teleportation and recorded node teleportation with teleportation rate $0.15$ where the latter corresponds to standard PageRank.
We test different flow models because they describe different dynamic processes on the network, lead to different community structures, and are therefore suitable for different applications.
In our evaluation, we consider two-level partitions with non-overlapping communities.
We have also tested hierarchical partitions, but did not see a substantial performance difference.
For comparison, we include degree centrality as a local measure, betweenness centrality as a global measure, and the three community-aware centrality scores modularity vitality \cite{modularity-vitality}, community hub-bridge \cite{ghalmane2019immunization}, and community-based centrality \cite{zhao2015community}. 
Modularity vitality calculates a node $u$'s importance, given a network $G$ and a partition $\mathsf{M}$, as the difference in modularity between the original network and partition and the network and partition with $u$ removed, $Q\left(G,\mathsf{M}\right) - Q\left(G - \left\{u\right\}, \mathsf{M} - \left\{u\right\} \right)$, where $Q$ is the modularity function.
Depending on whether deleting a node and its incident links increases or decreases the partitions modularity, the result can be positive or negative.
Following previous evaluations, we consider modularity vitality's absolute value \cite{community-aware-centrality-evaluation}.
Community hub-bridge determines a node $u$'s importance by considering its intra- and inter-community links, weighing them by $u$'s own community size and the number of other communities it links to, respectively, assigning high importance to nodes with many links in large communities and nodes with many links to a large number of communities, $\sum_{\mathsf{m} \in \mathsf{M}} \left|\mathsf{m}\right| \cdot k_u^\mathsf{m} + \operatorname{NNC}_u \cdot k_u^{\overline{\mathsf{m}}}$.
Here, $k_u^\mathsf{m}$ is the number of $u$'s neighbours in module $\mathsf{m}$, $\operatorname{NNC}_u$ is $u$'s number of neighbouring communities, and $k_u^{\overline{\mathsf{m}}}$ is the number of $u$'s neighbours outside of $\mathsf{m}$.
Community-based centrality calculates a node's importance as the number of connections it has to the different communities, weighted by the communities' relative sizes, $\sum_{\mathsf{m} \in \mathsf{M}} k_u^\mathsf{m} \frac{\left|\mathsf{m}\right|}{N}$.

\subsection{Evaluation with the Linear Threshold Model}
The linear threshold model simulates the spread and adoption of ideas and behaviours through a network and has previously been applied to evaluate the performance of community-aware centrality scores \cite{10.1007/978-3-030-93409-5_29}.
In the linear threshold model, nodes can be in either of two states, that is, they can be active or inactive.
At the beginning of the simulation, we activate an $x$-fraction of the nodes, selected as the nodes with the highest centrality according to a centrality measure; all other nodes begin inactive.
Then, during each time step of the simulation, the inactive nodes check what fraction of their neighbours is active, and get activated if that fraction is at least as high as a given threshold $t$.
This threshold can be uniform across all nodes, or it can be node-dependent.
Here, in absence of node-dependent threshold information in the data, we use the uniform threshold $t = 0.5$, and include further results for thresholds $t' = 0.4$ and $t'' = 0.6$ in the appendix.
The simulation continues until no more nodes get activated; then we count the influence of the initially active nodes in terms of the activation size, that is the fraction of active nodes, where a larger activation size means that the initially active nodes have more influence.

\begin{figure*}
    \centering
    \subfloat[\label{fig:sir-facebook-friends}]{%
        \includegraphics[width=.23\textwidth]{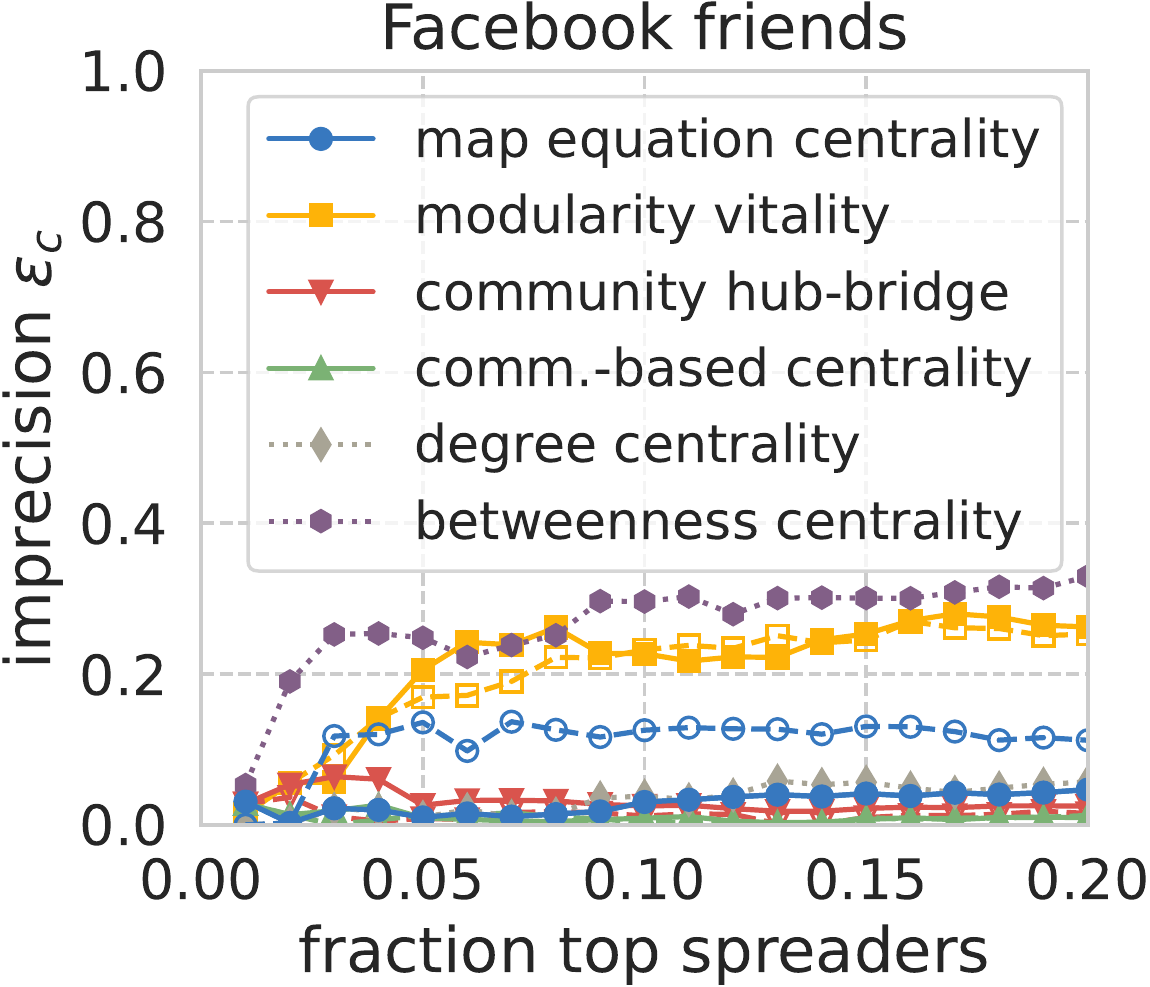}
    }\hfill
    \subfloat[\label{fig:sir-copenhagen}]{%
        \includegraphics[width=.23\textwidth]{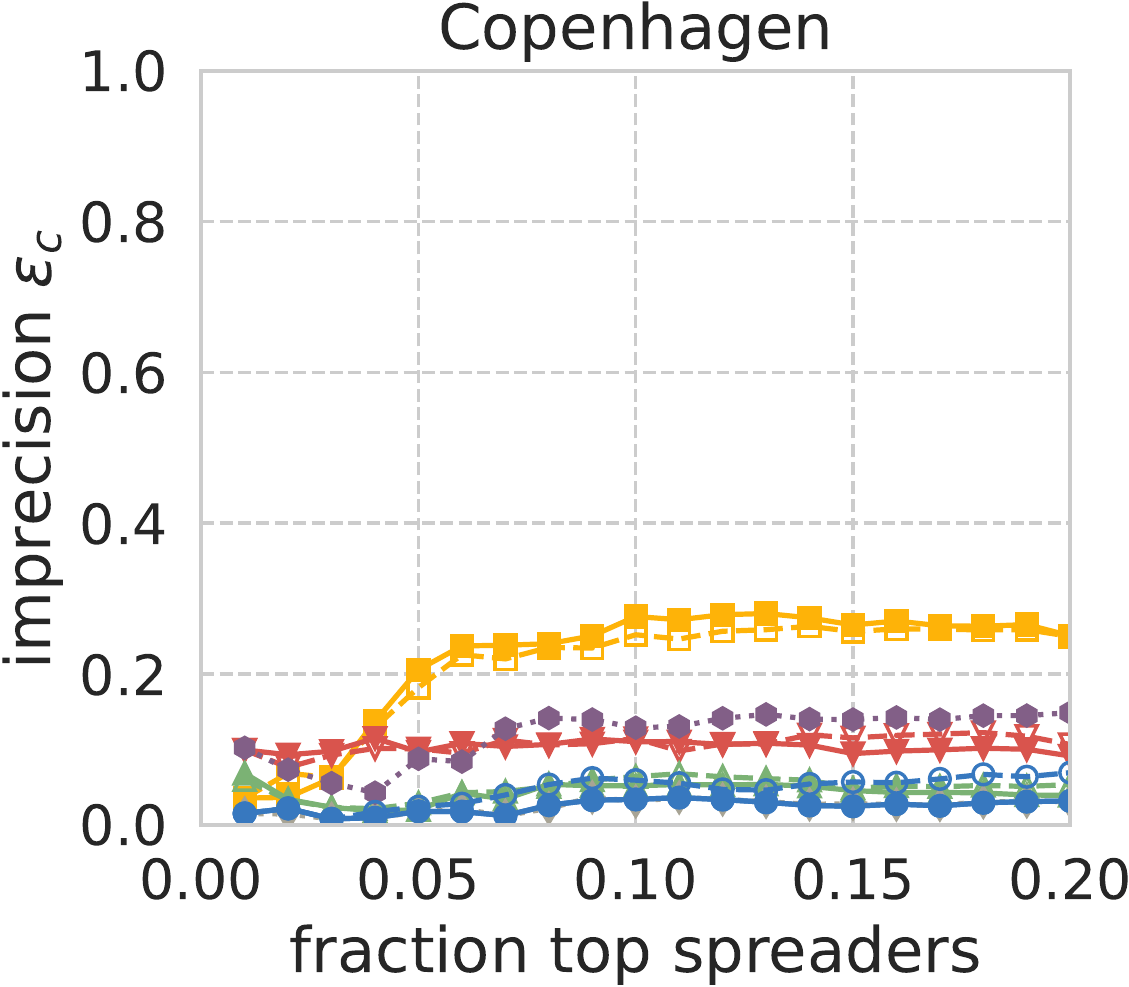}
    }\hfill
    \subfloat[\label{fig:sir-uni-email}]{%
        \includegraphics[width=.23\textwidth]{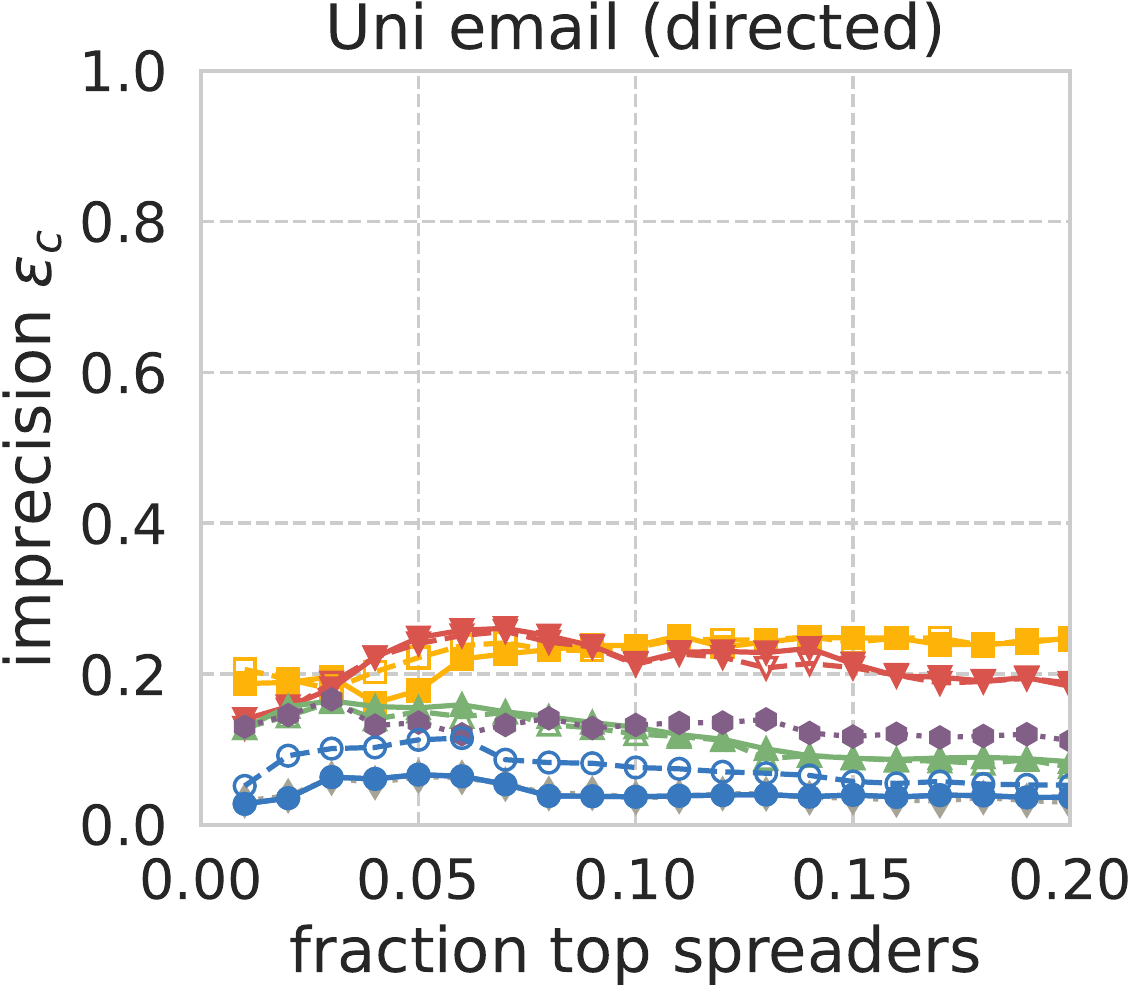}
    }\hfill
    \subfloat[\label{fig:sir-polblogs}]{%
        \includegraphics[width=.23\textwidth]{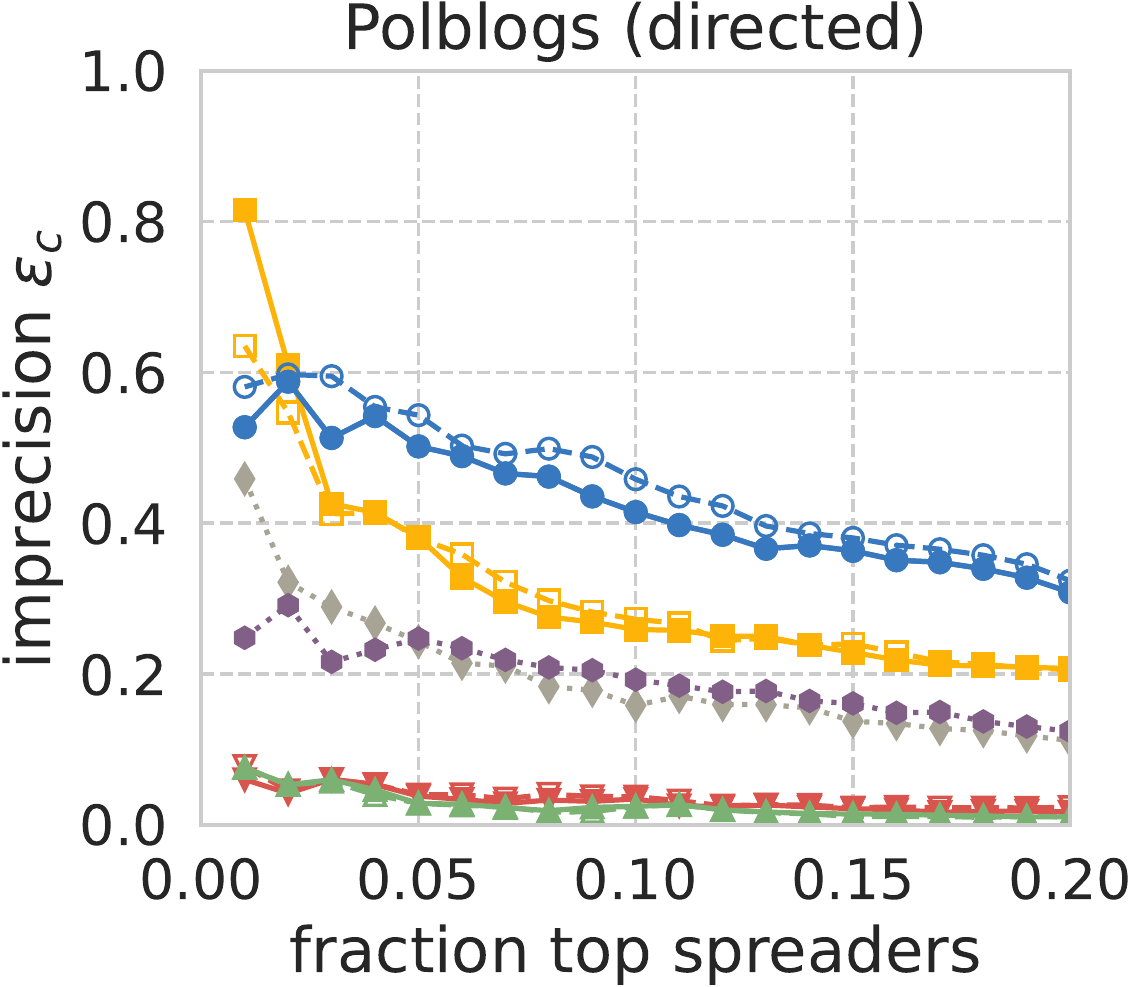}
    }\hfill
    \subfloat[\label{fig:sir-interactome-yeast}]{%
        \includegraphics[width=.23\textwidth]{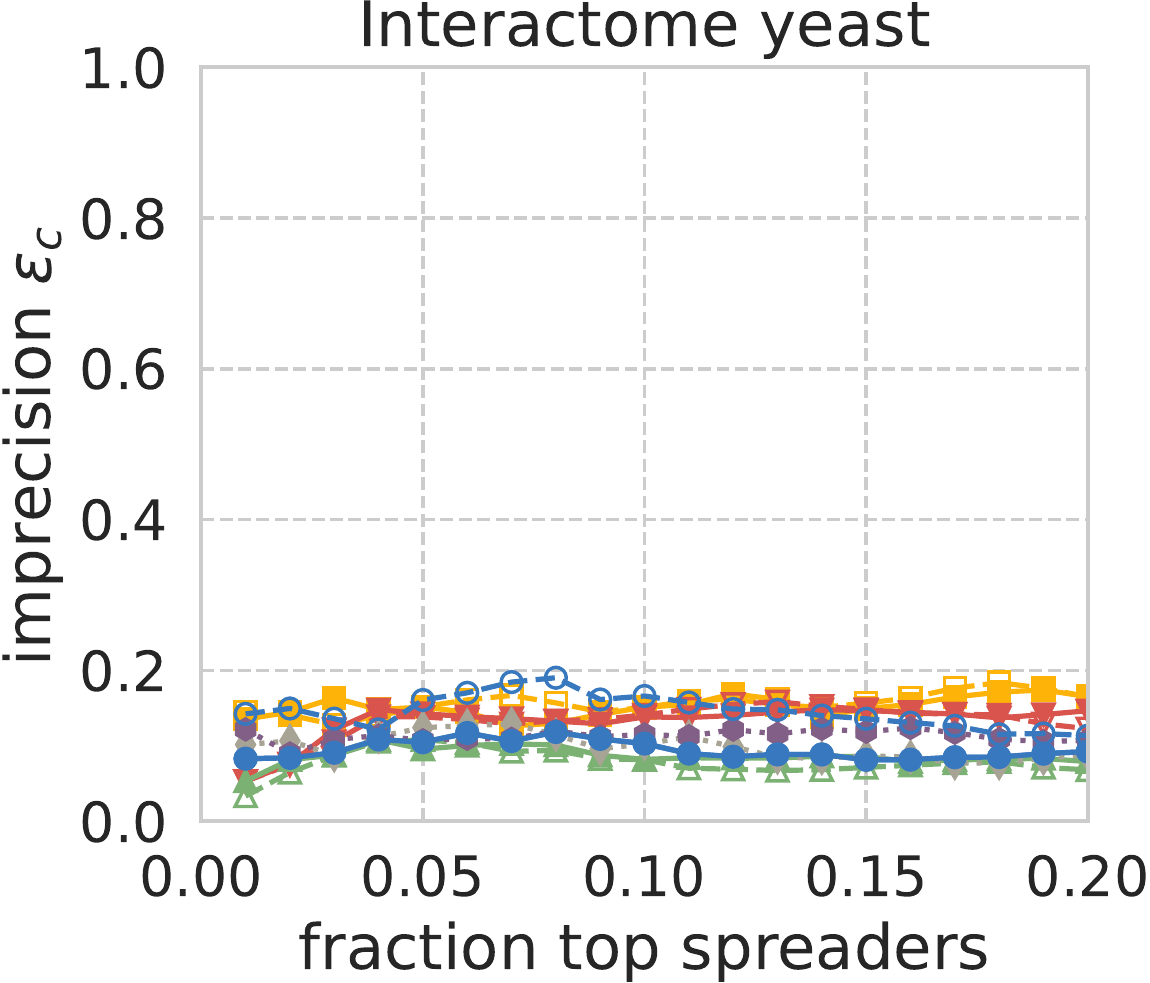}
    }\hfill
    \subfloat[\label{fig:sir-ego-facebook}]{%
        \includegraphics[width=.23\textwidth]{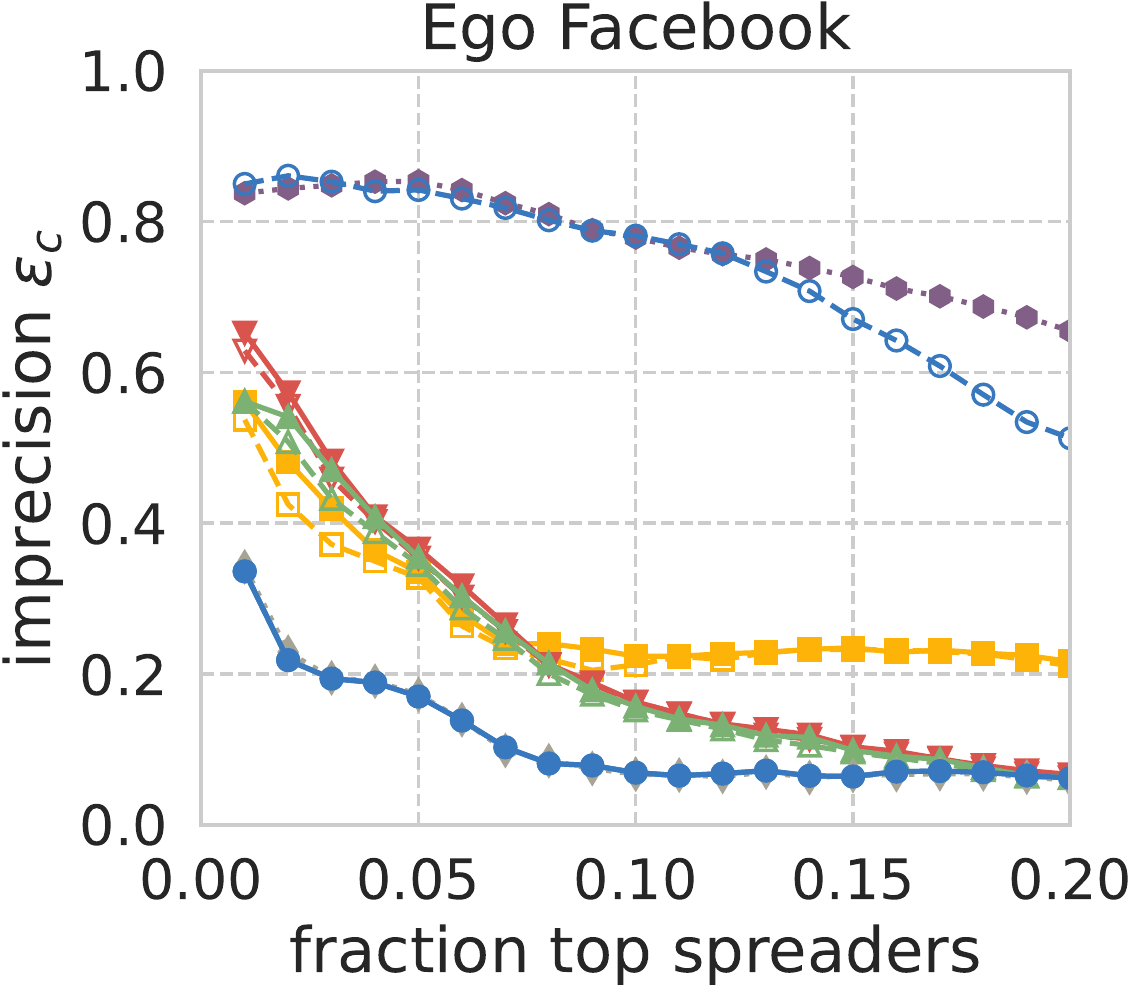}
    }\hfill
    \subfloat[\label{fig:sir-power}]{%
        \includegraphics[width=.23\textwidth]{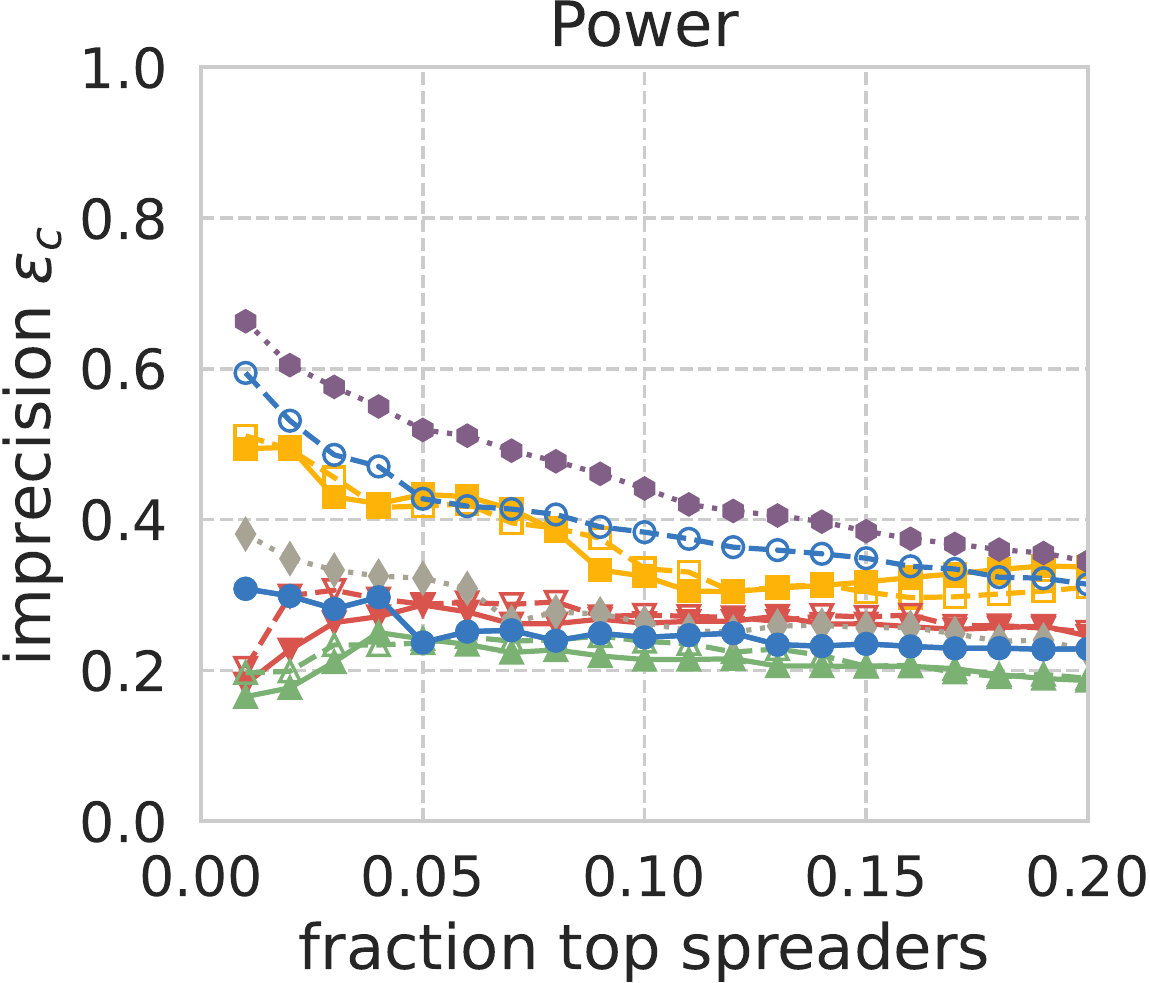}
    }\hfill
    \subfloat[\label{fig:sir-facebook-organizations}]{%
        \includegraphics[width=.23\textwidth]{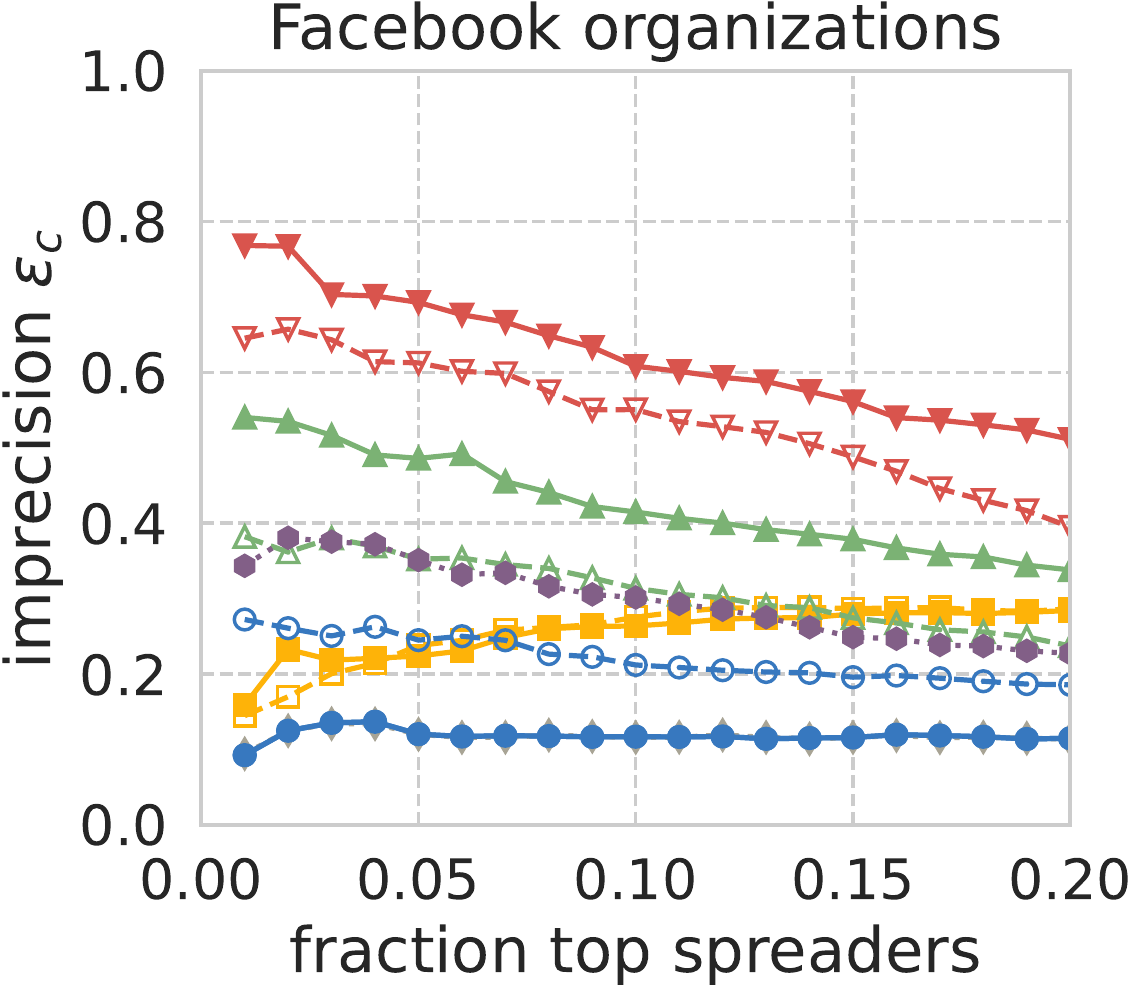}
    }\hfill
    \subfloat[\label{fig:sir-physics-collaborations}]{%
        \includegraphics[width=.23\textwidth]{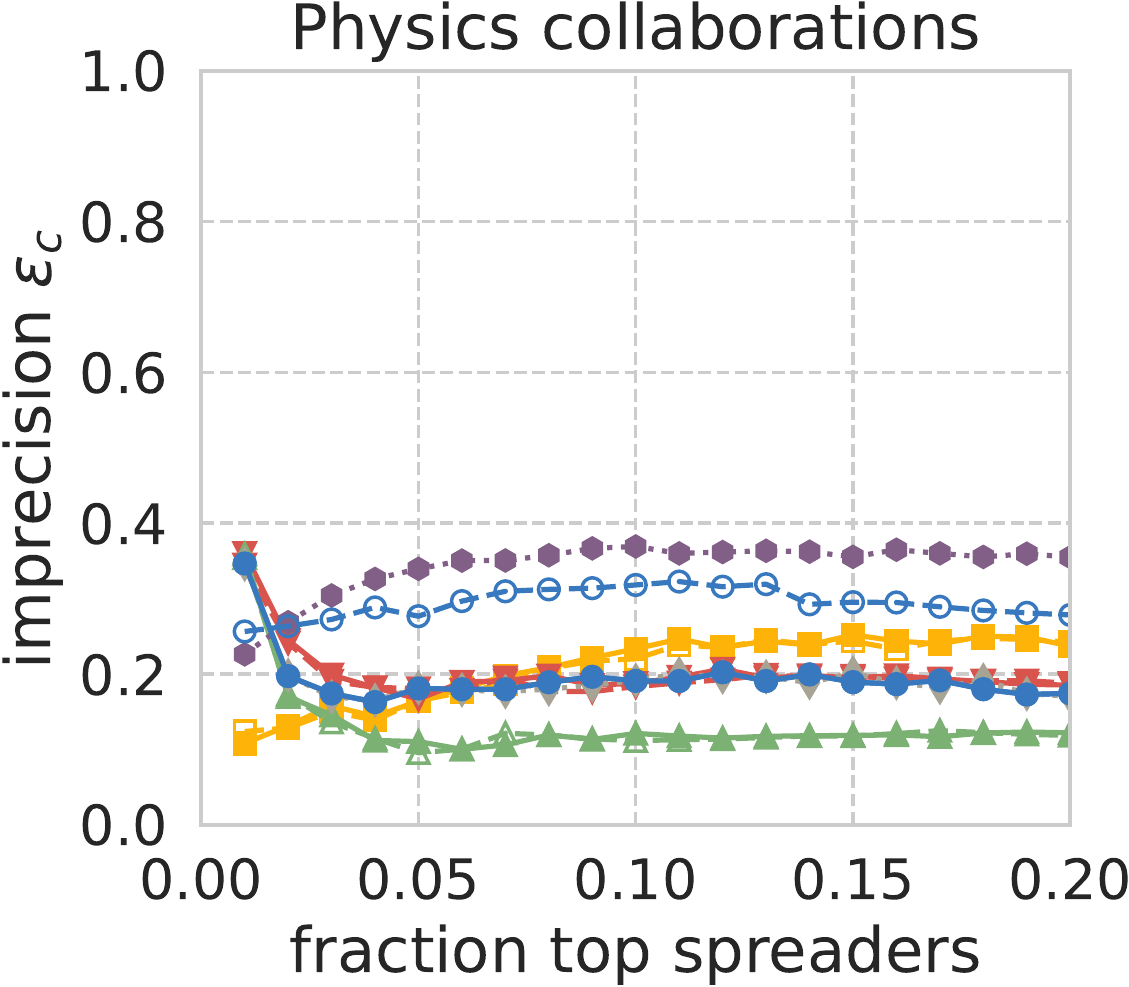}
    }\hfill
    \subfloat[\label{fig:sir-google}]{%
        \includegraphics[width=.23\textwidth]{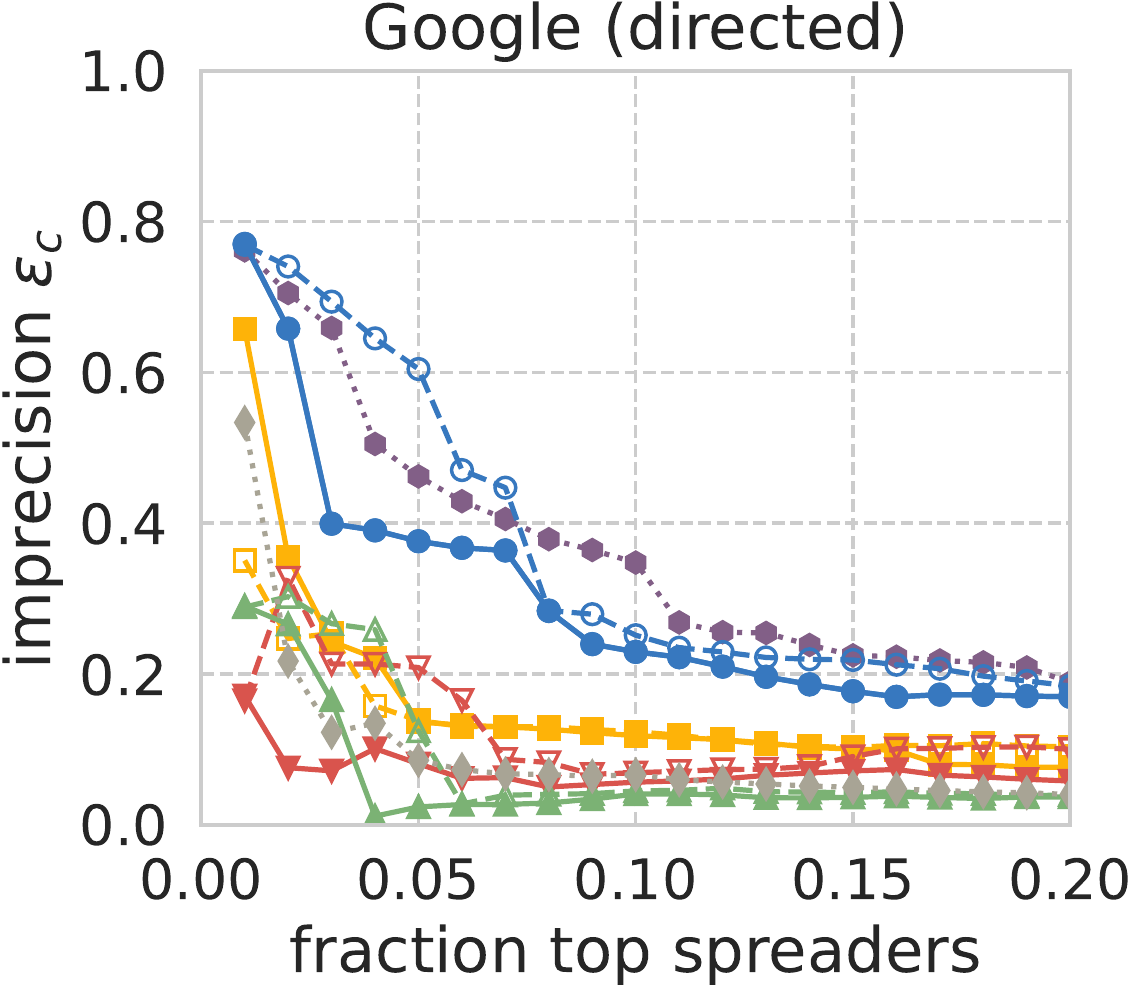}
    }\hfill
    \subfloat[\label{fig:sir-pgp}]{%
        \includegraphics[width=.23\textwidth]{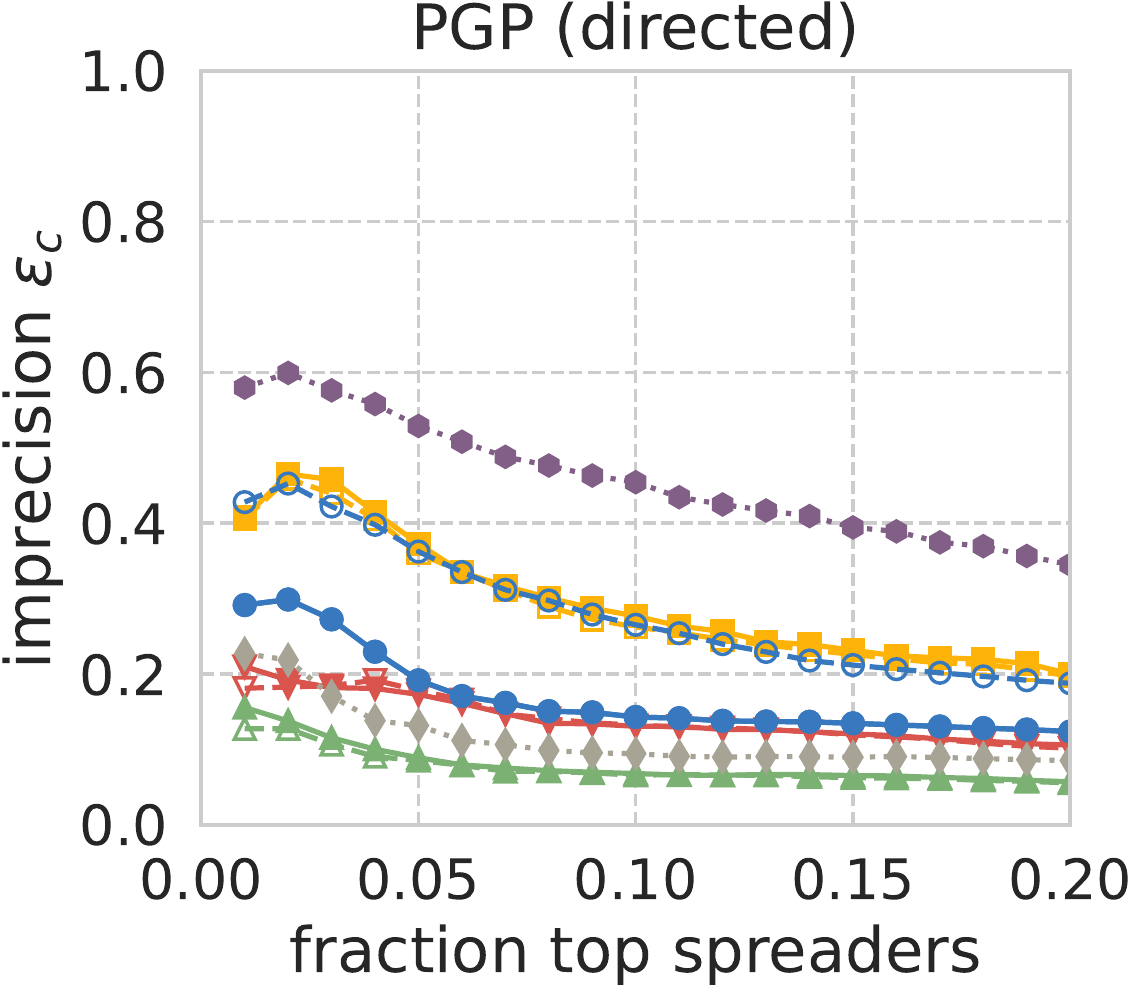}
    }\hfill
    \subfloat[\label{fig:sir-facebook-wall}]{%
        \includegraphics[width=.23\textwidth]{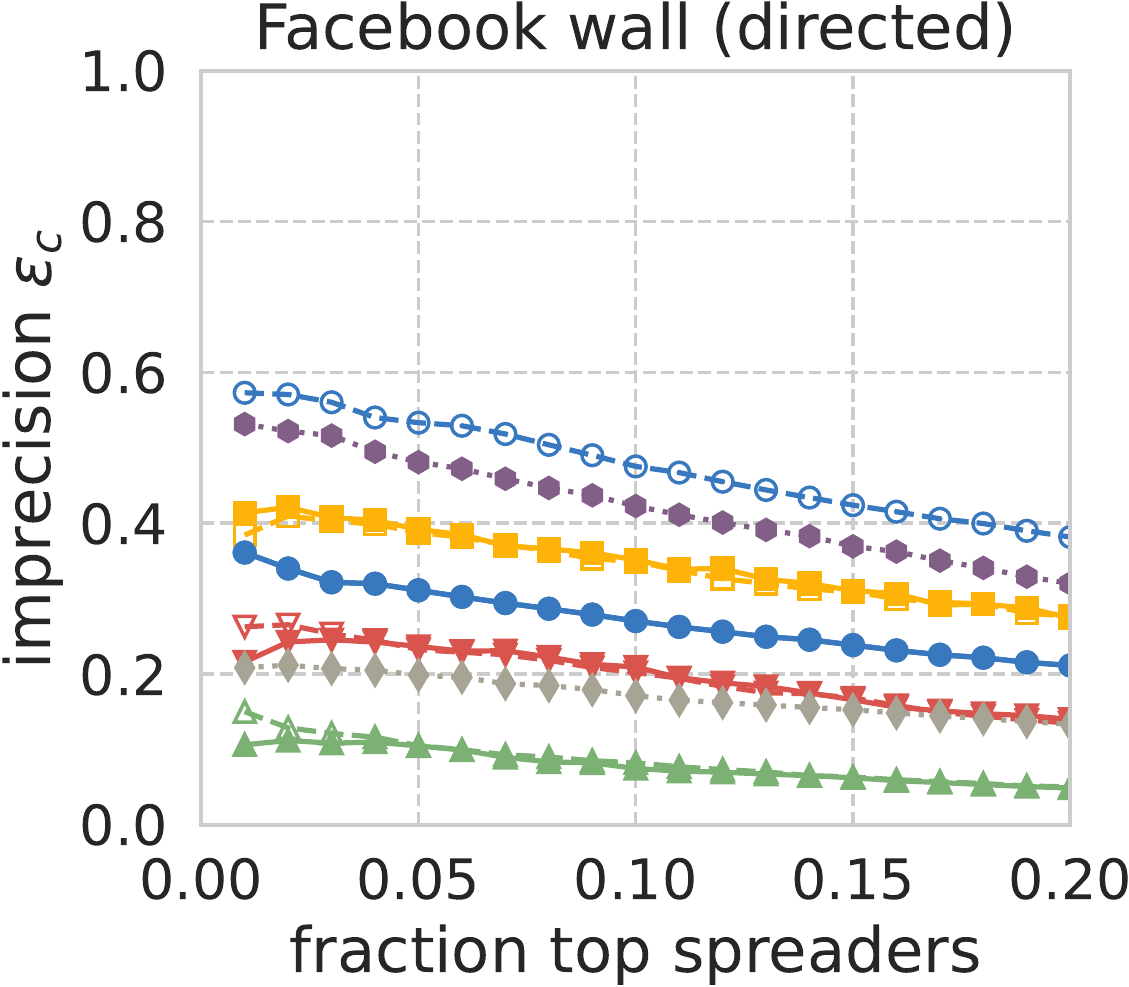}
    }
    \caption{Imprecision of map equation centrality, modularity vitality, community hub-bridge, community-based centrality, degree centrality, and betweenness centrality for identifying top spreaders in eight empirical networks. The curves show imprecision as a function of the fraction of top spreaders that are selected. A lower imprecision corresponds to more accurately identifying the top spreaders as determined with an SIR simulation. Community structures are identified with Infomap; solid lines use the unrecorded link teleportation flow model, dashed lines use recorded node teleportation. Degree and betweenness centrality do not rely on communities.}
    \label{fig:sir-results}
\end{figure*}

We find that, for large enough fractions of initially active nodes, map equation centrality outperforms the other measures when using the recorded link teleportation-based flow model in most cases, but has lower performance when based on unrecorded link teleportation.
Modularity vitality tends to outperform community hub-bridge, and community-based centrality, but is itself often outperformed by betweenness centrality, especially in the larger and directed networks.
For small fractions of initially active nodes, community hub-bridge and community-based centrality tend to perform better than map equation centrality and modularity vitality.
Further, the flow model choice has a larger effect on map equation centrality than on modularity vitality, community hub-bridge, and community-based centrality.
All of the measures tend to perform better when using the PageRank-based communities (Figs. \ref{fig:lt-0.5-facebook-friends} to \ref{fig:lt-0.5-facebook-wall}).
We describe the results in more detail in the appendix.

\subsection{Evaluation with the SIR Disease Spreading Model}
The second spreading process we use to evaluate map equation centrality's performance is a discrete-time SIR disease spreading simulation.
We follow the approach taken by Rajeh et al. \cite{community-aware-centrality-evaluation} to test how accurately the centrality measures identify influential nodes.

To estimate a node $u$'s influence, we calculate its spreading power, that is the expected number of nodes that get infected by a disease with the single initial spreader $u$:
Initially, only node $u$ is infected, all other nodes begin in the susceptible state, and the recovery time is set to 1 time step.
As long as there are infected nodes, the simulation continues.
Infected nodes infect their susceptible neighbours independently with probability $p_\text{th}$, then they recover.
Here, $p_\text{th}$ is the so-called epidemic threshold (\Tblref{tab:empirical-networks}) with $p_\text{th} = \frac{\left<k\right>}{\left<k^2\right> - \left<k\right>}$ where $\left<k\right> = \frac{1}{\left|V\right|} \sum_{v \in V} k_v$ and $\left<k^2\right> = \frac{1}{\left|V\right|} \sum_{v \in V} k_v^2$ are the first and second moment of the network's degree sequence, respectively \cite{wang2016predicting}.
When no infected nodes are left, the simulation ends, and we determine $u$'s spreading power as the number of recovered nodes.
Because of the stochasticity in the SIR model, we repeat the simulation 1000 times per node to calculate its expected spreading power.

Let $M_c$ and $M_\text{SIR}$ be the lists of nodes, ranked according to centrality score $c$, and their spreading power as determined with the SIR simulation, respectively.
Then, we measure the ability of centrality score $c$ to identify influential spreaders using the so-called imprecision function, $\epsilon_c\left(x\right) = 1 - \frac{M_c\left(x\right)}{M_\text{SIR}\left(x\right)}$ \cite{kitsak2010identification}.
Here, $M_c\left(x\right)$ and $M_\text{SIR}\left(x\right)$ are the average spreading power of the top $x$-fraction of nodes according to centrality score $c$ and the SIR simulation, respectively.
A smaller imprecision value corresponds to a better alignment between centrality score $c$ and spreading power.

In four of the tested networks, map equation centrality outperforms modularity vitality, community hub-bridge, and community-based centrality, performs second-to-third best in six networks, is worst or second-worst in the remaining two networks when based on unrecorded link teleportation.
The performance is often similar to degree centrality because the nodes' visit rates for unrecorded link teleportation are proportional to their degree in undirected networks \cite{PhysRevE.85.056107}.
Our community hub-bridge and community-based centrality implementations in directed networks consider nodes' outgoing links.
Because nodes with higher out degrees are expected to infect more nodes in the SIR model, our implementations may explain the measures' good performance.
In contrast, map equation centrality cares about the nodes' in-degree because the flow in the map equation framework is based on random walker transitions into the nodes.
To calculate degree and betweenness centrality, \texttt{networkx} considers the nodes' total degree.
With recorded node teleportation, map equation centrality does not perform as well and is often more similar to betweenness centrality.
Modularity vitality, community hub-bridge, and community-based centrality are less affected by the choice of flow model (Figs. \ref{fig:sir-facebook-friends} to \ref{fig:sir-facebook-wall}).
We describe the results in more detail in the appendix.

\begin{figure}
    \centering
    \subfloat[\label{fig:sir-copenhagen-modularity}]{%
        \includegraphics[width=.45\columnwidth]{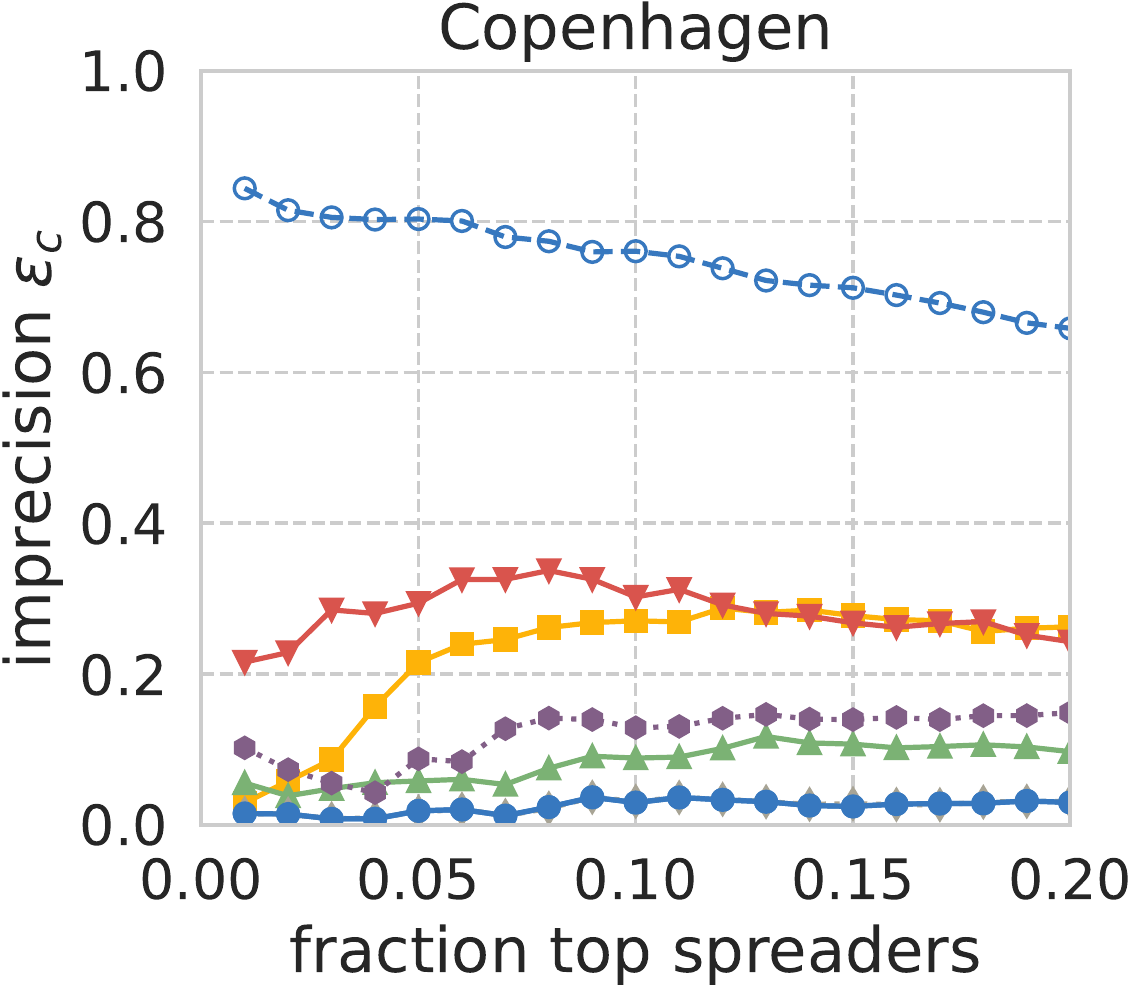}
    }
    \subfloat[\label{fig:sir-uni-email-modularity}]{%
        \includegraphics[width=.45\columnwidth]{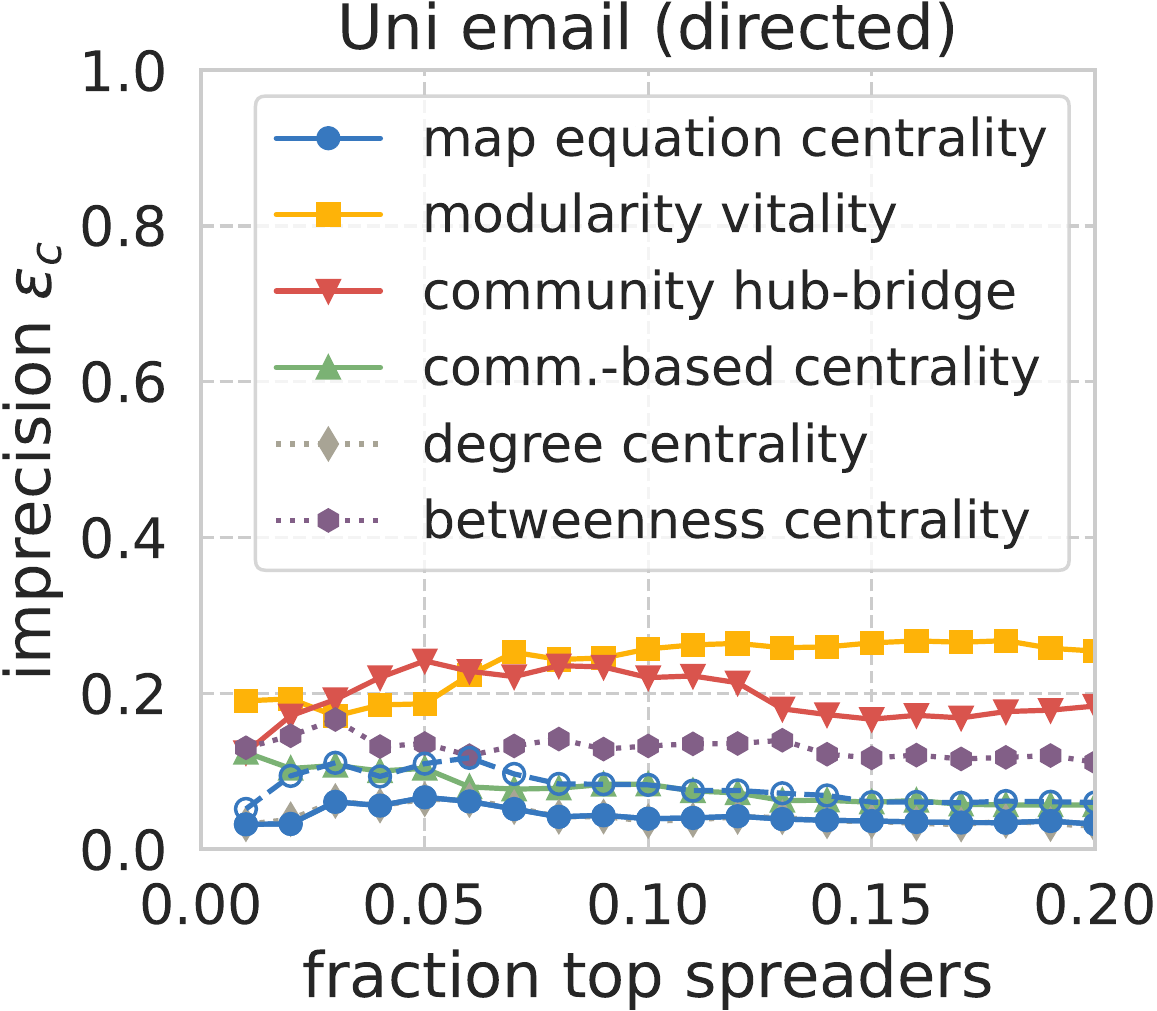}
    }\\
    \subfloat[\label{fig:sir-ego-facebook-modularity}]{%
        \includegraphics[width=.45\columnwidth]{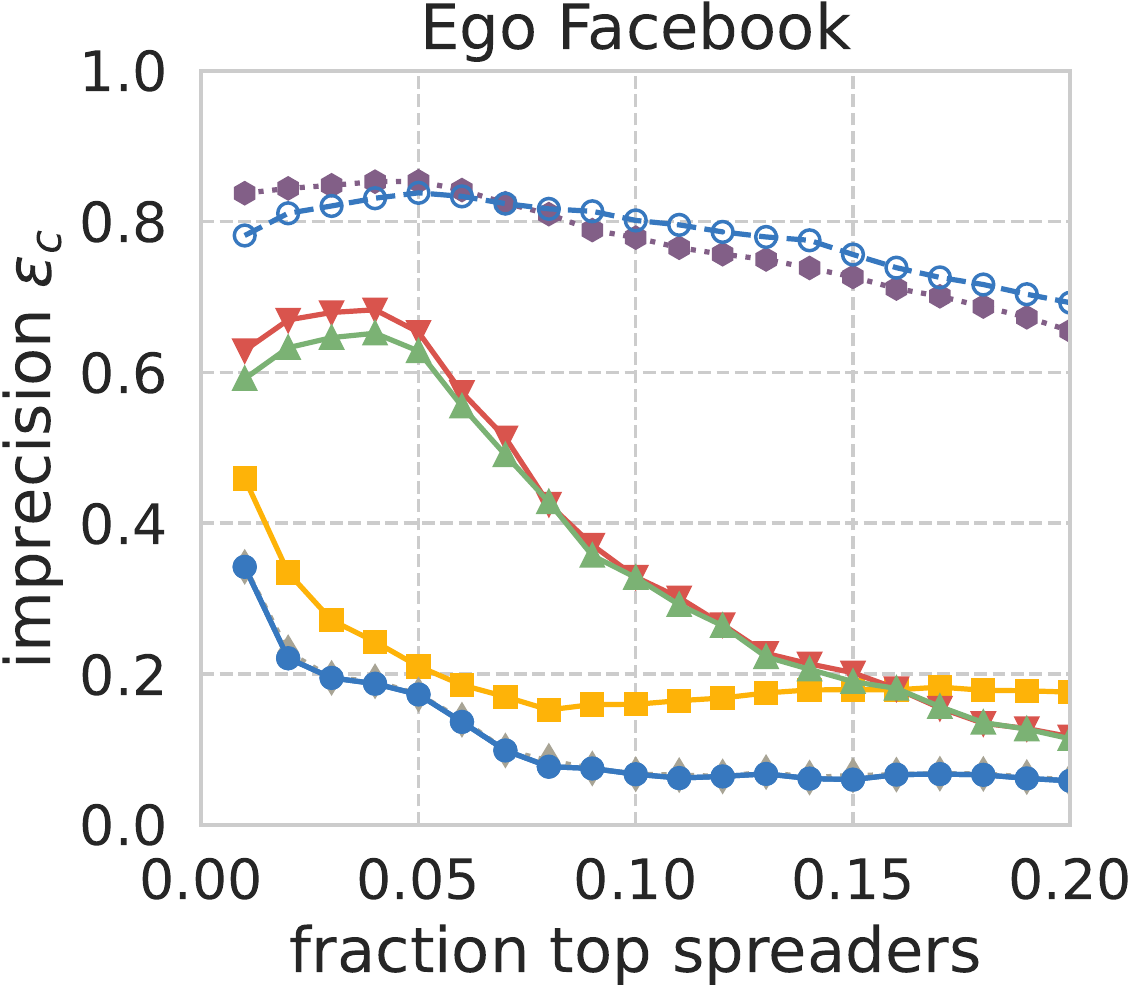}
    }
    \subfloat[\label{fig:sir-facebook-organizations-modularity}]{%
        \includegraphics[width=.45\columnwidth]{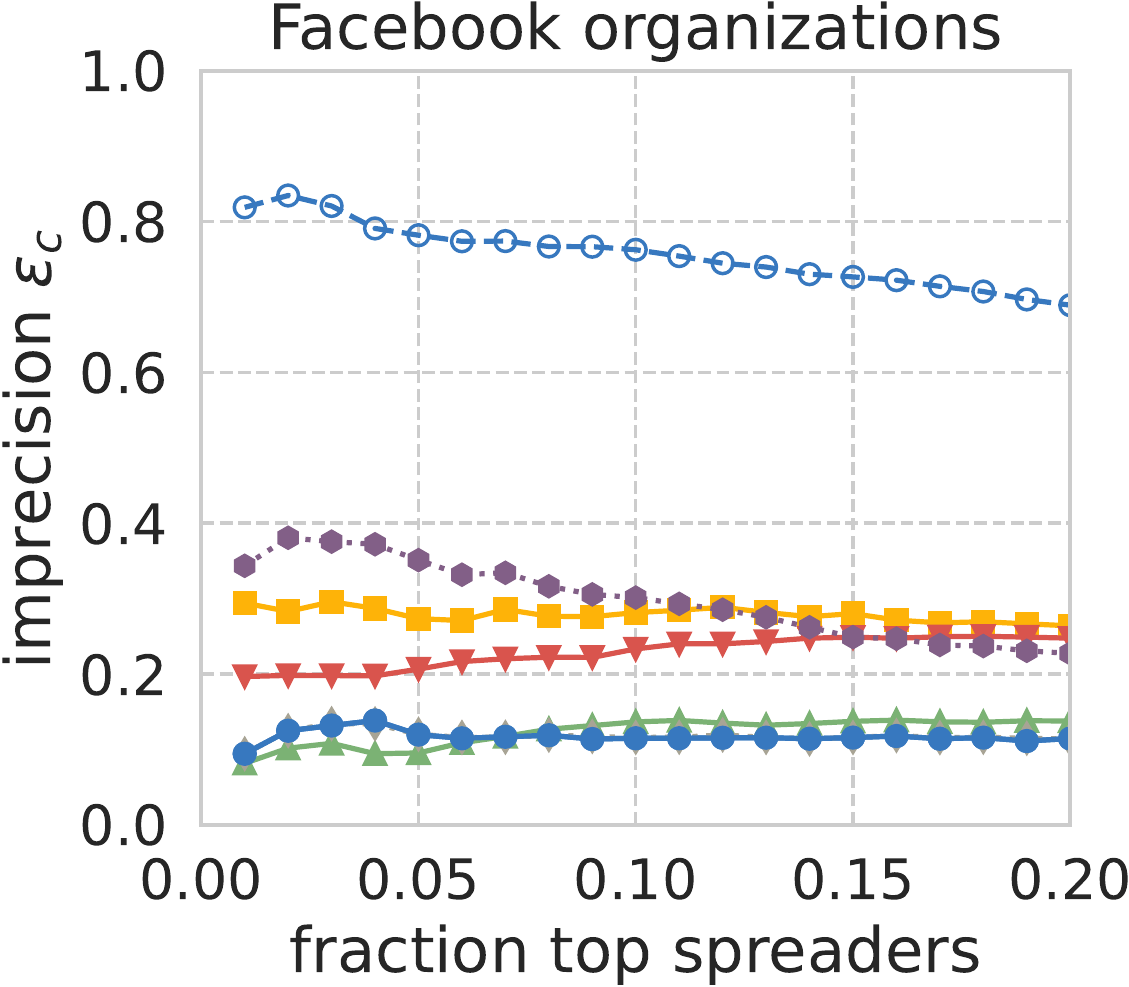}
    }
    \caption{Imprecision of map equation centrality, modularity vitality, community hub-bridge, community-based centrality, degree centrality, and betweenness centrality in four empirical networks based on partitions inferred thorugh modularity maximisation with the Louvain algorithm.}
    \label{fig:sir-results-modularity}
\end{figure}
To investigate whether map equation centrality is at an advantage because it is by definition faithful to the map equation, we have repeated our experiments in the four networks where map equation centrality performed best, using partitions based on modularity maximisation.
We infer the community structure in the Copenhagen, Uni email, Ego Facebook, and Facebook organizations networks, with the Louvain algorithm \cite{Blondel_2008}, using the \texttt{networkx} implementation, and proceed with highest-modularity partitions from 1000 runs with different seeds.
Louvain detects 13 (11) communities in the Copenhagen network, 20 (17) communities in the Uni email network, 17 (12) communities in the Ego Facebook network, and 13 (9) communities in the Facebook organizations network, where the numbers in parenthesis are the effective numbers of communities.
Overall, we find that map equation centrality with on unrecorded link teleportation and modularity vitality perform similar to before (Figs. \ref{fig:sir-copenhagen-modularity} to \ref{fig:sir-facebook-organizations-modularity}).
Whether community hub-bridge and community-based centrality perform better or worse depends on the network, and map equation centrality with recorded node teleportation performs worse than before.

To summarise, we found that none of the tested centrality scores outperforms all other scores in all networks, but none of the scores performed worst in all cases either.

\subsection{Distribution of Influential Nodes}
\begin{figure*}
    \centering
    \subfloat[\label{fig:perplexity-facebook-friends}]{%
        \includegraphics[width=.23\textwidth]{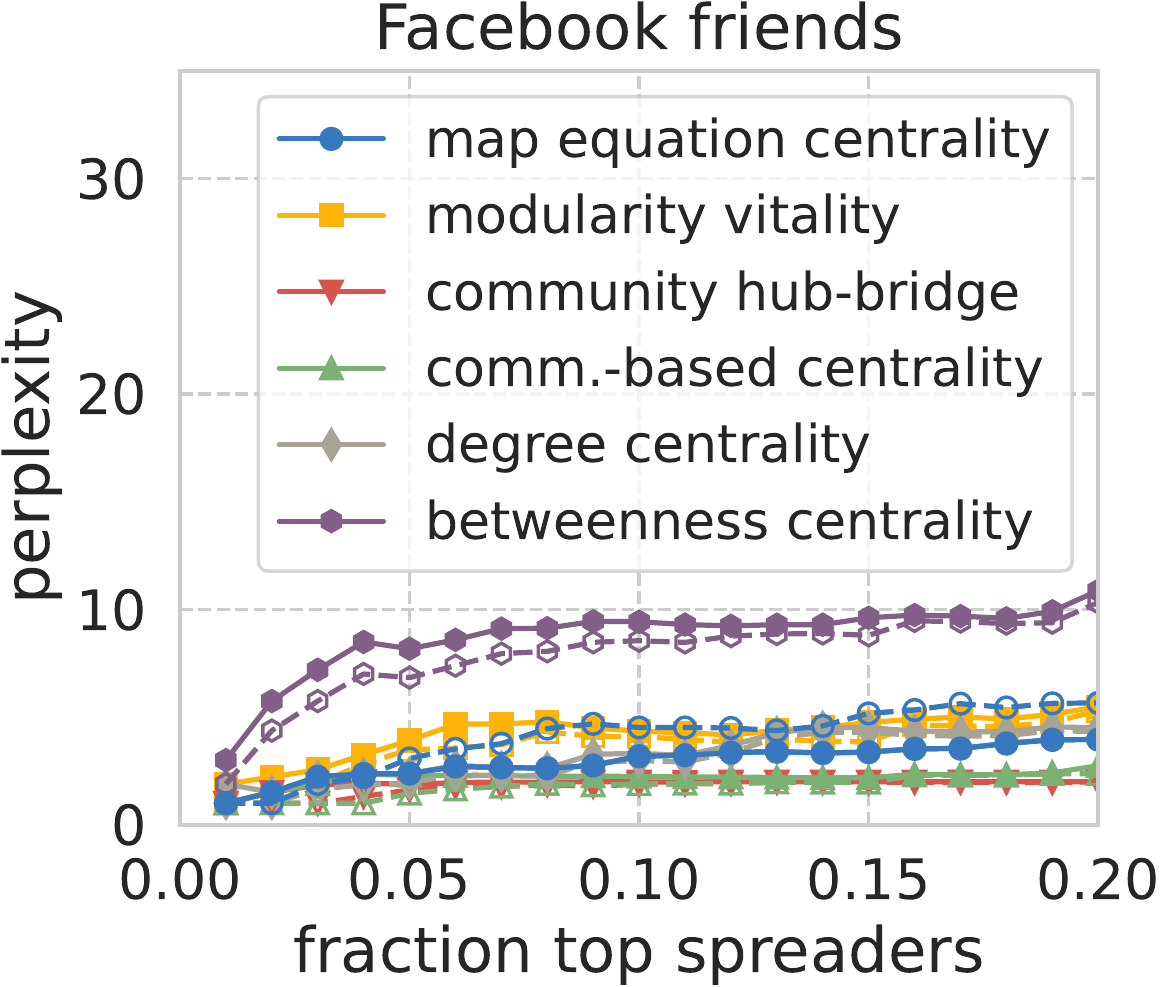}
    }\hfill
    \subfloat[\label{fig:perplexity-copenhagen}]{%
        \includegraphics[width=.23\textwidth]{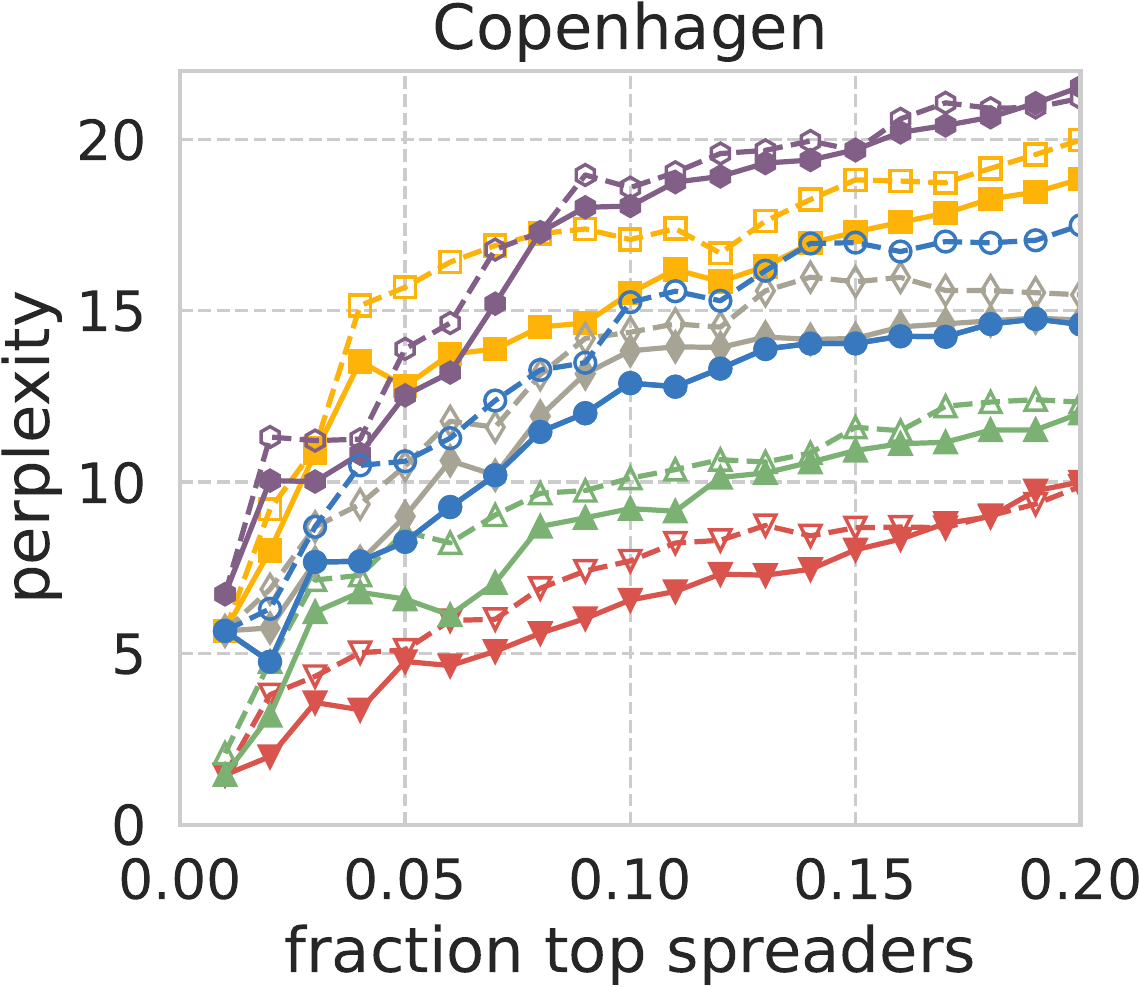}
    }\hfill
    \subfloat[\label{fig:perplexity-uni-email}]{%
        \includegraphics[width=.23\textwidth]{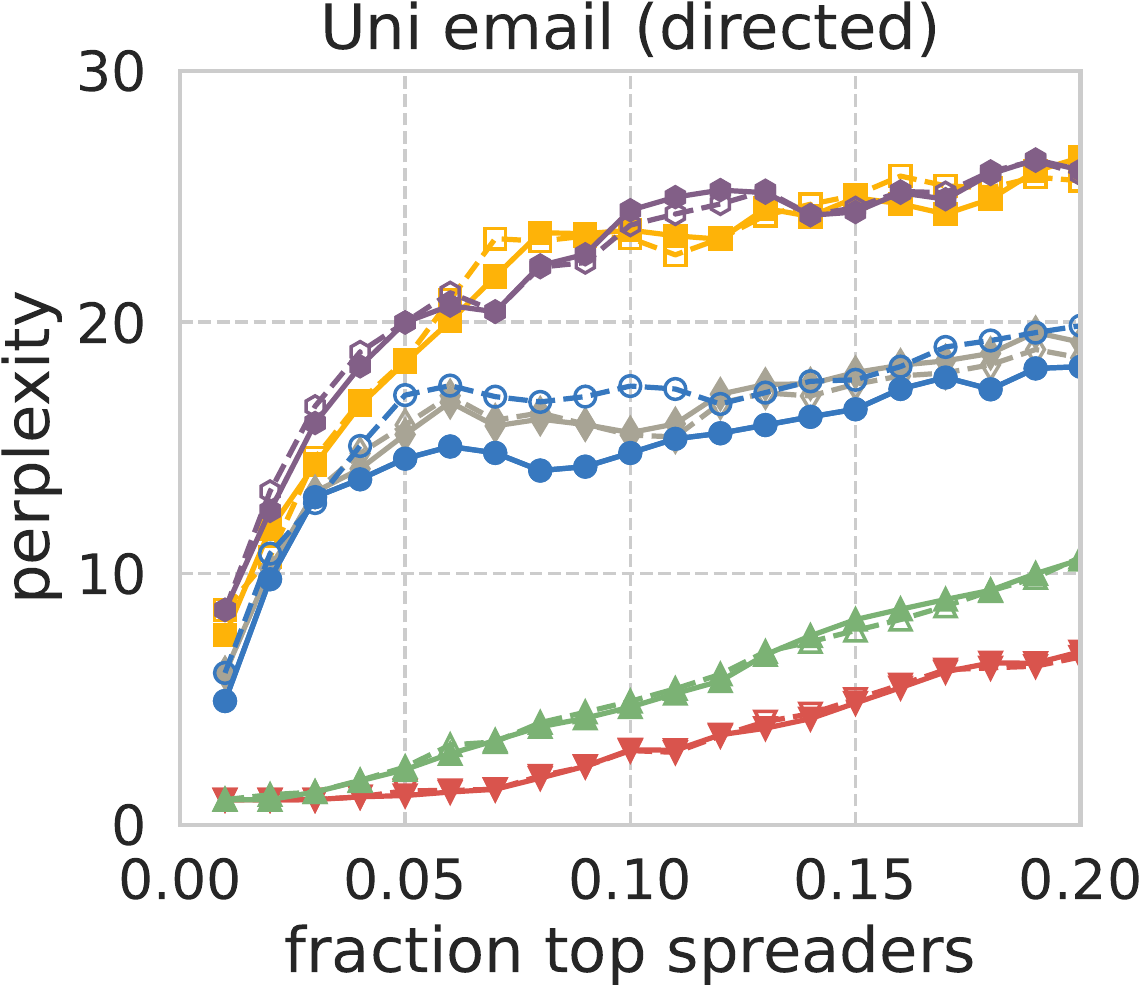}
    }\hfill
    \subfloat[\label{fig:perplexity-polblogs}]{%
        \includegraphics[width=.23\textwidth]{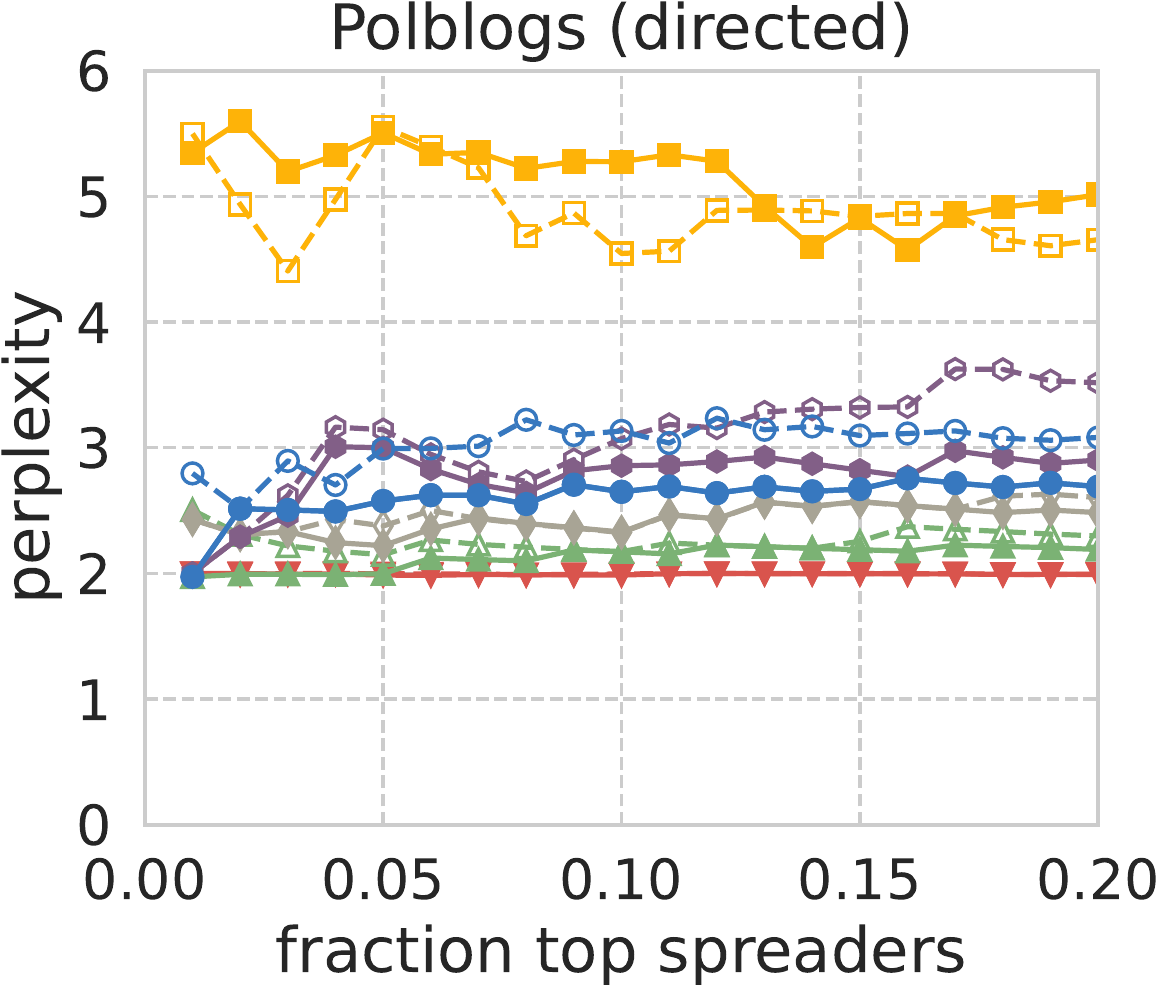}
    }\hfill
    \subfloat[\label{fig:perplexity-interactome-yeast}]{%
        \includegraphics[width=.23\textwidth]{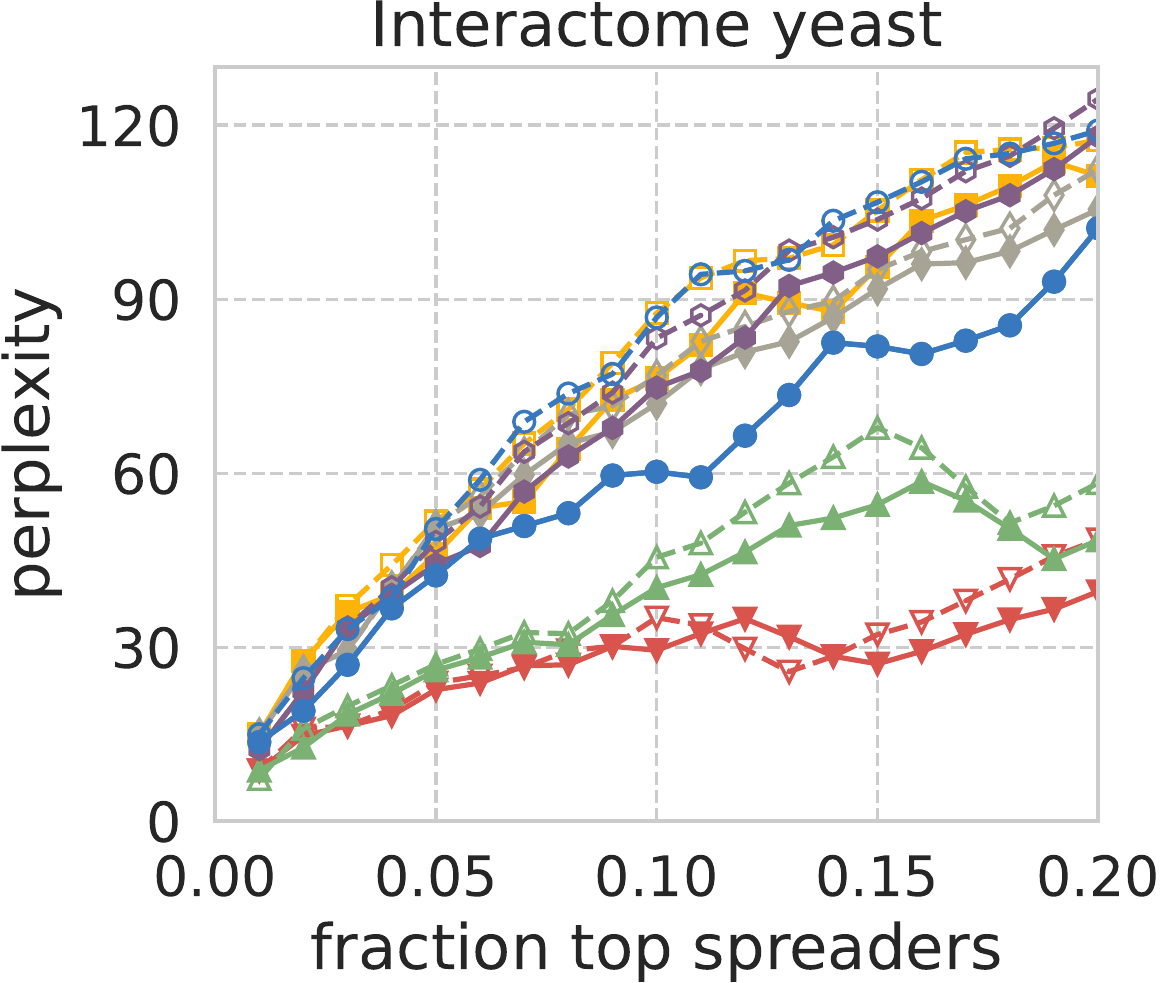}
    }\hfill
    \subfloat[\label{fig:perplexity-ego-facebook}]{%
        \includegraphics[width=.23\textwidth]{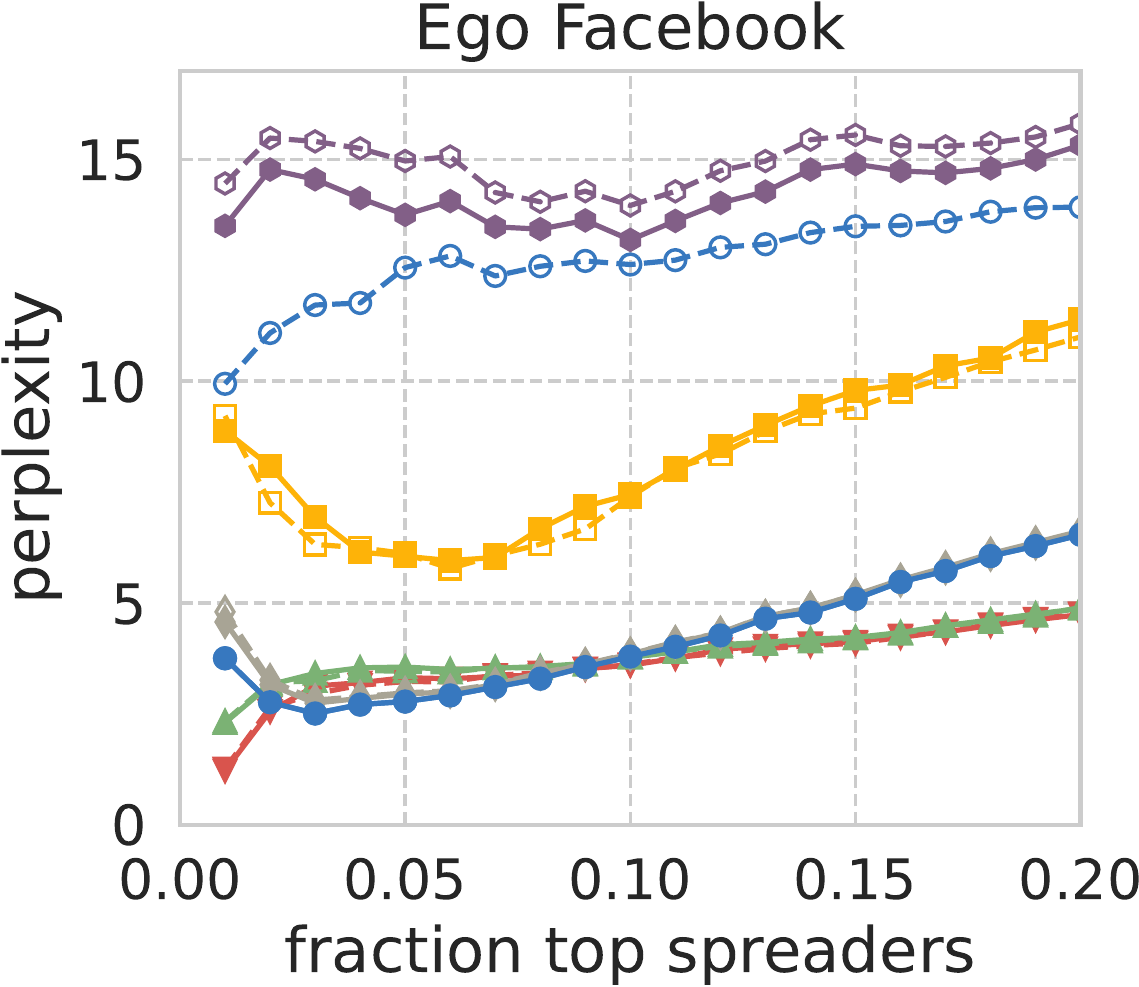}
    }\hfill
    \subfloat[\label{fig:perplexity-power}]{%
        \includegraphics[width=.23\textwidth]{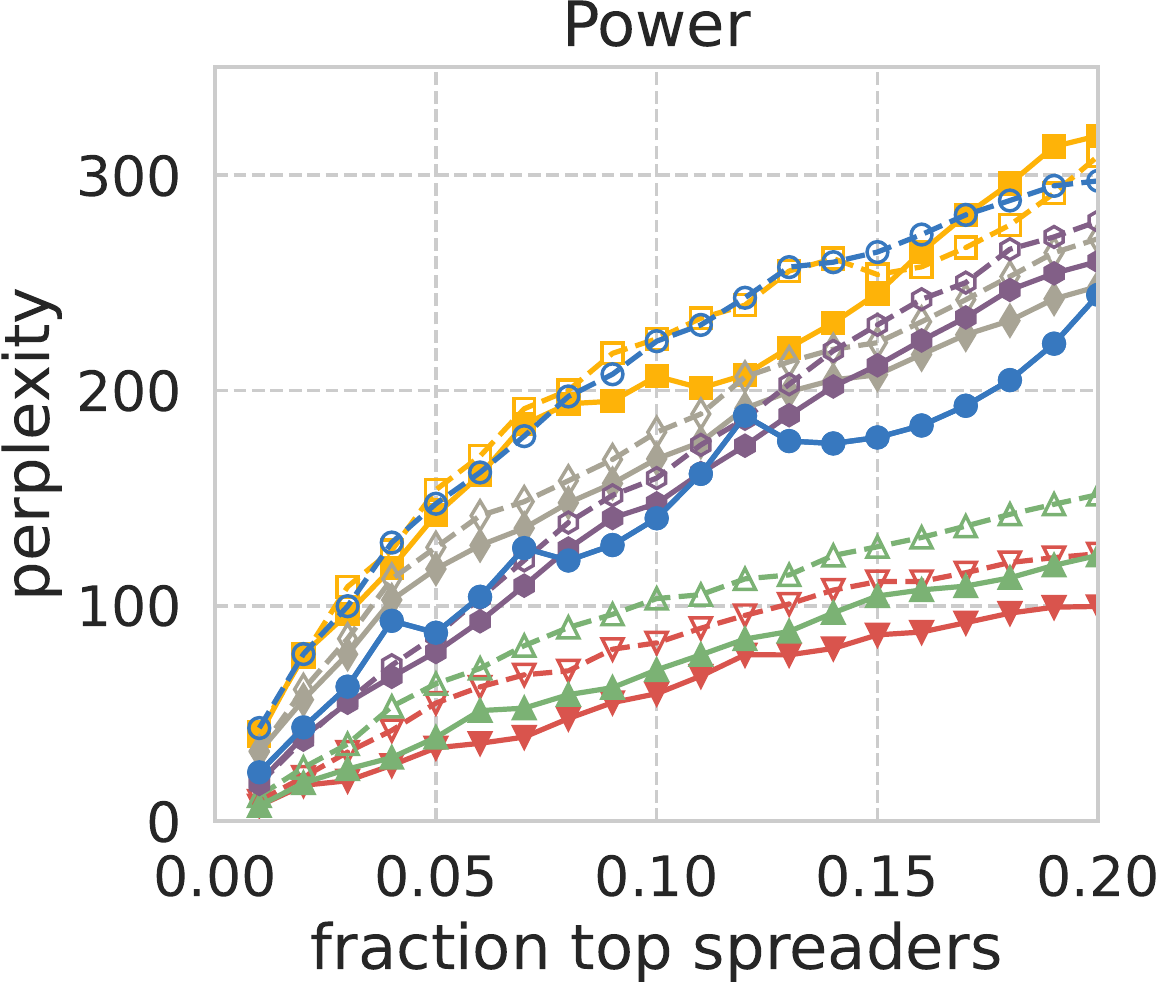}
    }\hfill
    \subfloat[\label{fig:perplexity-facebook-organizations}]{%
        \includegraphics[width=.23\textwidth]{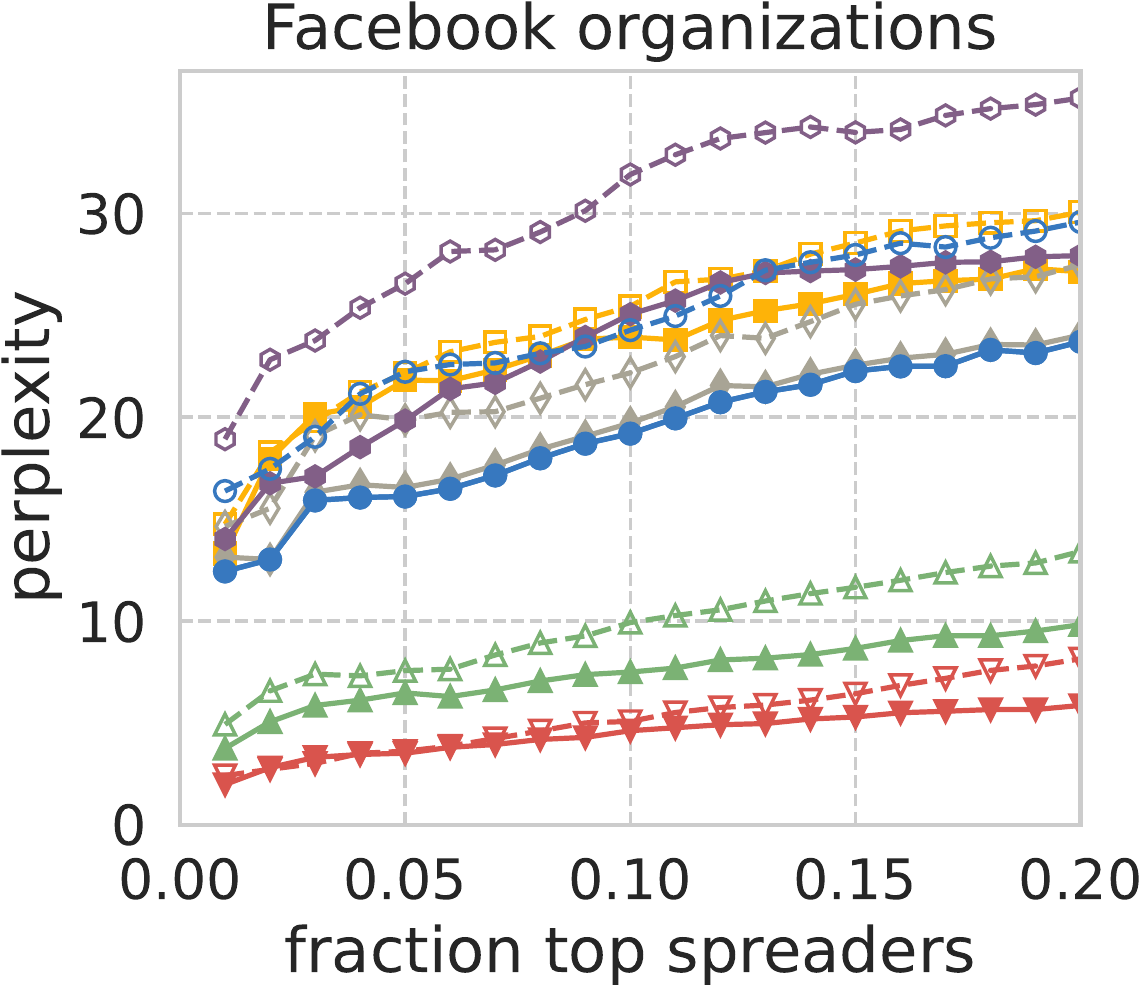}
    }\hfill
    \subfloat[\label{fig:perplexity-physics-collaborations}]{%
        \includegraphics[width=.23\textwidth]{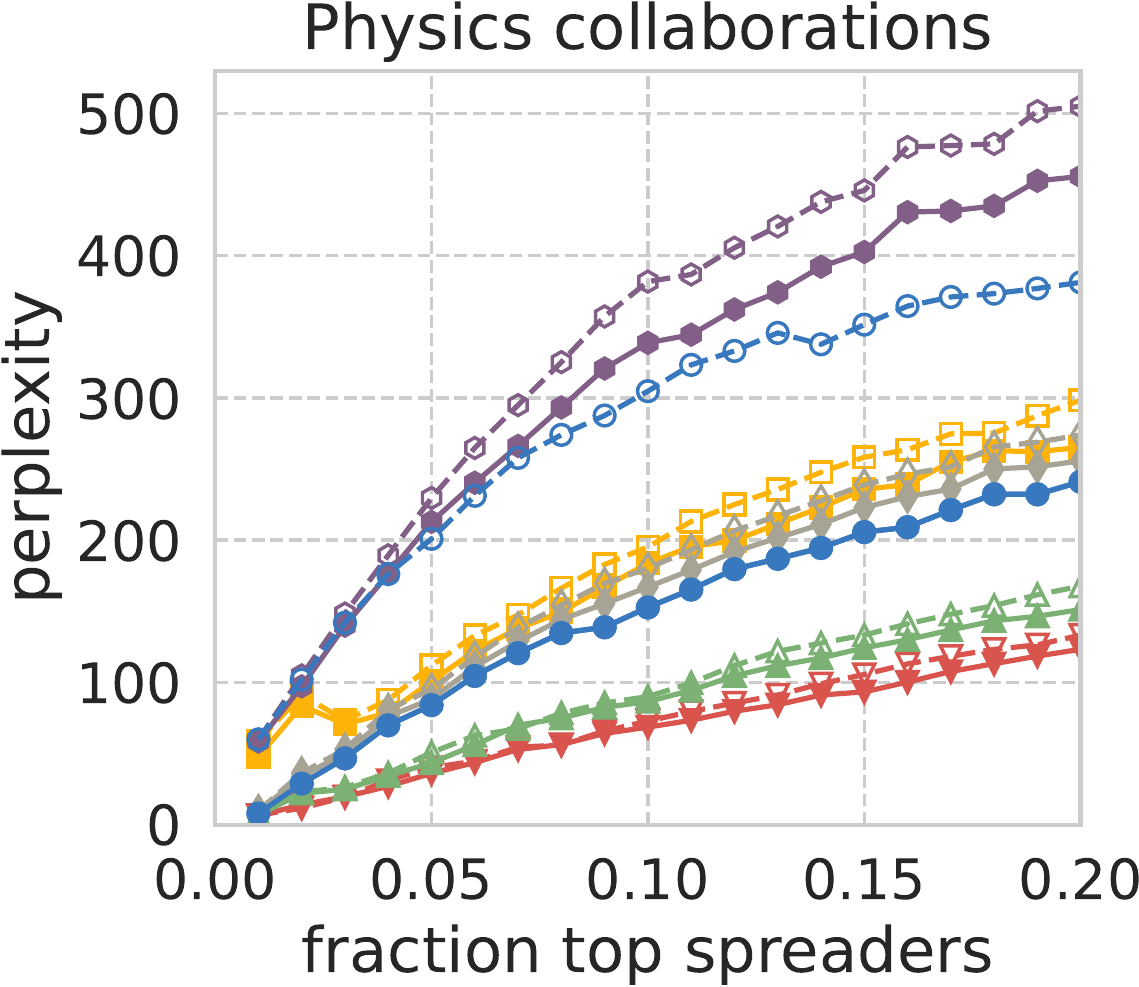}
    }\hfill
    \subfloat[\label{fig:perplexity-google}]{%
        \includegraphics[width=.23\textwidth]{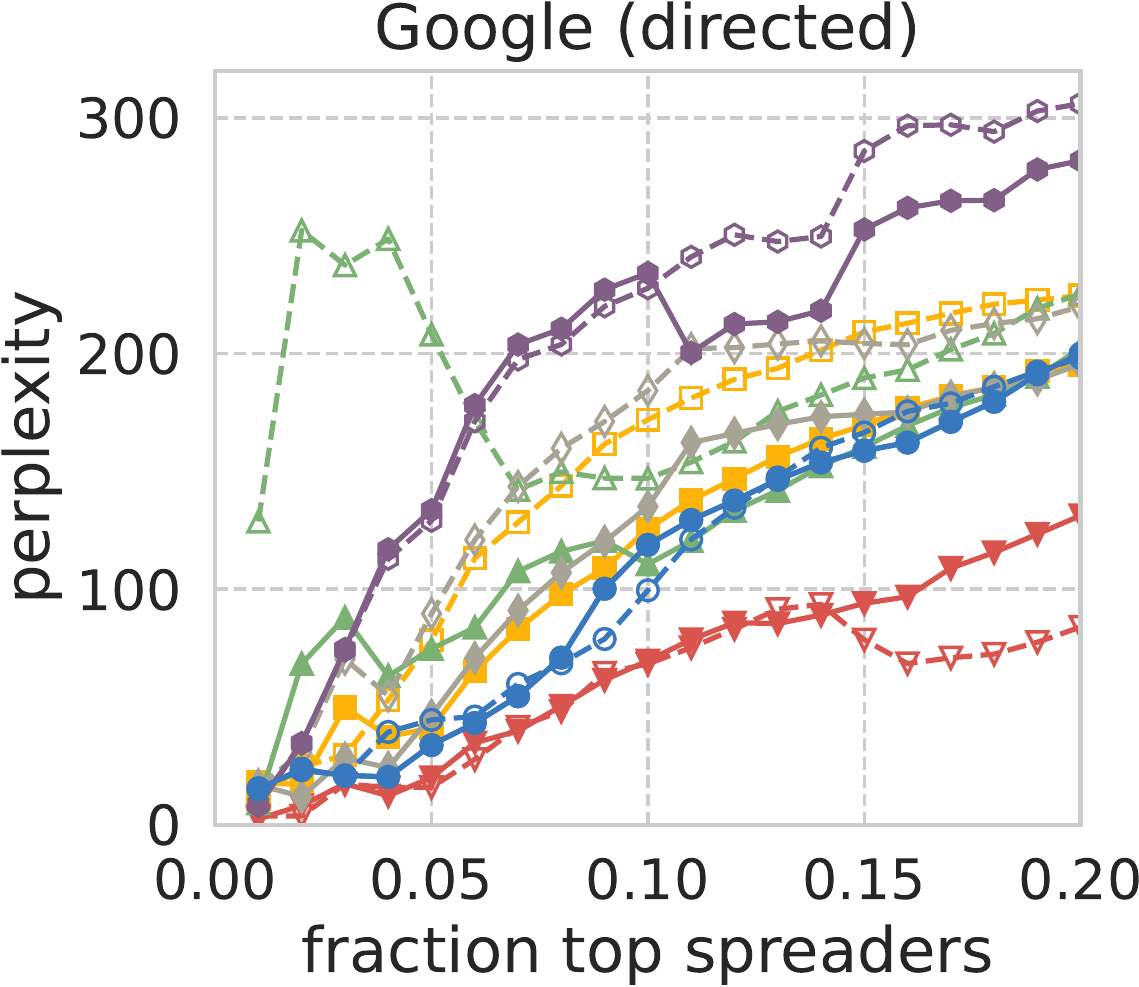}
    }\hfill
    \subfloat[\label{fig:perplexity-pgp}]{%
        \includegraphics[width=.23\textwidth]{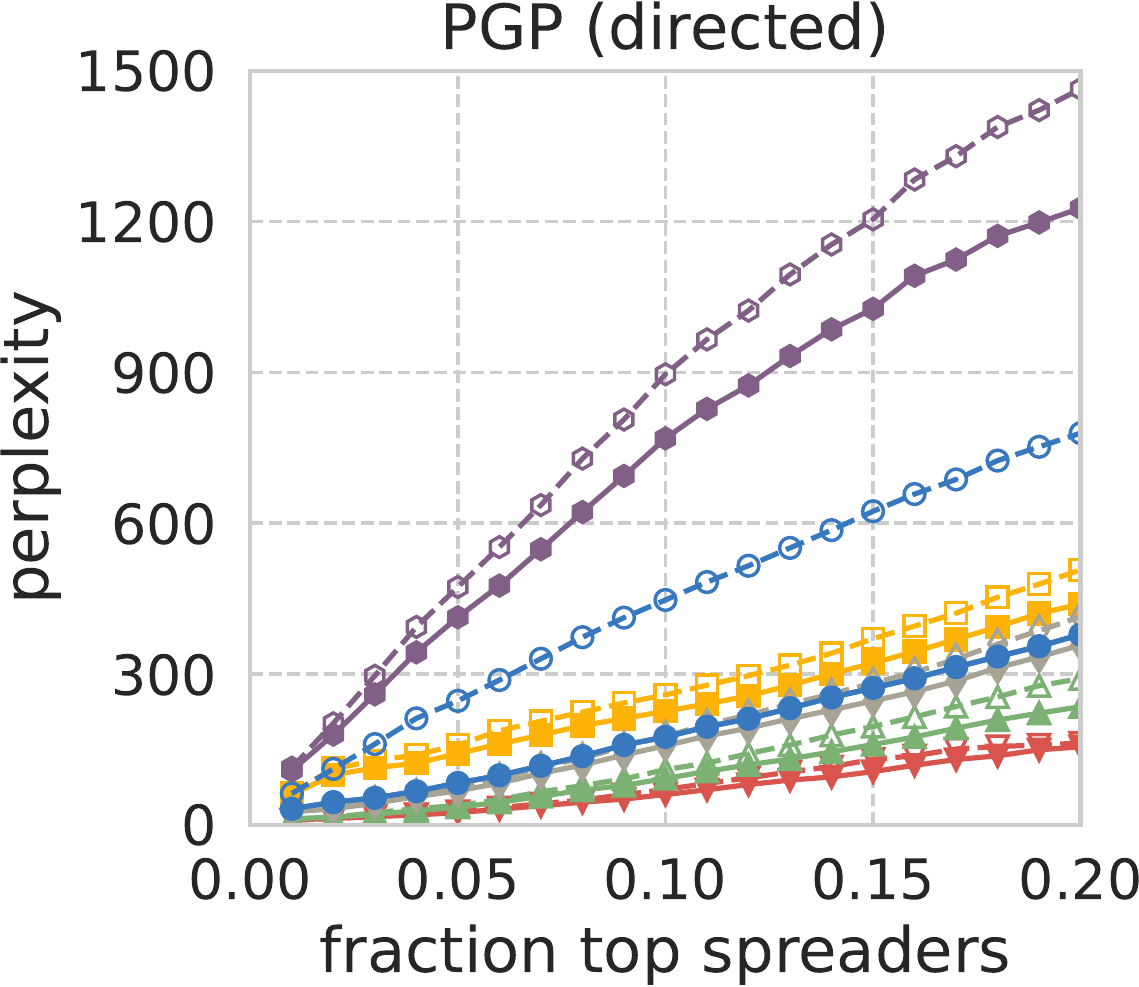}
    }\hfill
    \subfloat[\label{fig:perplexity-facebook-wall}]{%
        \includegraphics[width=.23\textwidth]{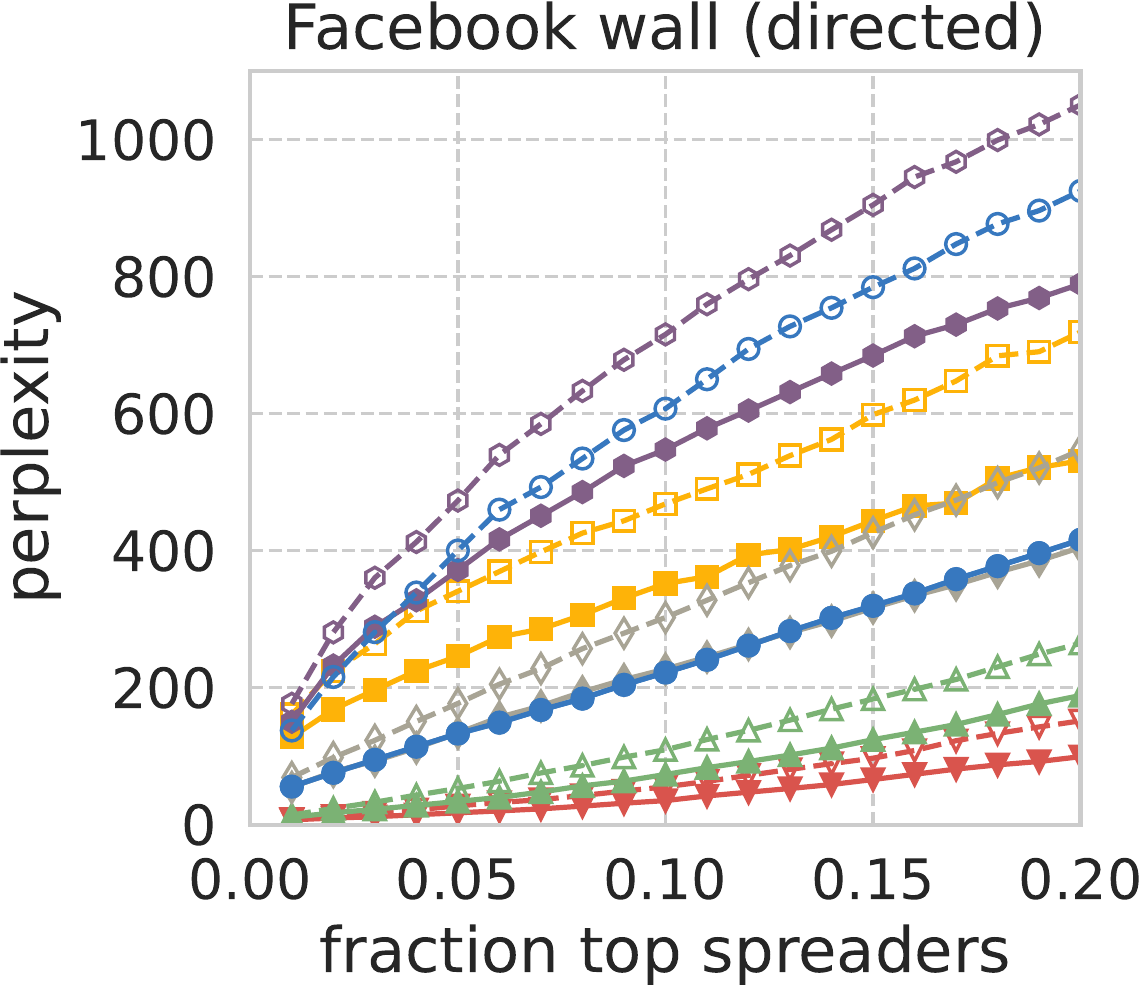}
    }
    \caption{Perplexity for the distribution of selected nodes as a function of the fraction of selected top spreaders for map equation centrality, modularity vitality, community hub-bridge, community-based centrality, degree centrality, and betweenness centrality in twelve empirical networks. Community structures are identified with Infomap; solid lines use the unrecorded link teleportation flow model, dashed lines use recorded node teleportation.}
    \label{fig:perplexity-results}
\end{figure*}
\noindent
To understand why unrecorded link teleportation facilitates more accurate identification of top spreaders in the SIR case while recorded node teleportation works better for the linear threshold model, we analyse how the top-ranked nodes are distributed across modules.
Let $\mathsf{M}$ be a partition of the nodes into modules, $\mathsf{m} \in \mathsf{M}$ be a module, and let $S$ be the set of selected nodes by some centrality measure.
Then, $\frac{\left|\textsf{m} \cap S\right|}{\left|S\right|}$ is the fraction of selected nodes in module $\mathsf{m}$; we calculate the perplexity for $S$ as $2^{H\left(S\right)}$ where $H\left(S\right) = - \sum_{\mathsf{m} \in \mathsf{M}} \frac{\left| \mathsf{m} \cap S \right|}{\left|S\right|} \log_2 \frac{\left| \mathsf{m} \cap S \right|}{\left|S\right|}$.
The perplexity corresponds to the effective number of same-size modules across which the selected nodes are distributed uniformly.
That is, a higher perplexity means that the selected nodes are more spread out across modules.

The nodes selected by map equation centrality are more spread with standard PageRank flow than with unrecorded link teleportation.
This is because PageRank with a teleportation rate of $r$ assigns an $r$-fraction of the flow to nodes uniformly, resulting in at least a flow of $\frac{r}{n}$ per node in a network with $n$ nodes.
Here, we used $r = 0.15$.
For modularity vitality, community hub-bridge, and community-based centrality, the difference between link and node teleportation is less pronounced, and in some settings even reversed.
Overall, community hub-bridge and community-based centrality have lower perplexity, selecting nodes that are less spread out across modules.
Map equation centrality and modularity vitality have substantially higher perplexity, spreading out the selected nodes more across communities (Figs. \ref{fig:perplexity-facebook-friends} to \ref{fig:perplexity-facebook-wall}).
To perform well in the SIR case, a centrality measure should select high-degree nodes because they have a higher opportunity to infect other nodes.
Conversely, under the linear threshold model, it is more important to spread out the selected nodes across tightly-knit communities to reach a high activation size, or high-density communities will stop the activation of nodes \cite{morris2000contagion}.

\section{Conclusion}
\label{sec:conclusion}
We have studied node importance from a community-detection perspective within the map equation framework and analytically derived a community-aware centrality score.
Our score exploits modular network structure, is agnostic to the chosen flow model, and assigns centrality scores to nodes based on their community embedding; to determine a node's centrality, it suffices to consider those nodes that belong to the same community.
In contrast, traditional centrality measures typically neglect local network structure and rely on node features or global patterns to determine node importance instead.
Community-aware centrality measures are often defined in an ad-hoc way, disconnected from the assumptions made by community-detection methods.
In contrast, map equation centrality is true to the map equation.
We have highlighted how map equation centrality discerns nodes indistinguishable to global centrality measures using a synthetic network. 
On a set of twelve real-world networks, map equation centrality often performs better than baseline methods in identifying influential nodes.

\begin{acknowledgements}
We would like to thank Anton Eriksson for helping with implementing map equation centrality in Infomap, and Jelena Smiljani\'c and the anonymous reviewers for comments that helped to improve the manuscript.
This work was partially supported by the Wallenberg AI, Autonomous Systems and Software Program (\href{https://wasp-sweden.org}{WASP}) funded by the Knut and Alice Wallenberg Foundation.
Martin Rosvall was supported by the Swedish Research Council, Grant No.\ 2016-00796.
\end{acknowledgements}

%

\appendix

\section{Generalisation for Sets of Nodes}
We generalise map equation centrality and derive the expression in \Eqnref{eqn:map-equation-centrality-sets} that can be used to calculate the combined centrality for sets of nodes $U$.
We follow the same approach as before, that is, we first derive an expression for the expected per-step codelength when silencing all nodes in $U$ while using the old coding scheme; then we derive an expression for for the expected per-step codelength when designing a new coding scheme that does not assign codewords to nodes in $U$ to start with.

Let $G = \left(V, E, \delta\right)$ be a network with nodes $V$, links $E$, weights $\delta$, $U \subseteq V$ be a set of nodes, and $p_U = \sum_{u \in U} p_u$ be the visit rate sum of nodes in $U$.
Further, for a module $\mathsf{m}$, let $p_{\mathsf{m} \cap U} = \sum_{u \in \mathsf{m} \cap U} p_u$ be the visit rate sum of nodes that are members in $\mathsf{m}$ and in $U$, and let $P_{\mathsf{m} \cap U} = \left\{ p_u \,|\, u \in \mathsf{m} \cap U \right\}$ be their set of visit rates.

We begin with the one-level partition $\mathsf{M}_1$ and obtain the expected per-step codelength for describing a random walk with nodes in $U$ silenced while using the old coding scheme.
Removing the silenced nodes from the summation in \Eqnref{eqn:map-equation-one-level}, we get
\begin{equation}
  L^U\left(G, \mathsf{M}_1\right) = - \sum_{v \in V \setminus U} p_v \log_2 p_v.
  \tag{A1}
  \label{eqn:contribution-one-level-same-code-sets}
\end{equation}
We obtain the codelength for a new coding scheme that does not assign codewords to nodes in $U$ by re-normalising the visit rates for the remaining nodes with $1-p_U$,
\begin{equation*}
  L^{U*}\left(G, \mathsf{M}_1\right) = - \sum_{v \in V \setminus U} p_v \log_2 \frac{p_v}{1 - p_U}.
  \tag{A2}
  \label{eqn:contribution-one-level-new-code-sets}
\end{equation*}
The difference between \Eqnref{eqn:contribution-one-level-same-code-sets} and \Eqnref{eqn:contribution-one-level-new-code-sets} is the joint map equation centrality score of the nodes in $U$ under $\mathsf{M}_1$,
\begin{align}
  \lambda\left(G, \mathsf{M}_1, U\right) & = L^U\left(G, \mathsf{M}_1\right) - L^{U*}\left(G, \mathsf{M}_1\right) \nonumber \\
  & = - \left(1-p_U\right) \log_2 \left(1 - p_U\right).
  \tag{A3}
\end{align}

For two-level partitions, we begin by rewriting the map equation (\Eqnref{eqn:map-equation-two-level}) to distinguish explicitly between modules that have an overlap with $U$ and those that do not,
\begin{align}
  & L\left(G,\mathsf{M}\right) = \overbracket{q H\left(Q\right)}^{\text{index level}} \tag{A4} \\
  & + \overbracket{\sum_{\substack{\mathsf{m} \in \mathsf{M} \\ \mathsf{m} \cap U = \emptyset}} p_\mathsf{m} H\left(P_\mathsf{m}\right)}^{\text{no overlap with $U$}} - \overbracket{\sum_{\substack{\mathsf{m} \in \mathsf{M} \\ \mathsf{m} \cap U \neq \emptyset}} \sum_{p \in P_\mathsf{m}} p \log_2 \frac{p}{p_\mathsf{m}}}^{\text{overlap with $U$}}. \nonumber
\end{align}
The codelength for describing a random walk in partition $\mathsf{M}$ with nodes in $U$ silenced when using the old coding scheme is
\begin{align}
  & L^U\left(G, \mathsf{M}\right) = \overbracket{q H\left(Q\right)}^{\text{index level}} \tag{A5} \label{eqn:contribution-two-level-old-code-sets} \\
  & + \overbracket{\sum_{\substack{\mathsf{m} \in \mathsf{M} \\ \mathsf{m} \cap U = \emptyset}} p_\mathsf{m} H\left(P_\mathsf{m}\right)}^{\text{no overlap with $U$}} - \overbracket{\sum_{\substack{\mathsf{m} \in \mathsf{M} \\ \mathsf{m} \cap U \neq \emptyset}} \sum_{\substack{p \in P_\mathsf{m} \setminus P_{\mathsf{m} \cap U}}} p \log_2 \frac{p}{p_\mathsf{m}}}^{\text{overlap with $U$}}. \nonumber
\end{align}
With a new code that does not assign codewords to nodes in $U$ and that normalises accordingly, the codelength is
\begin{align}
  & L^{U*}\left(G,\mathsf{M}\right) = \overbracket{q H\left(Q\right)}^{\text{index level}} + \overbracket{\sum_{\substack{\mathsf{m} \in \mathsf{M} \\ \mathsf{m} \cap U = \emptyset}} p_\mathsf{m} H\left(P_\mathsf{m}\right)}^{\text{no overlap with $U$}} \nonumber \\
  & - \overbracket{\sum_{\substack{\mathsf{m} \in \mathsf{M} \\ \mathsf{m} \cap S \neq \emptyset}} \sum_{\substack{p \in P_\mathsf{m} \setminus P_{\mathsf{m} \cap U}}} p \log_2 \frac{p}{p_\mathsf{m} - p_{\mathsf{m} \cap U}}}^{\text{overlap with $U$}}. \tag{A6}
  \label{eqn:contribution-two-level-new-code-sets}
\end{align}
The difference between \Eqnref{eqn:contribution-two-level-old-code-sets} and \Eqnref{eqn:contribution-two-level-new-code-sets} is the joint map equation centrality of the nodes in $U$ under $\mathsf{M}$,
\begin{align}
  & \lambda\left(G, \mathsf{M}, U\right) = L^U\left(G,\mathsf{M}\right) - L^{U*}\left(G,\mathsf{M}\right) \nonumber \\
  & = - \sum_{\mathclap{\mathsf{m} \in \mathsf{M}, \mathsf{m} \cap U \neq \emptyset}} \left(p_\mathsf{m} - p_{\mathsf{m} \cap U}\right) \log_2 \frac{p_\mathsf{m} - p_{\mathsf{m} \cap U}}{p_\mathsf{m}}. \tag{A7}
\end{align}

\section{Descriptions of Linear Threshold Model Results}
In the Facebook friends network, initially all measures perform similarly well.
Modularity vitality, community hub-bridge, and community-based centrality outperform map equation centrality between $x = 0.02$ and $0.04$; beyond $x = 0.04$, map equation centrality performs best, followed by betweenness centrality, degree centrality, modularity vitality, community-based centrality, and community hub-bridge (\Figref{fig:lt-0.5-facebook-friends}).

In the Copenhagen network, up to $x = 0.03$, all scores perform equally well, between $x = 0.03$ and $x = 0.05$, community hub-bridge and community-based centrality perform slightly better than map equation centrality and modularity vitality.
Beyond $x = 0.05$ and up to $x = 0.13$, map equation centrality performs best; for $x \geq 0.13$, modularity vitality performs best and reaches an activation size of $1$ (\Figref{fig:lt-0.5-copenhagen}).

In the Uni email network, initially, community hub-bridge and community-based centrality slightly outperform the other measures.
Then, at $x = 0.06$, map equation centrality and degree centrality reach an activation size of nearly $1$, followed by betweenness centrality at $x = 0.07$, modularity vitality at $x = 0.09$, community-based centrality at $x = 0.12$, and community hub-bridge at $x = 0.15$ (\Figref{fig:lt-0.5-uni-email}).

In the Polblogs network, community-based centrality, community hub-bridge, and betweenness centrality perform best, followed by degree centrality, modularity vitality, and finally map equation centrality (\Figref{fig:lt-0.5-polblogs}).

In the Interactome yeast network, map equation centrality performs best, followed by degree centrality, and the remaining measures which have similar performance in this case (\Figref{fig:lt-0.5-interactome-yeast}).

In the Ego Facebook network, all four measures have similar performance up to $x = 0.05$, beyond which map equation centrality dominates, followed by betweenness centrality, community-based centrality and community hub-bridge, and modularity vitality (\Figref{fig:lt-0.5-ego-facebook}).

In the Power network, map equation centrality performs best, followed by degree centrality, modularity vitality, community-based centrality, community hub-bridge, and betweenness centrality (\Figref{fig:lt-0.5-power}).

In the Facebook organizations network, community hub-bridge and community-based centrality perform best up to $x = 0.05$.
Beyond that, map equation centrality performs best; modularity vitality, betweenness centrality, degree centrality, and community-based centrality have similar performance, and community hub-bridge performs weakest with some distance.
From $x = 0.14$, betweenness centrality performs slightly better than map equation centrality (\Figref{fig:lt-0.5-facebook-organizations}).

In the Physics collaborations map equation centrality outperforms the other measures over the whole tested range, followed by betweenness centrality, degree centrality, modularity vitality, and community hub-bridge and community-based centrality (\Figref{fig:lt-0.5-physics-collaborations}).

In the Google network, betweenness centrality performs best while map equation centrality performs weakest in this scenario. 
The remaining measures have similar performance, but none clearly wins against the others (\Figref{fig:lt-0.5-google}).

In the PGP network, betweenness centrality outperforms the remaining measures, followed by map equation centrality, degree centrality, modularity vitality, community hub-bridge, and community-based centrality (\Figref{fig:lt-0.5-pgp}).

Finally, in the Facebook wall network, initially, map equation centrality based on unrecorded link teleportation and degree centrality perform best, followed by community-based centrality, community hub-bridge, betweenness centrality, and modularity vitality.
Beyond $x = 0.04$, map map equation centrality with recorded node teleporation and betweenness centrality perform best, followed by modularity vitality, degree centrality, community hub-bridge, and community-based centrality (\Figref{fig:lt-0.5-facebook-wall}).

\section{Descriptions of the SIR Model Results}
In the Facebook friends network, map equation centrality, degree centrality, community hub-bridge, and community-based centrality are nearly tied with an imprecision up to approximately $0.05$, identifying the top spreaders accurately.
Modularity vitality initially performs similarly well, but achieves imprecision values between around $0.2$ and $0.3$ beyond $x = 0.05$ (\Figref{fig:sir-facebook-friends}).

In the Copenhagen network, map equation centrality and degree centrality outperform the other measures.
Community-based centrality performs slightly worse than map equation centrality, followed by community hub-bridge, betweenness centrality, and then modularity vitality (\Figref{fig:sir-copenhagen}).

In the Uni email network, map equation centrality and degree centrality outperform the other measures across the tested range of $x$-values, followed by community-based centrality, betweenness centrality, and modularity vitality and community hub-bridge, the latter two performing similarly in this scenario (\Figref{fig:sir-uni-email}).

In the Polblogs network, community-based centrality and community hub-bridge perform best, followed by degree and betweenness centrality, modularity vitality, and finally map equation centrality (\Figref{fig:sir-polblogs}).

In the Interactome yeast network, all measures perform similarly well, while community-based centrality and map equation centrality slightly outperform the rest (\Figref{fig:sir-interactome-yeast}).

In the Ego Facebook network, map equation centrality and degree centrality again outperform the other measures.
Initially and up to $x \approx 0.08$, modularity vitality, community hub-bridge, and community-based centrality show similar performance.
Beyond $x \approx 0.08$, modularity vitality's performance remains stable at an imprecision of around $0.2$ while community hub-bridge and community-based centrality improve and perform as well as map equation centrality at $x \approx 0.2$.
Map equation centrality based on recorded node teleportation and betweenness centrality perform substantially worse than the other measures in this scenario with imprecision values roughly between $0.9$ down to $0.5$ (\Figref{fig:sir-ego-facebook}).

In the Power network, community-based centrality performs best, followed by map equation centrality, community hub-bridge, degree centrality, modularity vitality, and finally betweenness centrality (\Figref{fig:sir-power}).

In the Facebook organizations network, map equation centrality and degree centrality outperform the other measures with a stable imprecision around $0.1$.
Modularity vitality performs second-best, with increasing imprecision as $x$ increases, followed by betweenness centrality, community-based centrality, and community hub-bridge (\Figref{fig:sir-facebook-organizations}).

In the Physics collaborations network, modularity vitality initially performs best, but with slightly decreasing performance as $x$ increases.
Map equation centrality, degree centrality, community hub-bridge, and community-based centrality initially perform similarly, all with an imprecision of around $0.35$, but outperform modularity vitality beyond $x \approx 0.05$, with community-based centrality performing best (\Figref{fig:sir-physics-collaborations}).

In the Google network, up to $x = 0.03$, community hub-bride performs best.
Beyond that, community-based centrality performs best, followed by degree centrality, community hub-bridge, modularity vitality, map equation centrality, and finally betweenness centrality (\Figref{fig:sir-google}).

In the PGP network, community-based centrality outperforms the other measures, followed by degree centrality.
Community hub-bridge performs third-best, followed by map equation centrality, betweenness centrality, and modularity vitality. In this scenario, node teleportation-based map equation centrality performs nearly identical to modularity vitality.
Beyond $x \approx 0.05$, community hub-bridge and map equation centrality are nearly tied (\Figref{fig:sir-pgp}).

Finally, in the Facebook wall network, community-based centrality outperforms the other measures, followed by degree centrality, community hub-bridge, map equation centrality, modularity vitality, and betweenness centrality.
Here, map equation centrality when using recorded node teleportation performs considerably worse compared to unrecorded link teleportation (\Figref{fig:sir-facebook-wall}).

\clearpage
\onecolumngrid

\section{Further results for the linear threshold model}
\label{appx:lt}
\label{appx:lt}
\begin{figure*}[h!]
    \centering
    \subfloat[\label{fig:lt-0.4-facebook-friends}]{%
        \includegraphics[width=.23\textwidth]{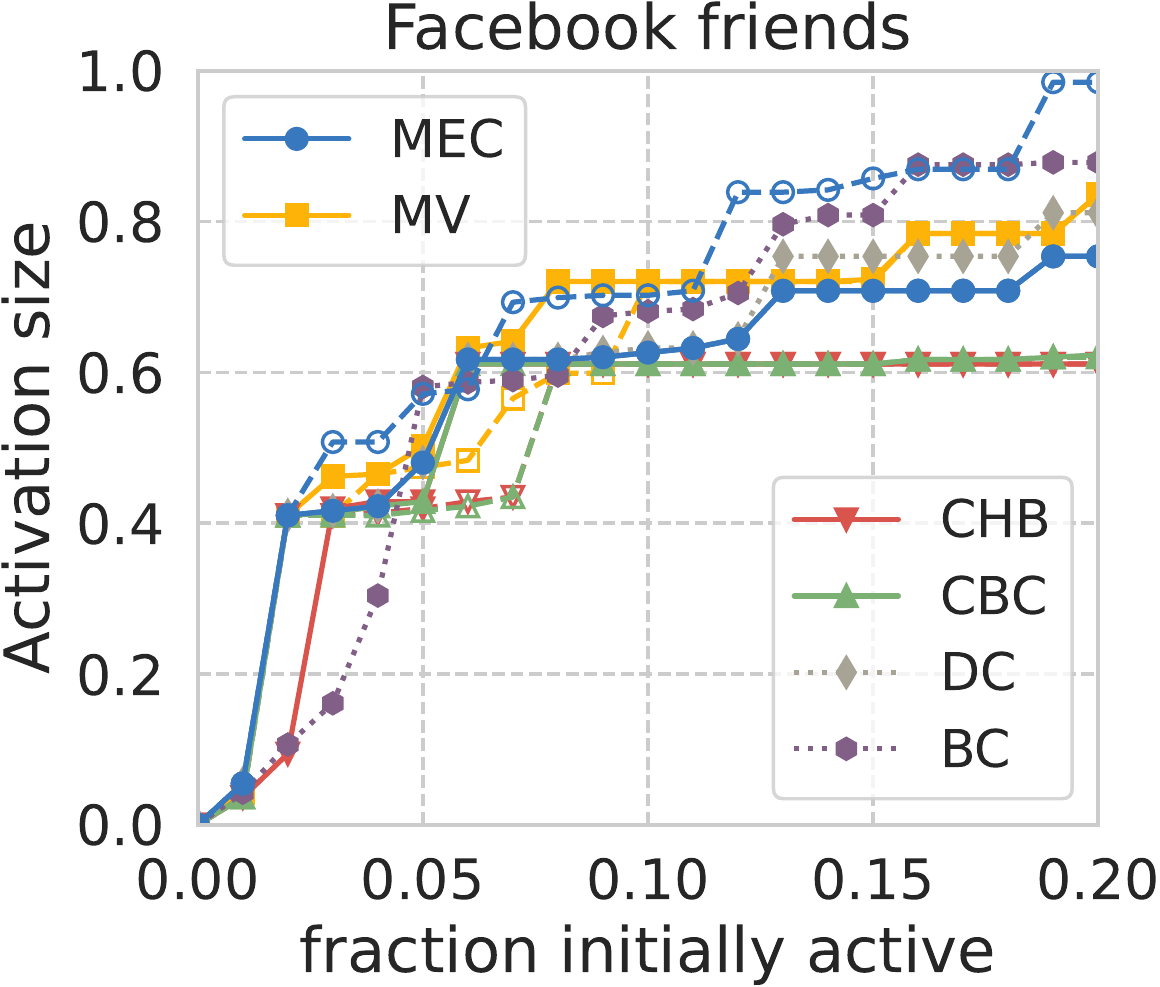}
    }\hfill
    \subfloat[\label{fig:lt-0.4-copenhagen}]{%
        \includegraphics[width=.23\textwidth]{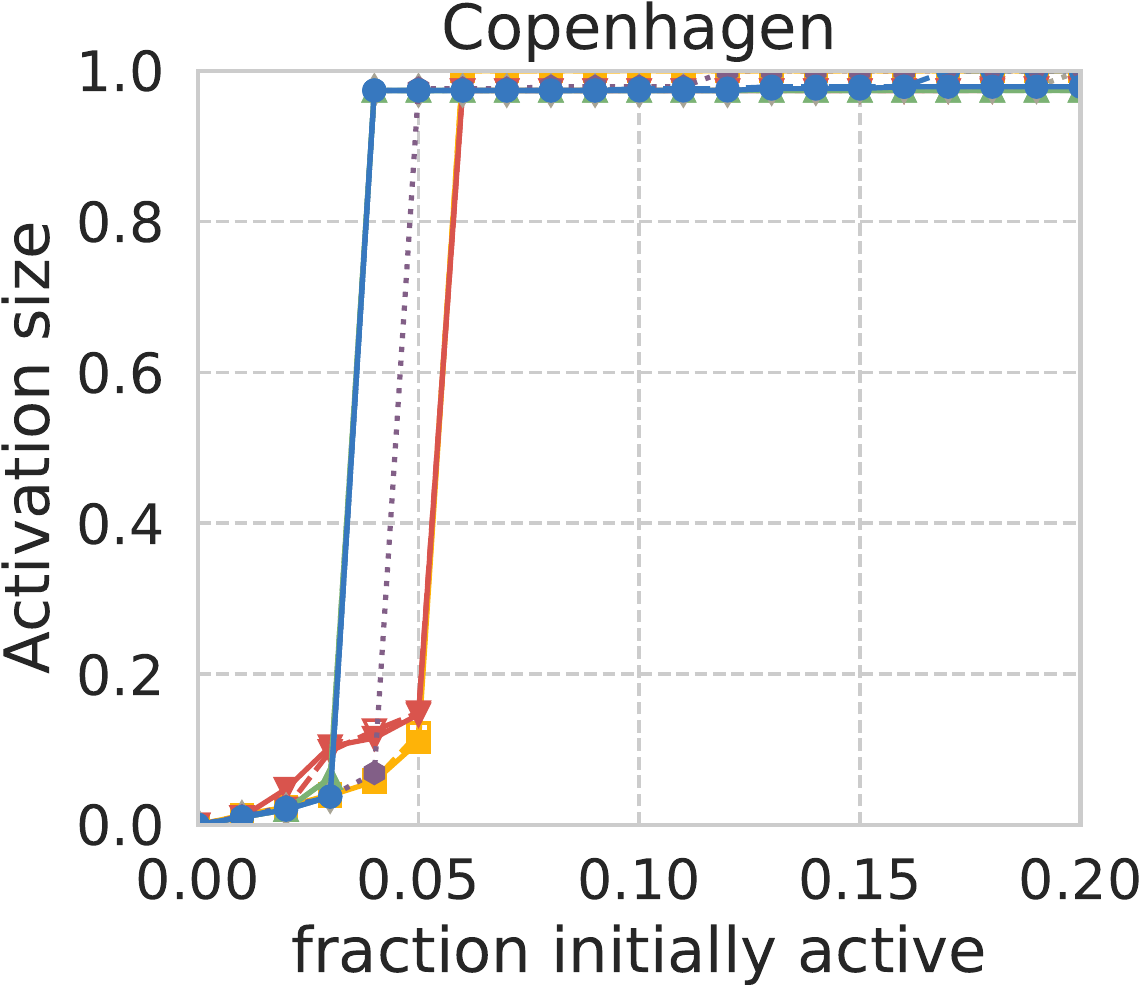}
    }\hfill
    \subfloat[\label{fig:lt-0.4-uni-email}]{%
        \includegraphics[width=.23\textwidth]{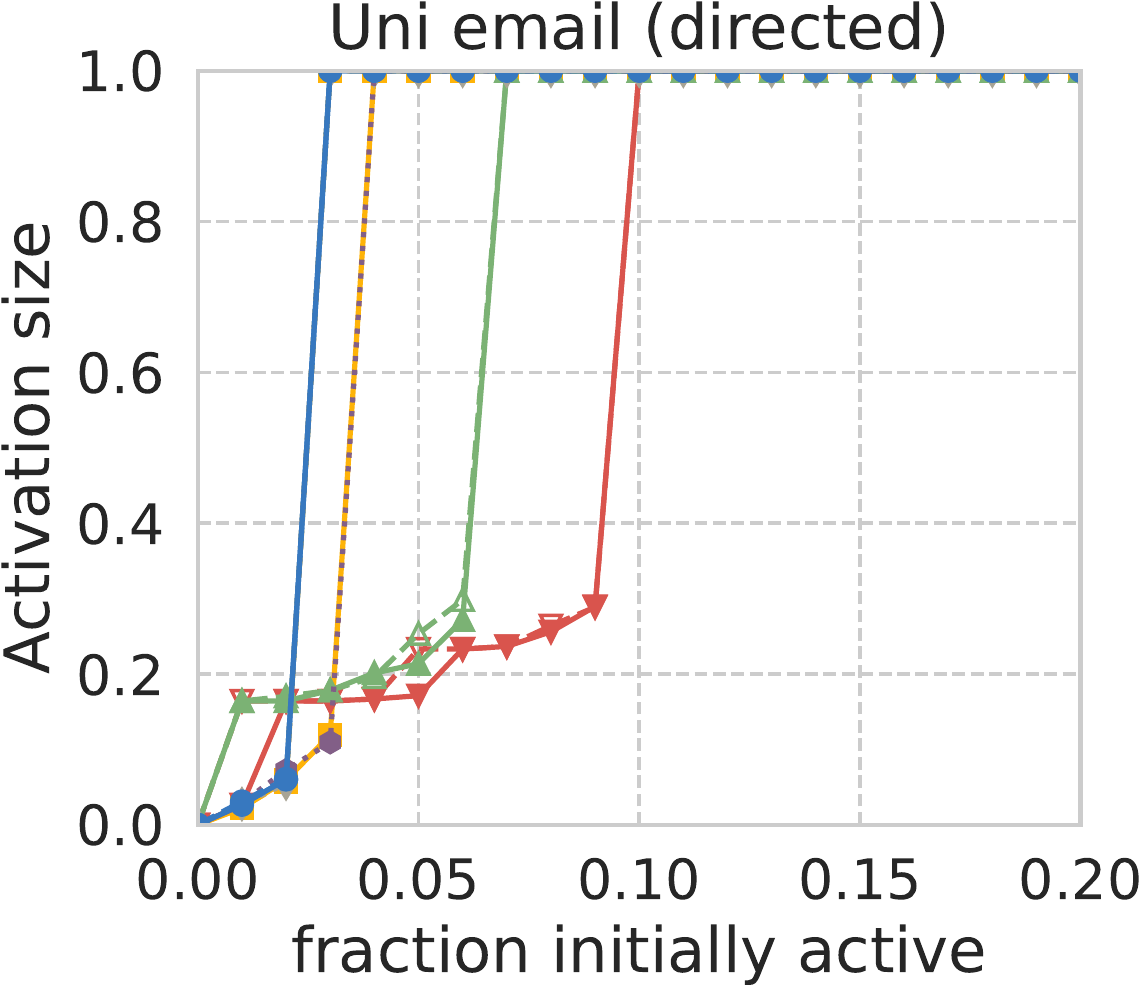}
    }\hfill
    \subfloat[\label{fig:lt-0.4-polblogs}]{%
        \includegraphics[width=.23\textwidth]{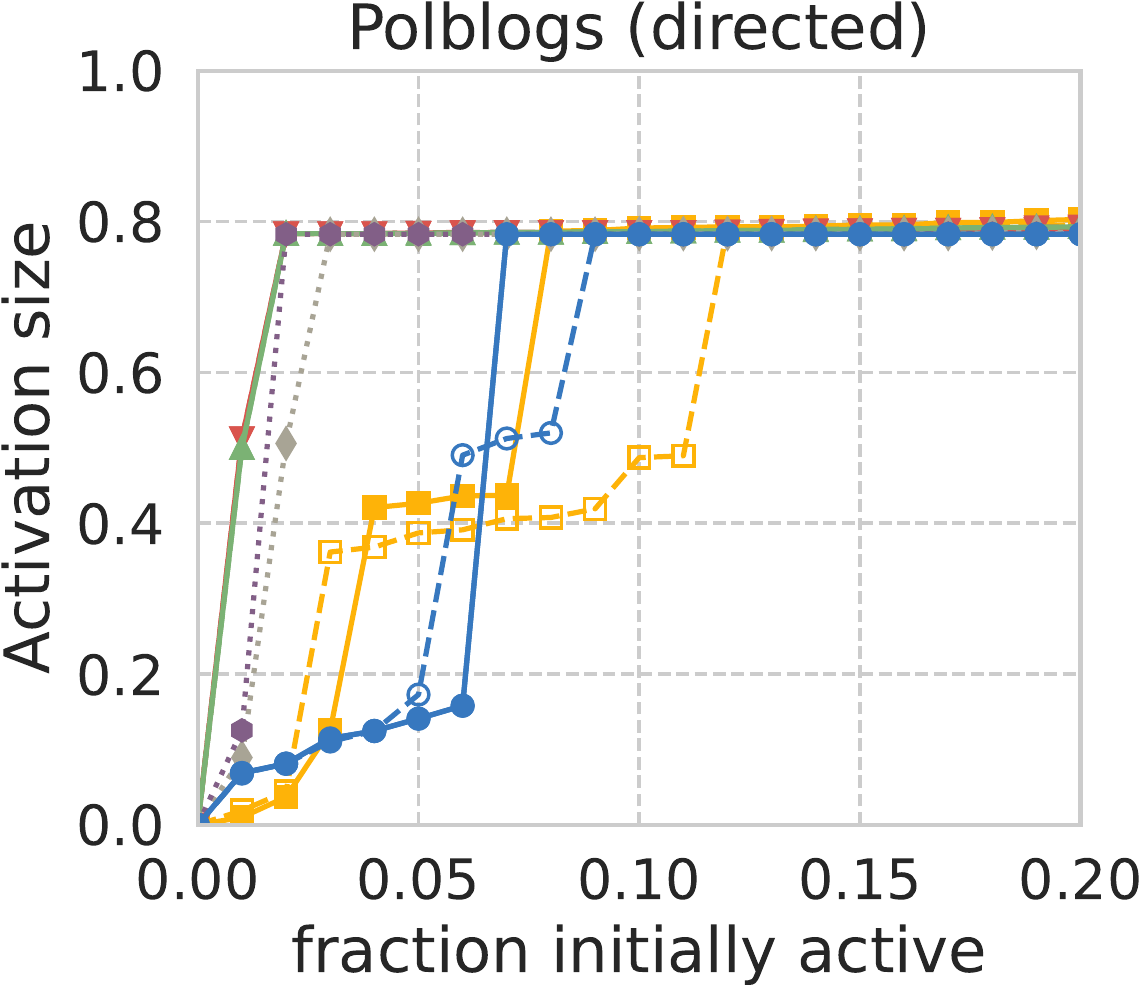}
    }\hfill
    \subfloat[\label{fig:lt-0.4-interactome-yeast}]{%
        \includegraphics[width=.23\textwidth]{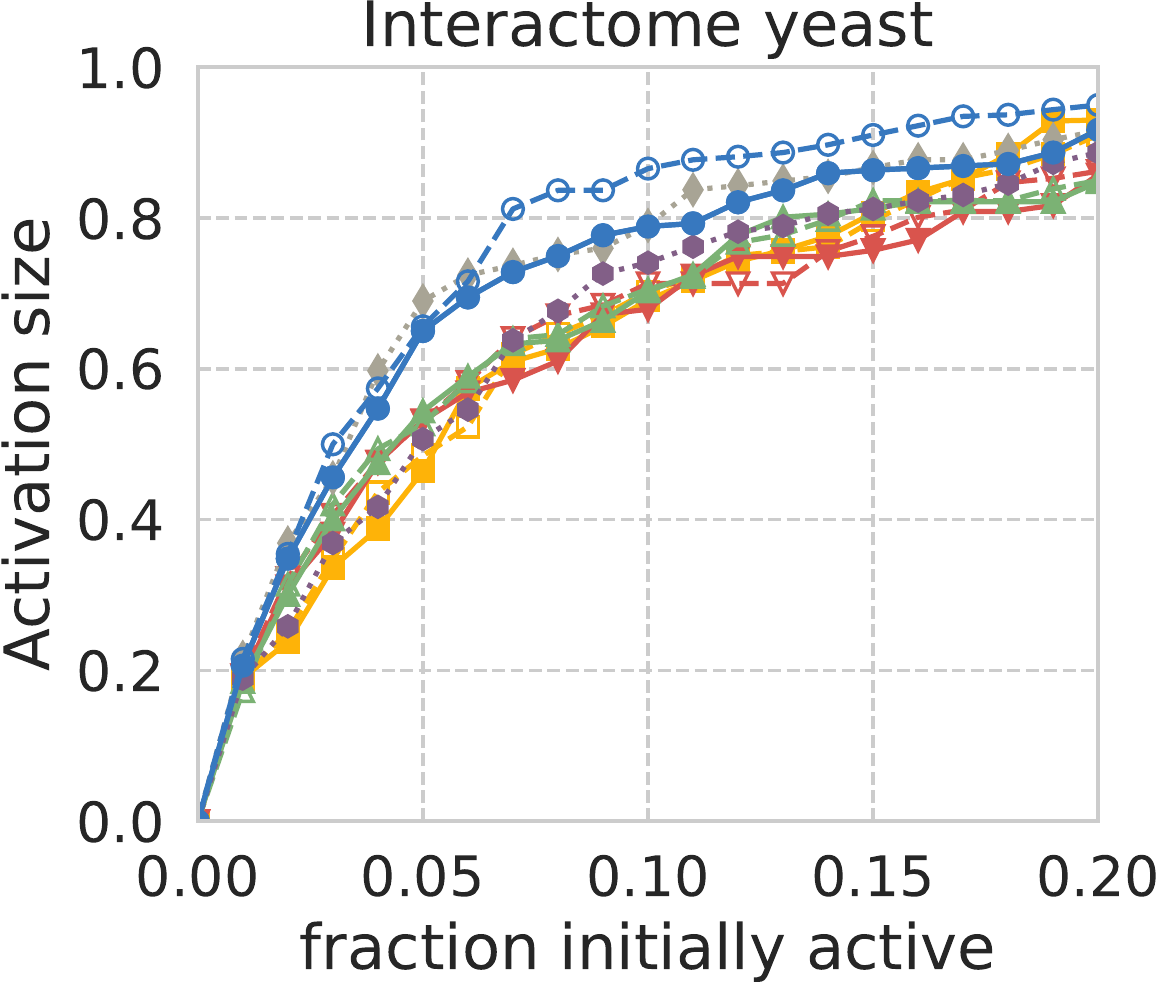}
    }\hfill
    \subfloat[\label{fig:lt-0.4-ego-facebook}]{%
        \includegraphics[width=.23\textwidth]{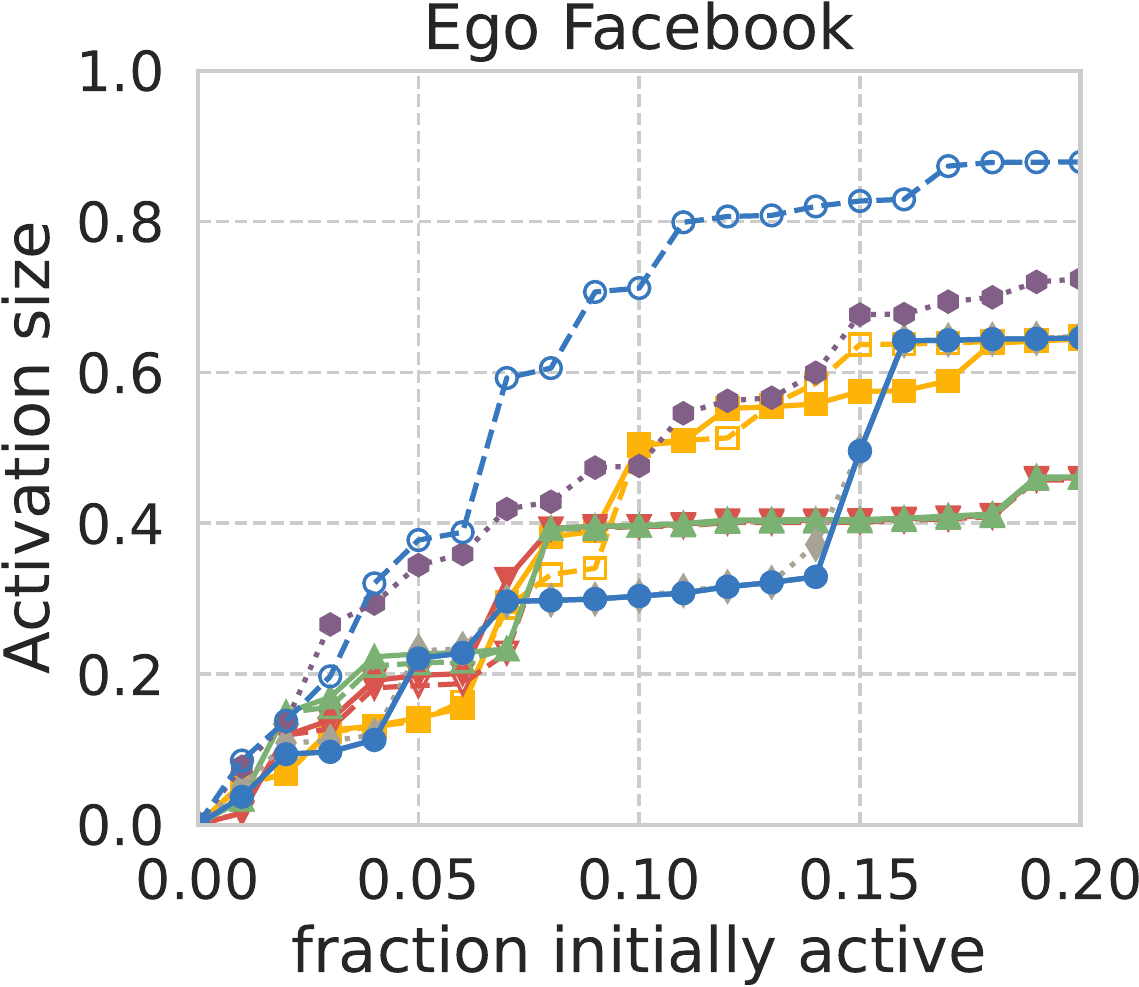}
    }\hfill
    \subfloat[\label{fig:lt-0.4-power}]{%
        \includegraphics[width=.23\textwidth]{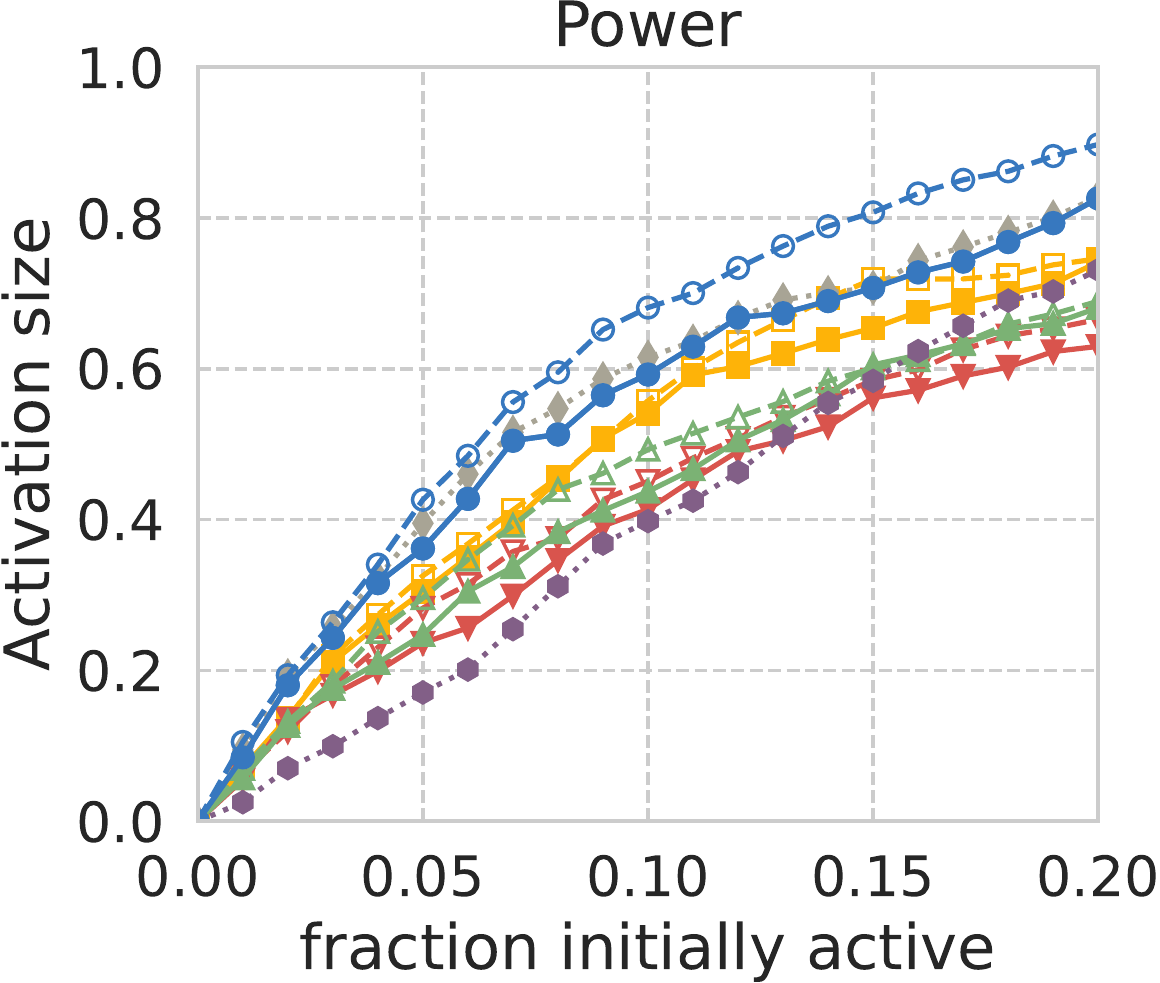}
    }\hfill
    \subfloat[\label{fig:lt-0.4-facebook-organizations}]{%
        \includegraphics[width=.23\textwidth]{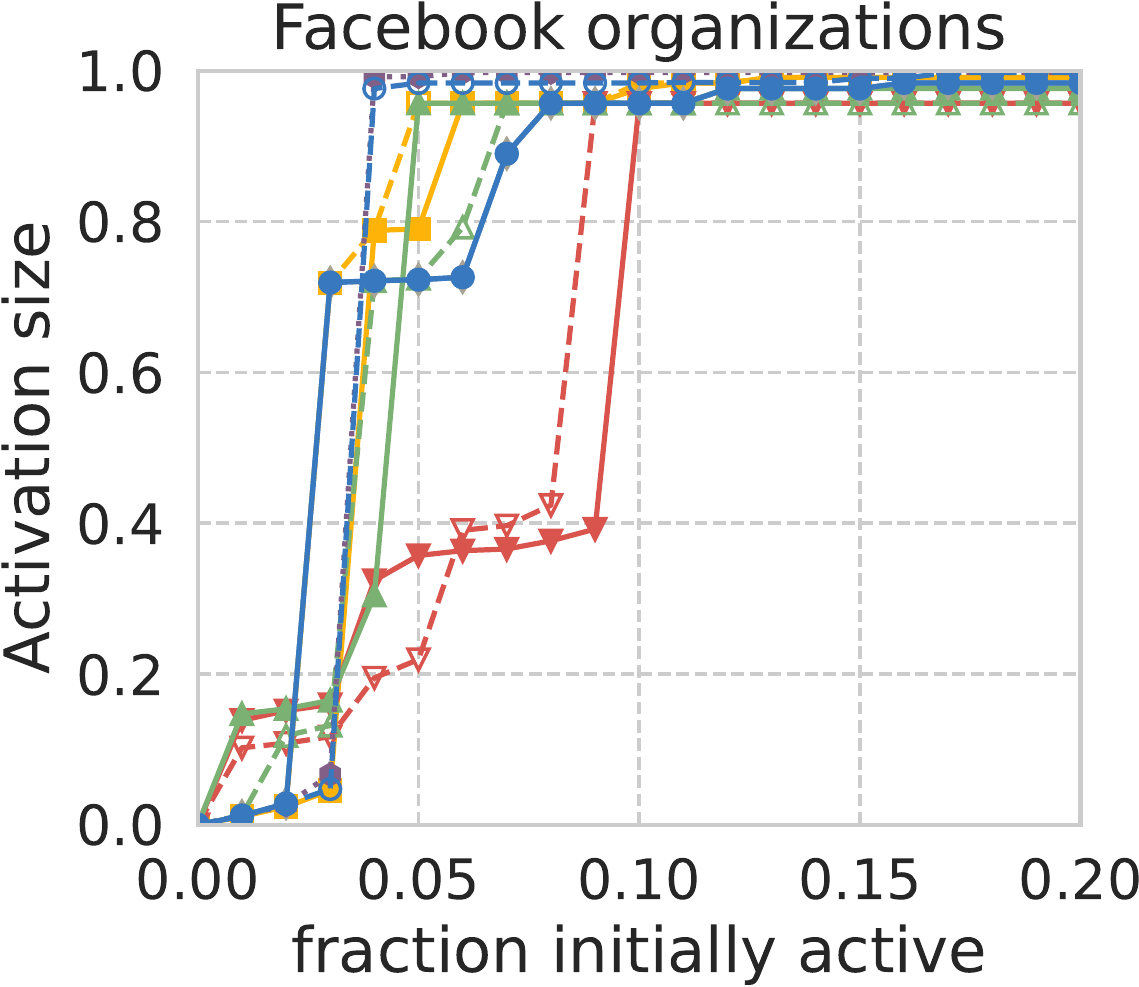}
    }\hfill
    \subfloat[\label{fig:lt-0.4-physics-collaborations}]{%
        \includegraphics[width=.23\textwidth]{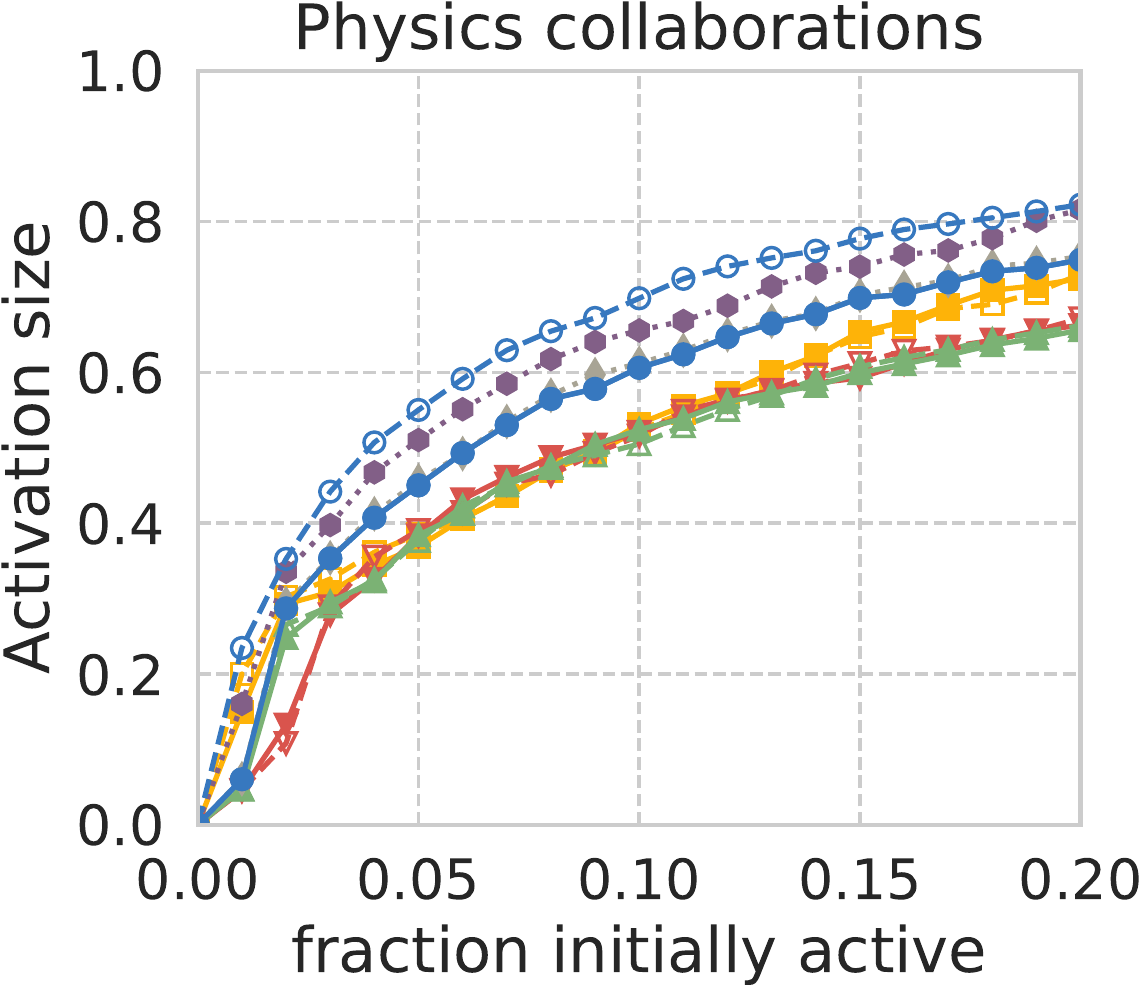}
    }\hfill
    \subfloat[\label{fig:lt-0.4-google}]{%
        \includegraphics[width=.23\textwidth]{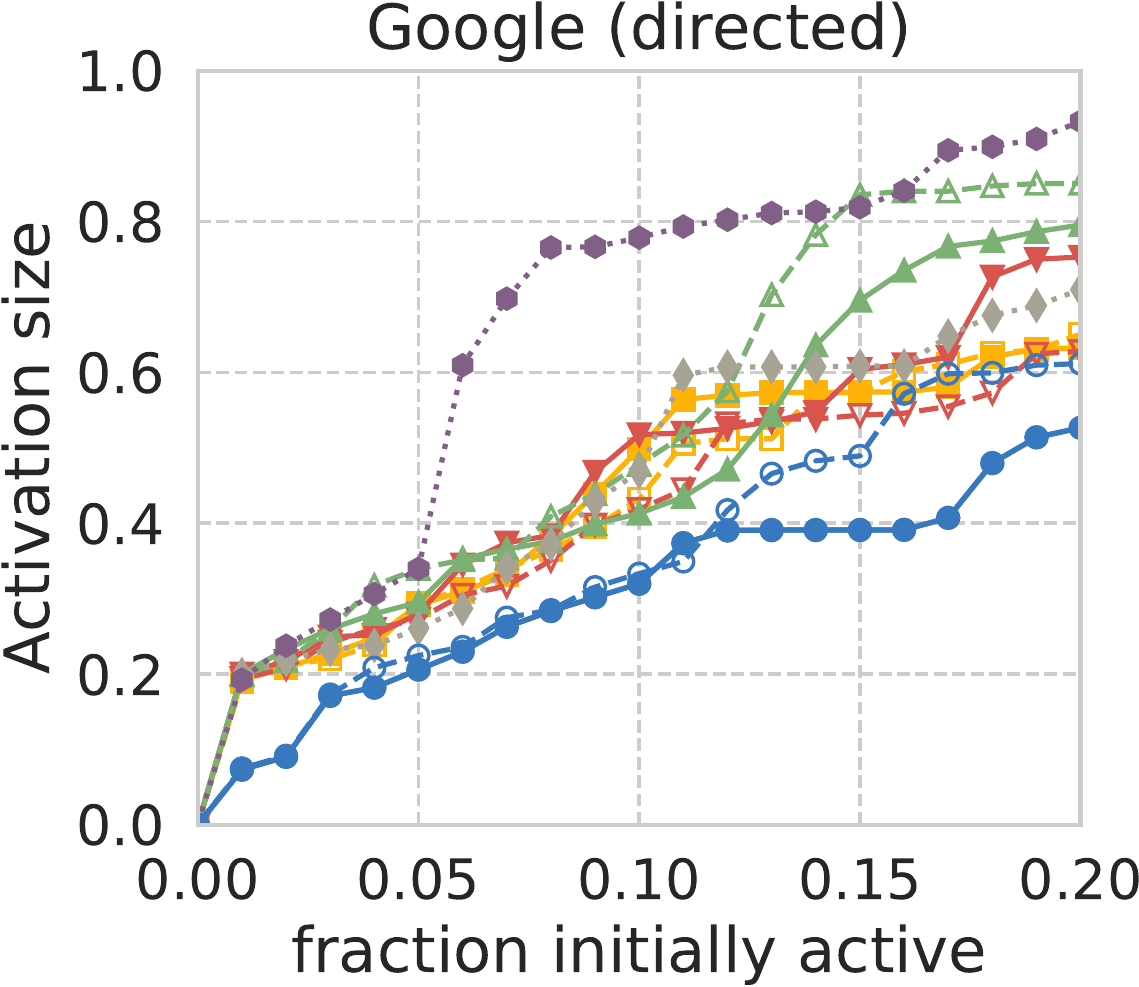}
    }\hfill
    \subfloat[\label{fig:lt-0.4-pgp}]{%
        \includegraphics[width=.23\textwidth]{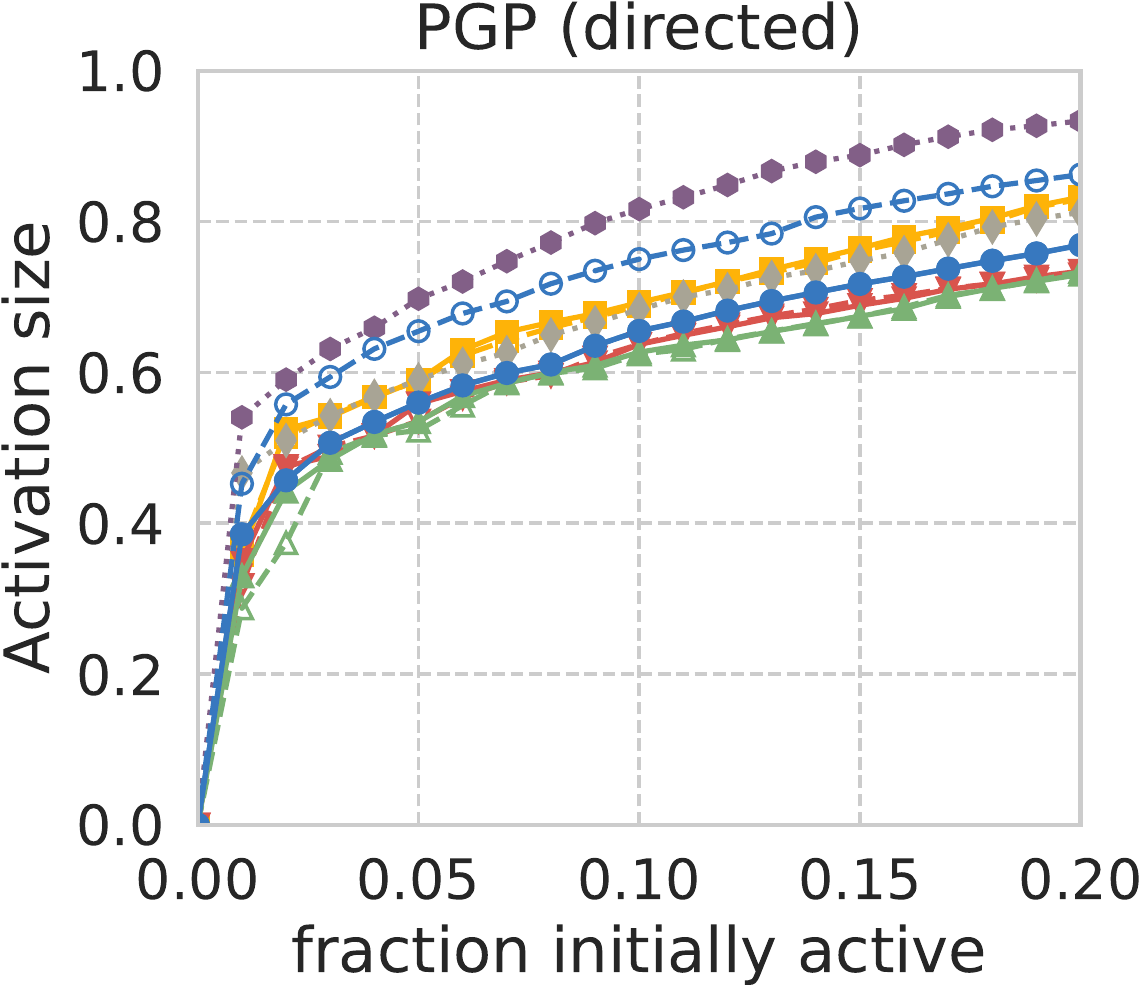}
    }\hfill
    \subfloat[\label{fig:lt-0.4-facebook-wall}]{%
        \includegraphics[width=.23\textwidth]{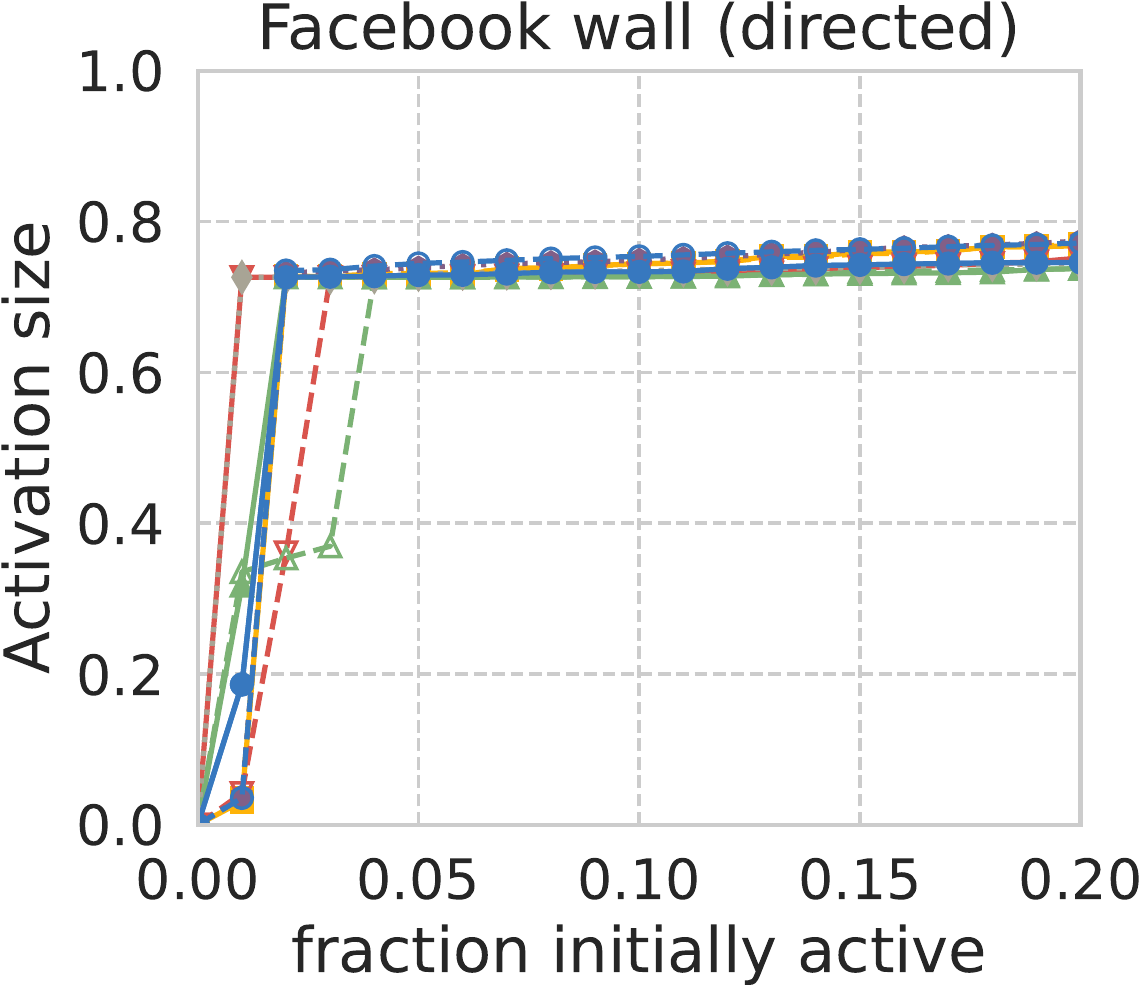}
    }
    \caption{
        Activation size for map equation centrality (MEC), modularity vitality (MV), community hub-bridge (CHB), community-based centrality (CBC), degree centrality (DC), and betweenness centrality (BC) in twelve empirical networks under the linear threshold model with threshold $t' = 0.4$. Community structures are identified with Infomap; solid lines use the unrecorded link teleportation flow model, dashed lines use recorded node teleportation.
        }
    \label{fig:empirical-results-lt-0.4}
\end{figure*}

\begin{figure*}[h!]
    \centering
    \subfloat[\label{fig:lt-0.6-facebook-friends}]{%
        \includegraphics[width=.23\textwidth]{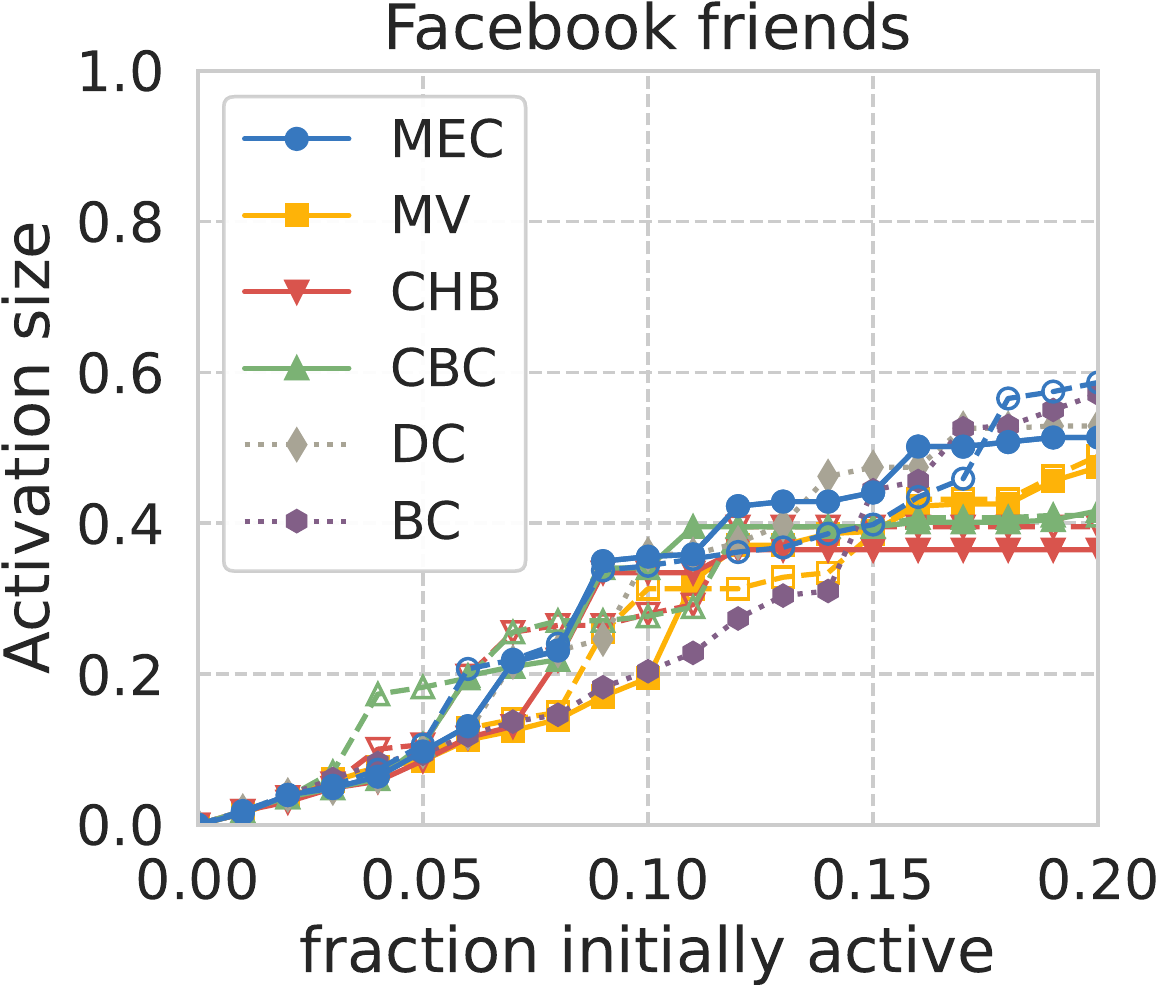}
    }\hfill
    \subfloat[\label{fig:lt-0.6-copenhagen}]{%
        \includegraphics[width=.23\textwidth]{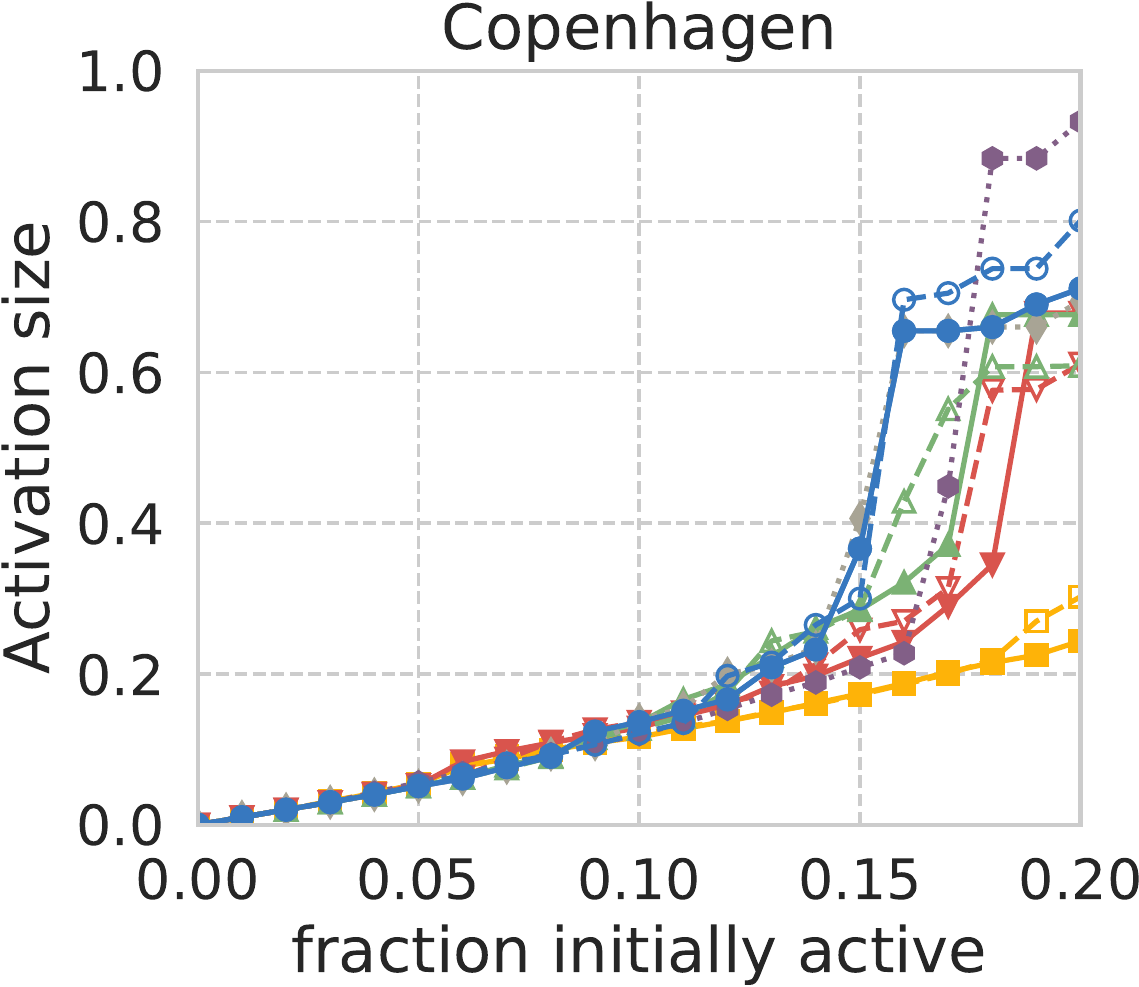}
    }\hfill
    \subfloat[\label{fig:lt-0.6-uni-email}]{%
        \includegraphics[width=.23\textwidth]{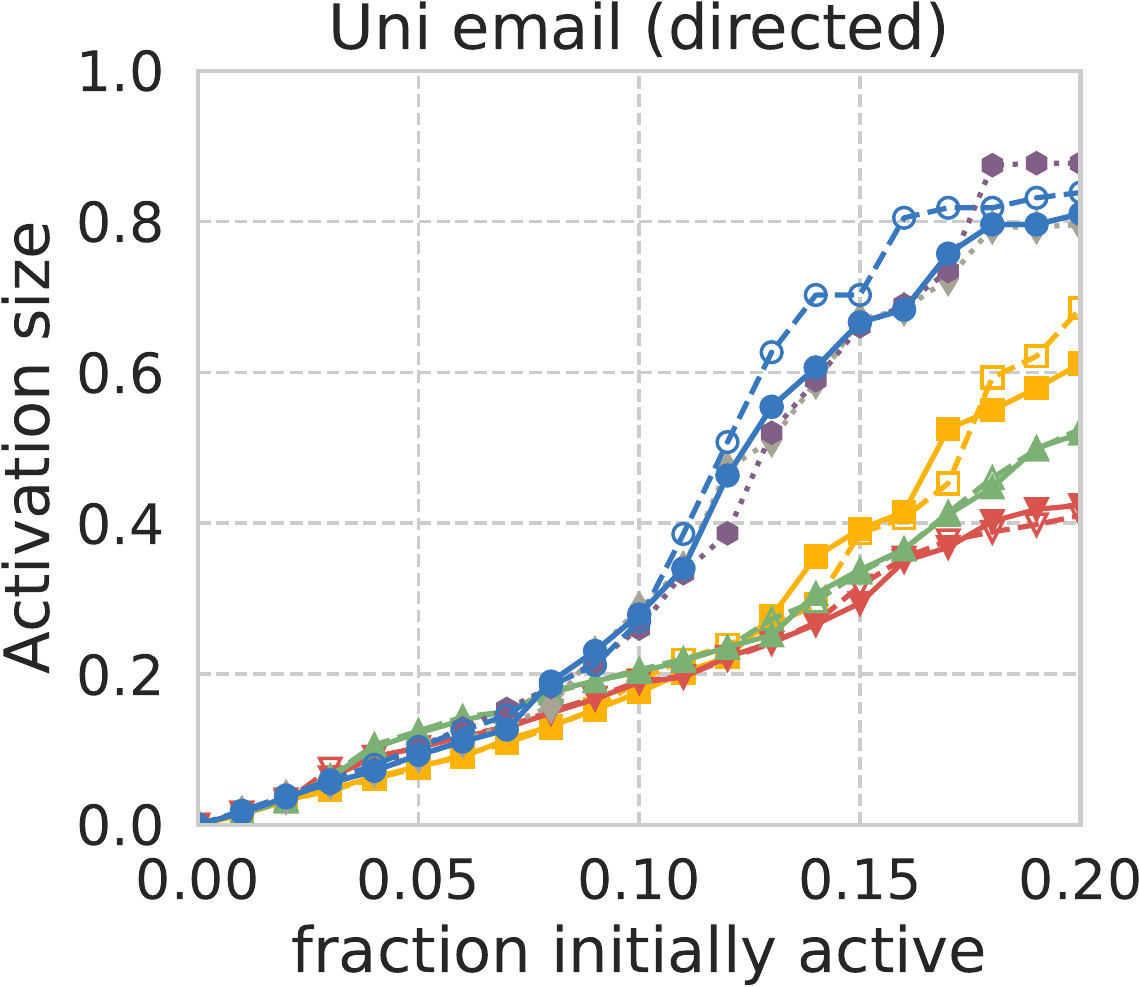}
    }\hfill
    \subfloat[\label{fig:lt-0.6-polblogs}]{%
        \includegraphics[width=.23\textwidth]{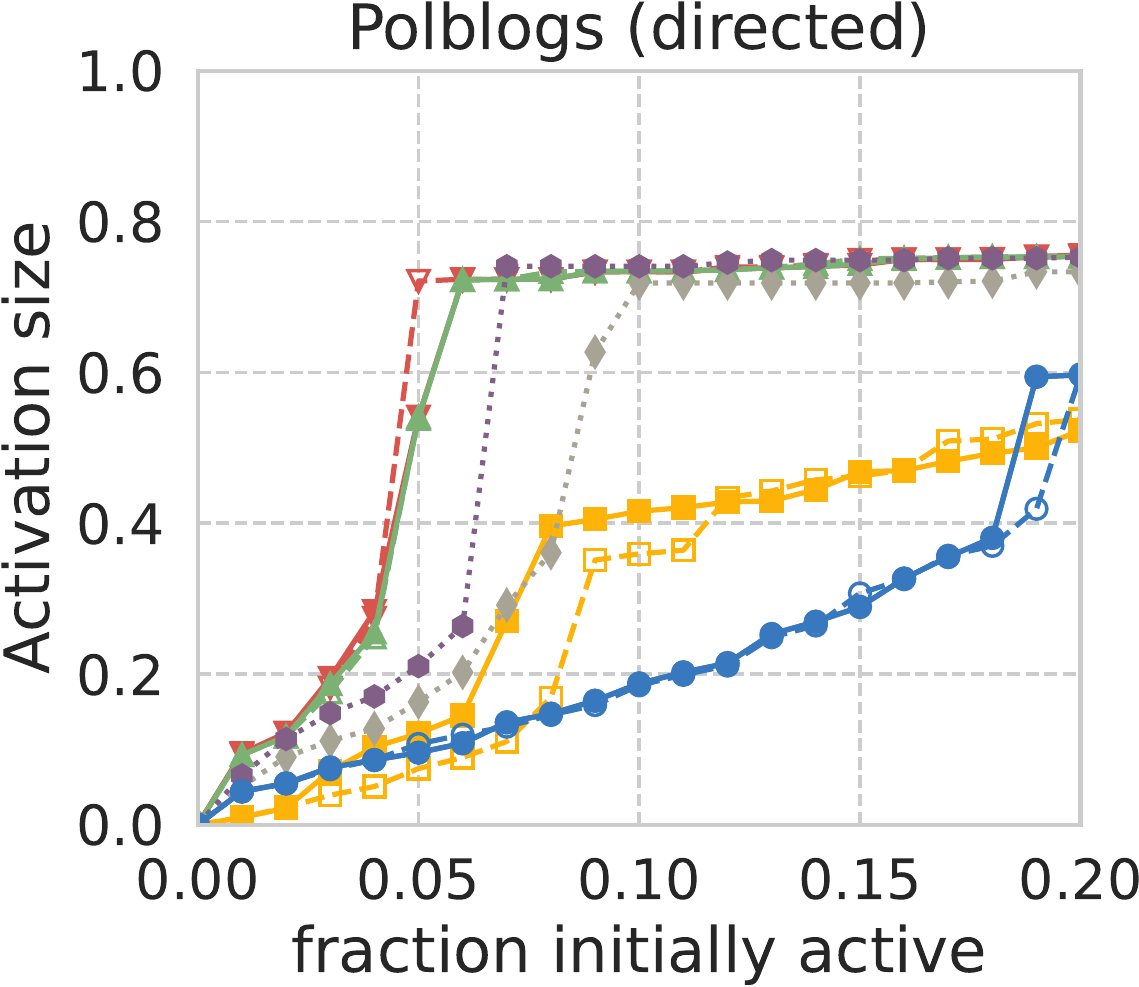}
    }\hfill
    \subfloat[\label{fig:lt-0.6-interactome-yeast}]{%
        \includegraphics[width=.23\textwidth]{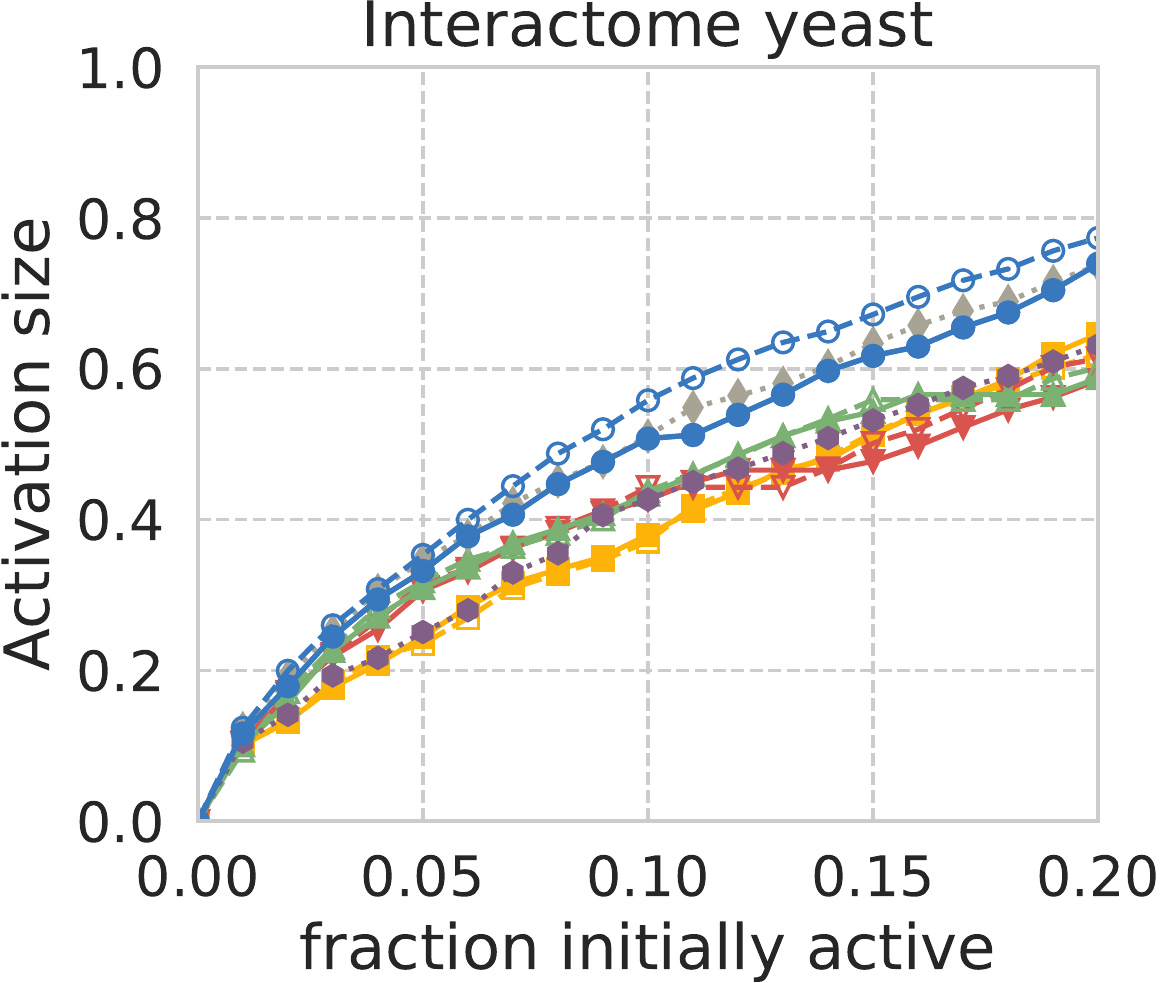}
    }\hfill
    \subfloat[\label{fig:lt-0.6-ego-facebook}]{%
        \includegraphics[width=.23\textwidth]{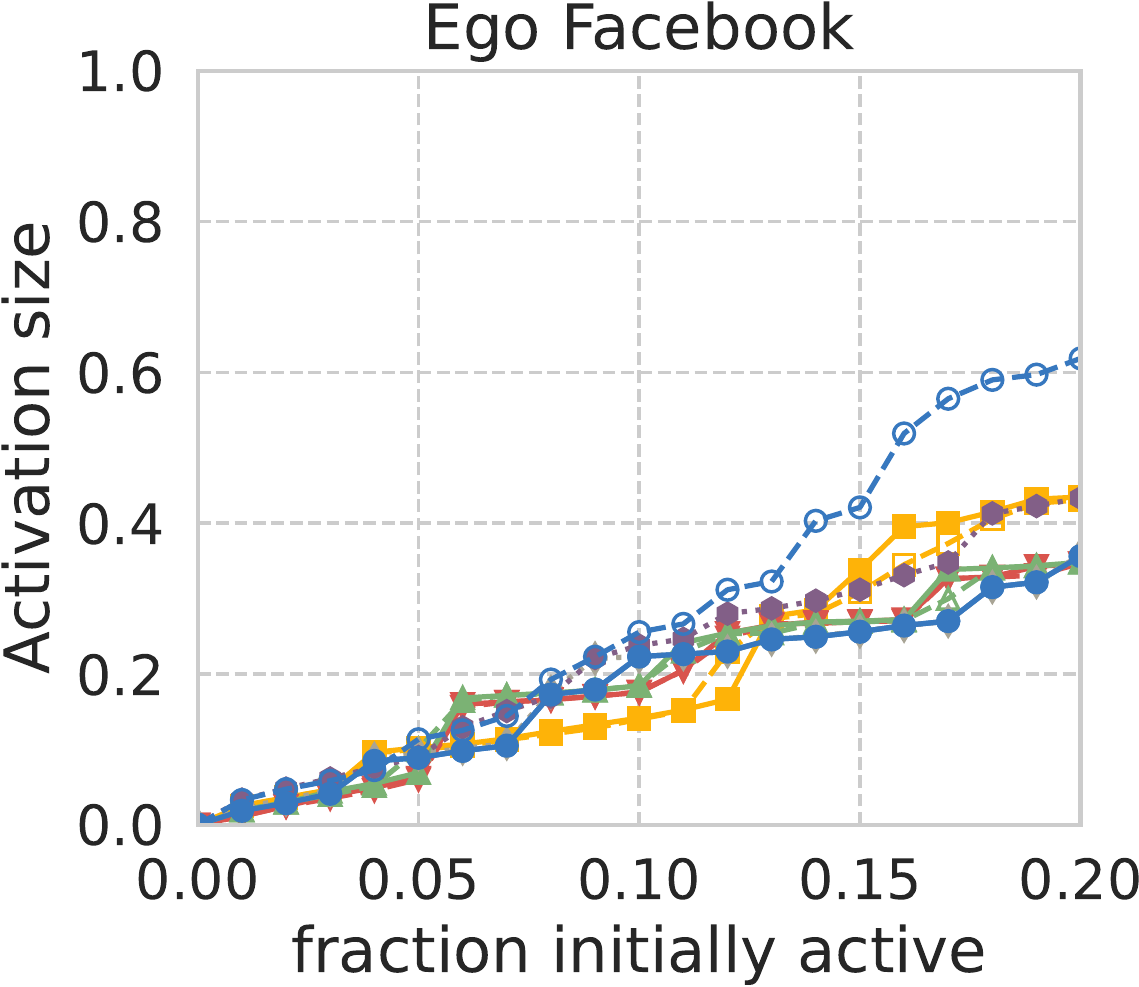}
    }\hfill
    \subfloat[\label{fig:lt-0.6-power}]{%
        \includegraphics[width=.23\textwidth]{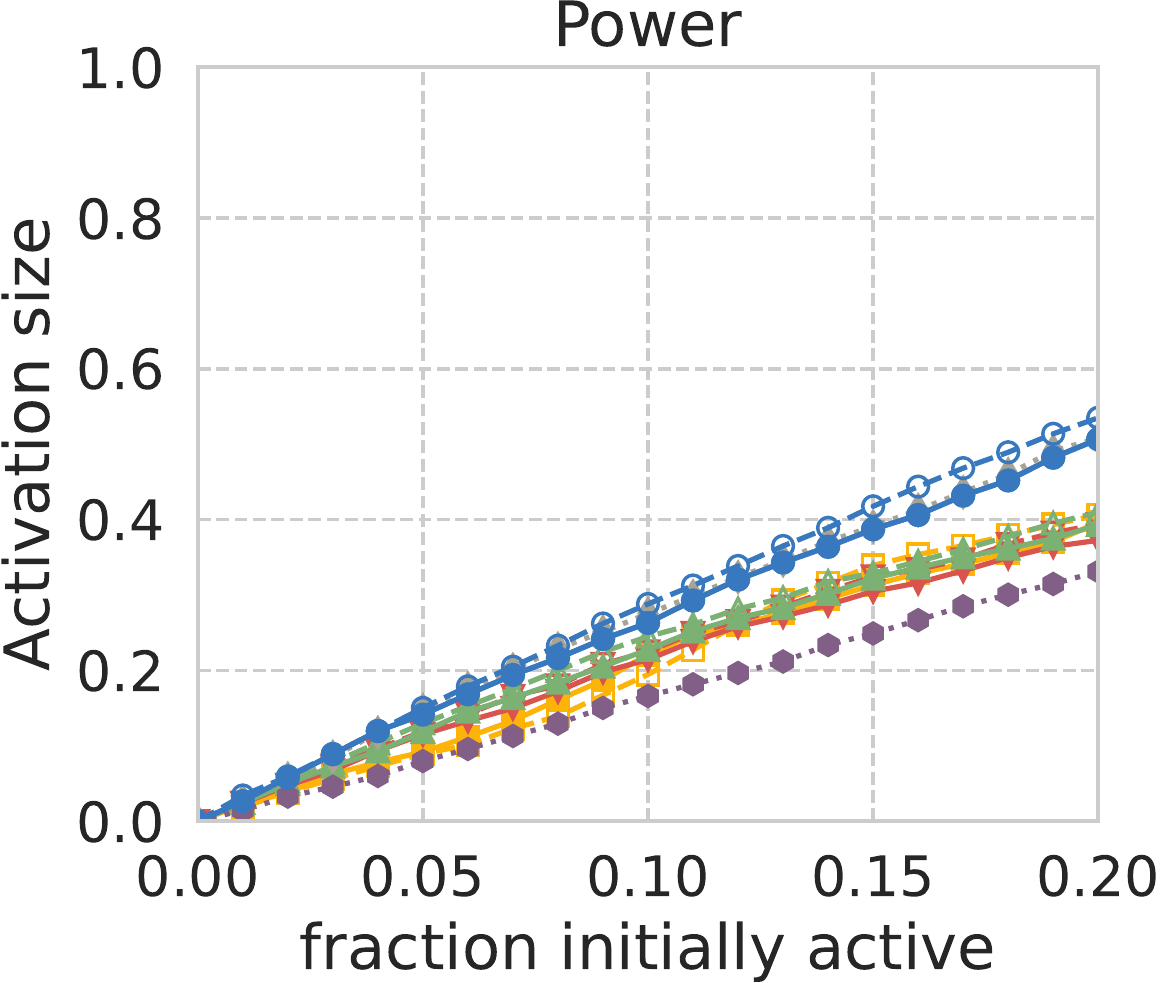}
    }\hfill
    \subfloat[\label{fig:lt-0.6-facebook-organizations}]{%
        \includegraphics[width=.23\textwidth]{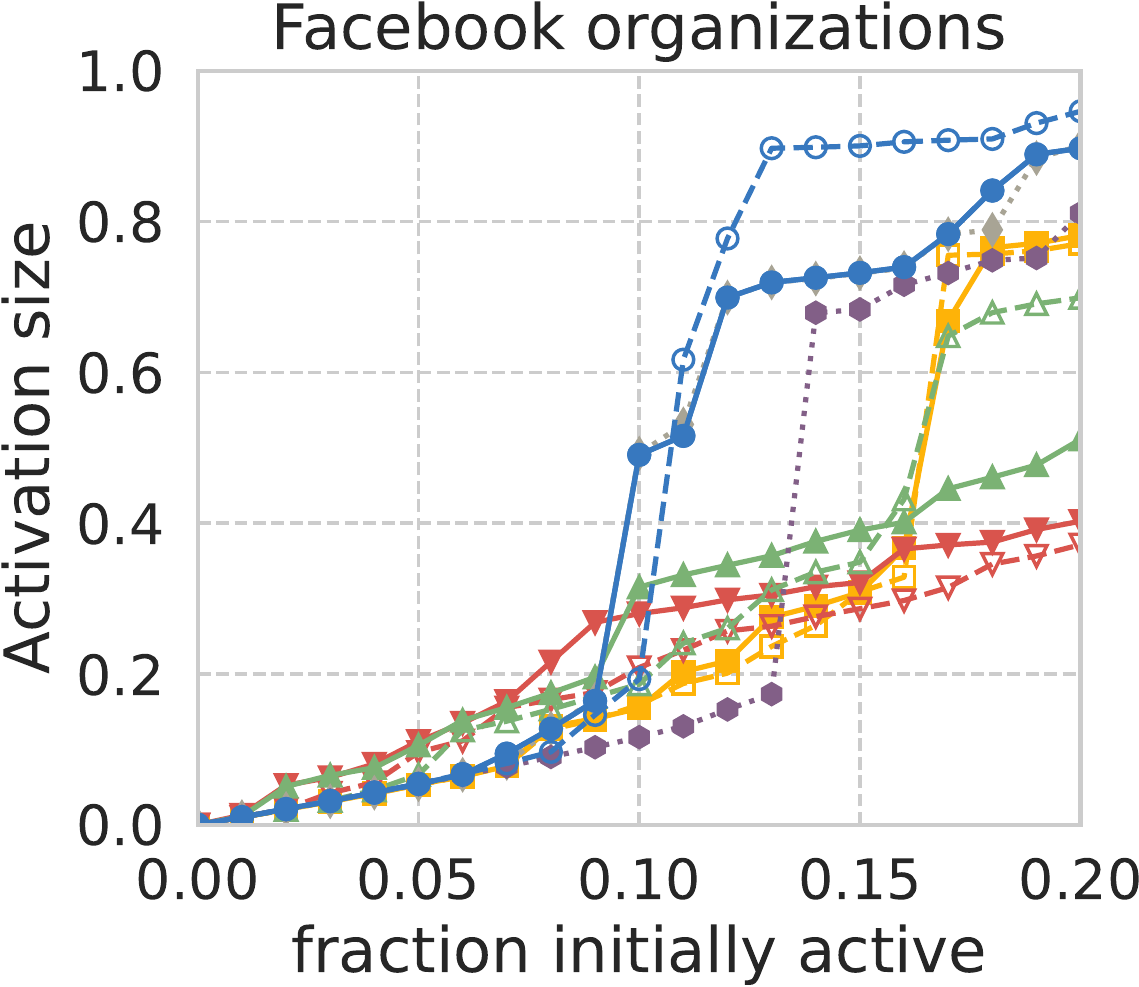}
    }\hfill
    \subfloat[\label{fig:lt-0.6-physics-collaborations}]{%
        \includegraphics[width=.23\textwidth]{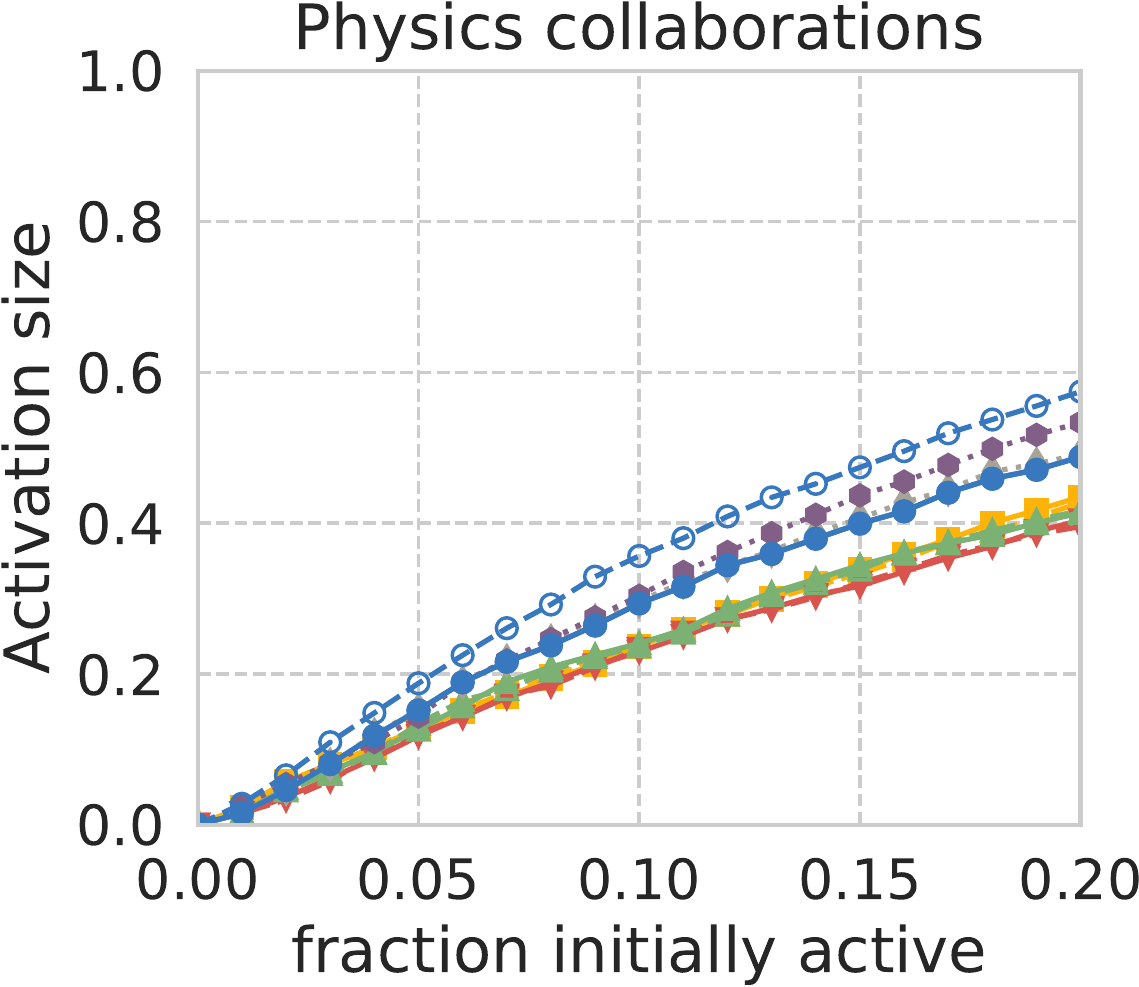}
    }\hfill
    \subfloat[\label{fig:lt-0.6-google}]{%
        \includegraphics[width=.23\textwidth]{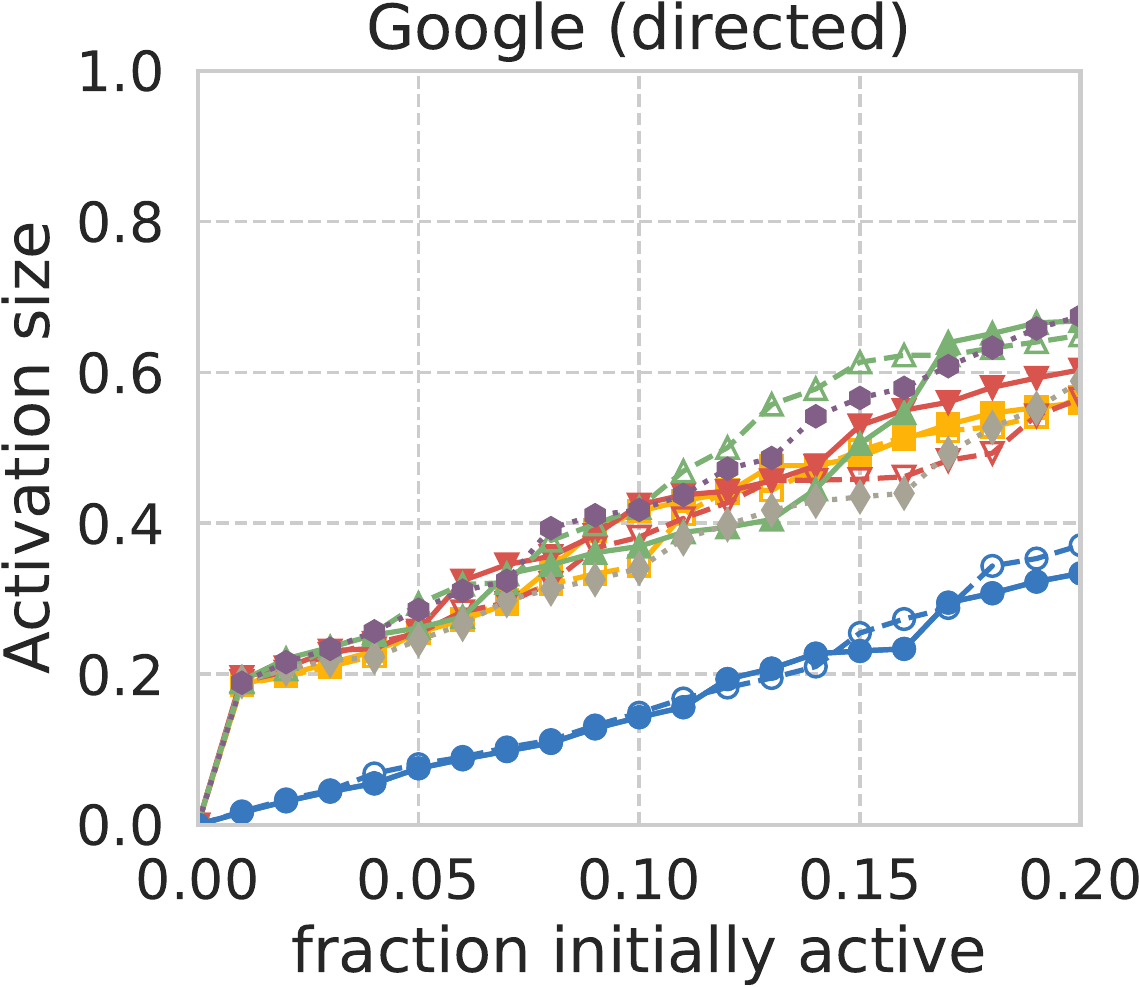}
    }\hfill
    \subfloat[\label{fig:lt-0.6-pgp}]{%
        \includegraphics[width=.23\textwidth]{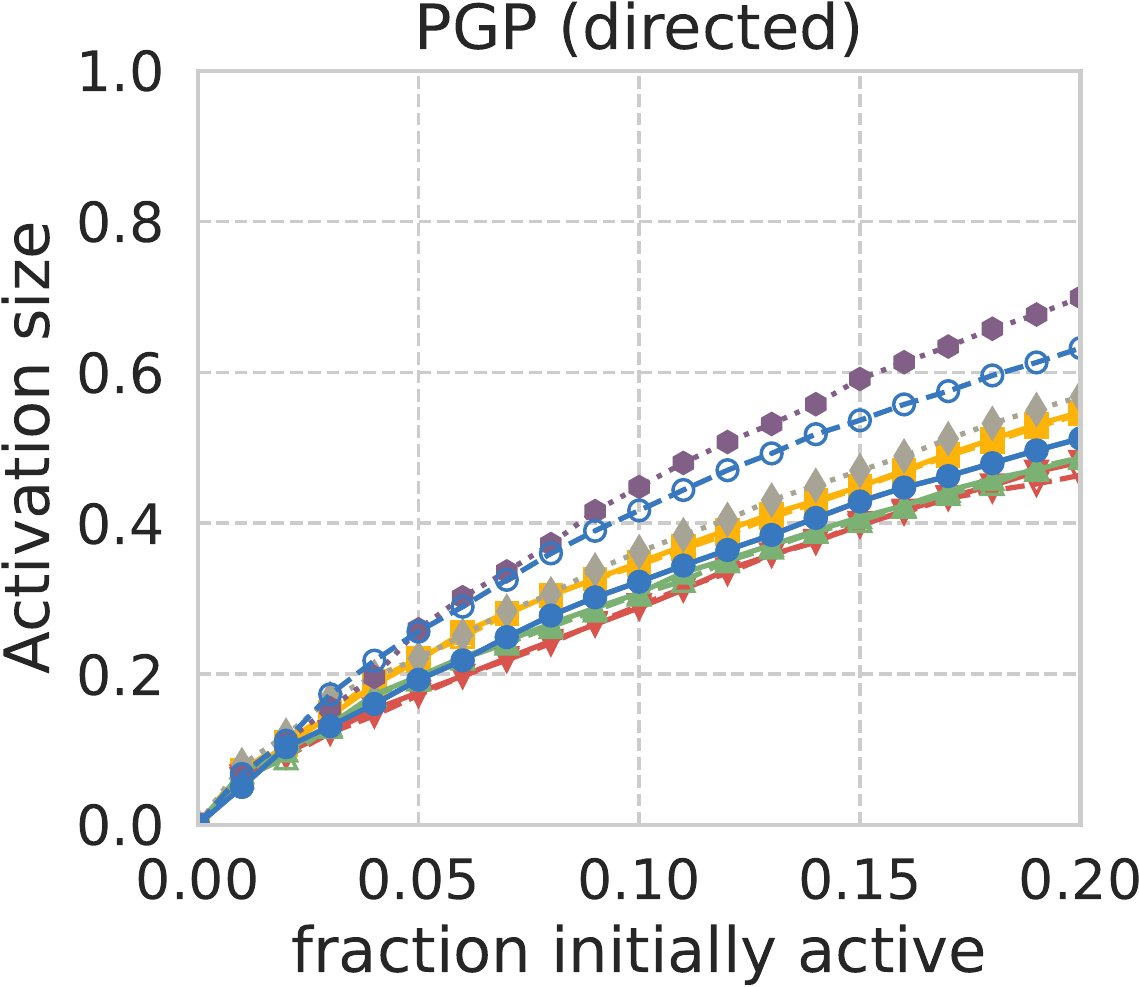}
    }\hfill
    \subfloat[\label{fig:lt-0.6-facebook-wall}]{%
        \includegraphics[width=.23\textwidth]{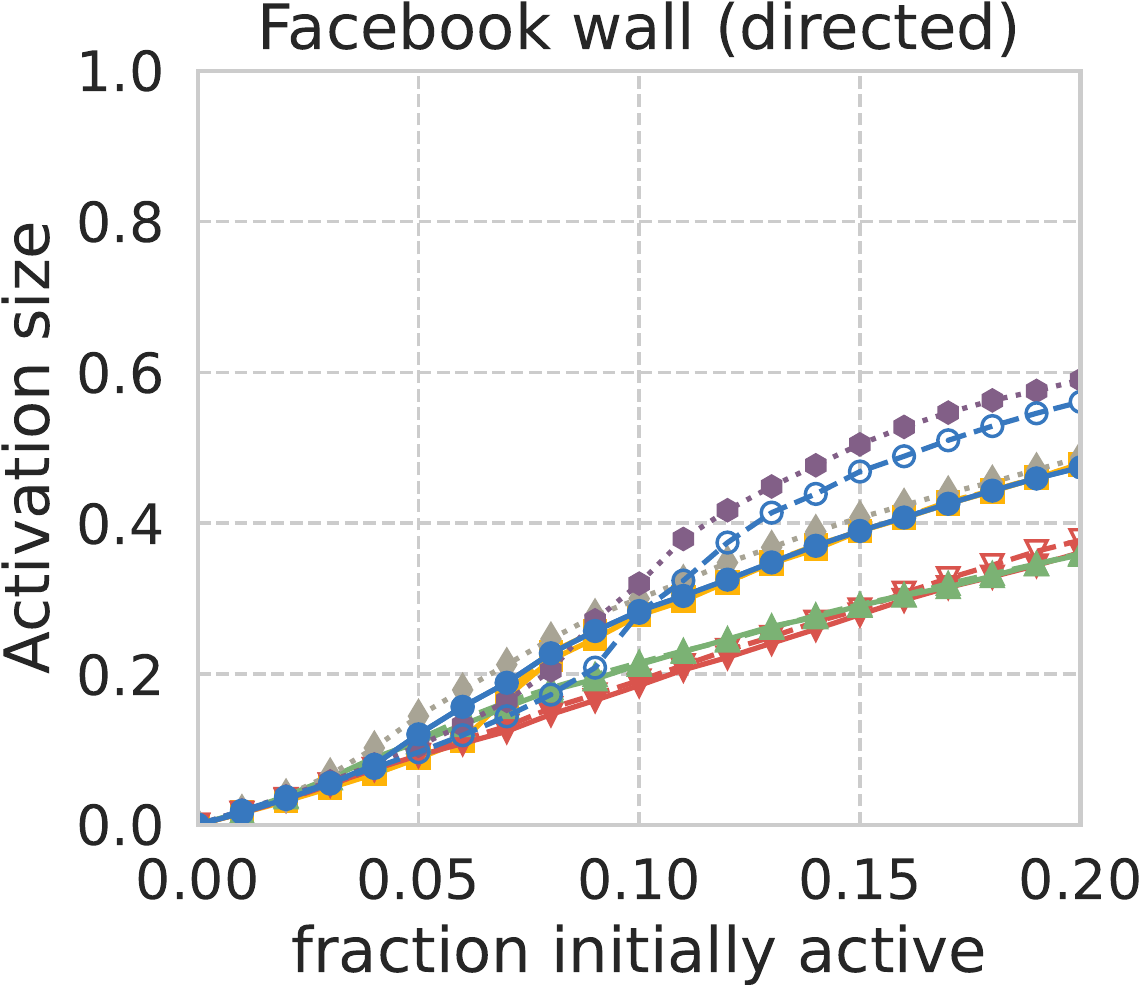}
    }
    \caption{
    Activation size for map equation centrality (MEC), modularity vitality (MV), community hub-bridge (CHB), community-based centrality (CBC), degree centrality (DC), and betweenness centrality (BC) in twelve empirical networks under the linear threshold model with threshold $t'' = 0.6$. Community structures are identified with Infomap; solid lines use the unrecorded link teleportation flow model, dashed lines use recorded node teleportation.
    }
    \label{fig:empirical-results-lt-0.6}
\end{figure*}

Further results for the linear threshold model with thresholds $t' = 0.4$ and $t'' = 0.6$ are shown in \Figref{fig:empirical-results-lt-0.4} and \Figref{fig:empirical-results-lt-0.6}, respectively.


\end{document}